\documentclass[twocolumn,twocolappendix,numberedappendix,appendixfloats]{openjournal}
\usepackage{graphicx}
\usepackage{dcolumn}
\usepackage{bm}
\usepackage[colorlinks,linkcolor=blue,citecolor=blue,urlcolor=blue ]{hyperref}
\usepackage{amsmath,amssymb,natbib,latexsym}
\usepackage[T1]{fontenc}
\usepackage[utf8]{inputenc}

\usepackage{bbm}
\usepackage{float}
\usepackage{url}
\usepackage[normalem]{ulem}
\usepackage{enumerate}
\usepackage[makeroom]{cancel}
\usepackage{array}
\usepackage{booktabs}
\usepackage{mathtools} 
\usepackage{eso-pic}

\usepackage{xcolor}
\usepackage{comment}
\usepackage{xspace}
\usepackage{lineno}

\providecommand{\mnras}{Mon. Not. R. Astron. Soc.}
\providecommand{\apjs}{Astrophys J}
\providecommand{\apjl}{Astrophys J Lett}
\providecommand{\aj}{Astron J}
\providecommand{\aap}{A\&A}
\providecommand{\pasj}{Publ. Astron. Soc. Jpn.}
\providecommand{\jcap}{J. Cosmol. Astropart. Phys.}

\newcommand{\code}[1]{\texttt{#1}\xspace}

\newcommand{\healpix}{\textsc{HEALPix}\xspace}

\newcommand{\nside}{\texttt{nside}\xspace}

\newcommand{\treecorr}{\textsc{TreeCorr}\xspace}
\newcommand{\healsparse}{\textsc{healsparse}\xspace}

\newcommand{\decasu}{\texttt{decasu}\xspace}
\newcommand{\maximask}{\textsc{MaxiMask}\xspace}

\newcommand{\ngal}{$n_{\rm gal}$ }

\newcommand{\griz}{\emph{griz} }
\newcommand{\fracdet}{\texttt{fracdet}\xspace}
\newcommand{\onextwo}{1$\times$2pt\xspace}
\newcommand{\twoxtwo}{2$\times$2pt\xspace}
\newcommand{\threextwo}{3$\times$2pt\xspace}
\newcommand{\maglim}{\textsc{MagLim}\xspace}
\newcommand{\maglimpp}{\textsc{MagLim++}\xspace}
\newcommand{\redmagic}{\textsc{redMaGiC}\xspace}
\newcommand{\isd}{\code{ISD}}
\newcommand{\enet}{\code{ENet}}

\newcommand{\goldfoot}{\textsc{DES Y6 Gold}\xspace}

\newcommand{\seed}{\textsc{seed}\xspace}
\newcommand{\baselinemask}{\maglim \textsc{baseline}\xspace}
\newcommand{\systmask}{\textsc{systematics}\xspace}
\newcommand{\jointmask}{\textsc{joint}\xspace}

\newcommand{\decals}{$\texttt{DECaLS}$\xspace}

\newcommand{\mrm}[1]{\textcolor{blue}{MRM: #1}}

\newcolumntype{Z}{>{\setbox0=\hbox\bgroup}c<{\egroup}@{}}

\newcommand{\appendixcite}[1]{\hyperref[#1]{\textcolor{blue}{Appendix \ref*{#1}}}}

\usepackage{graphicx}	
\usepackage{amsmath}	
\usepackage{comment}
\usepackage{multirow}

\begin{document}

\AddToShipoutPictureBG*{%
	\AtPageUpperLeft{%
		\hspace*{18.25cm}%
		\raisebox{-9.4\baselineskip}{%
			\makebox[0pt][l]{\textnormal{DES-2024-0878}}
			
}}}%

\AddToShipoutPictureBG*{%
	\AtPageUpperLeft{%
		\hspace*{20.45cm} 
		\raisebox{-10.5\baselineskip}{%
			\makebox[0pt][r]{\textnormal{FERMILAB-PUB-25-0630-PPD}}
}}}%

\title[DES Y6 Footprint and Systematics]{Dark Energy Survey Year 6 Results: improved mitigation of spatially varying observational systematics with masking for the \maglimpp lens sample }
\shorttitle{Dark Energy Survey Year 6 Masking Systematics}
\shortauthors{Rodríguez-Monroy \& Weaverdyck}


\author{M.~Rodr\'{i}guez-Monroy}\email{martin.rodriguez@inv.uam.es}
\affiliation{Instituto de F\'{i}sica Te\'{o}rica UAM/CSIC, Universidad Aut\'{o}noma de Madrid, 28049 Madrid, Spain}
\affiliation{Laboratoire de physique des 2 infinis Ir\`ene Joliot-Curie, CNRS Universit\'e Paris-Saclay, Bât. 100, F-91405 Orsay Cedex, France}
\affiliation{Ruhr University Bochum, Faculty of Physics and Astronomy, Astronomical Institute,
German Centre for Cosmological Lensing, 44780 Bochum, Germany}

\author{N.~Weaverdyck}\email{nweaverdyck@lbl.gov}
\affiliation{Lawrence Berkeley National Laboratory, 1 Cyclotron Road, Berkeley, CA 94720, USA}
\affiliation{Berkeley Center for Cosmological Physics, Department of Physics, University of California, Berkeley, CA 94720, USA}

\author{J.~Elvin-Poole}
\affiliation{Department of Physics and Astronomy, University of Waterloo, 200 University Ave W, Waterloo, ON N2L 3G1, Canada}

\author{I.~Sevilla-Noarbe}
\affiliation{Centro de Investigaciones Energ\'eticas, Medioambientales y Tecnol\'ogicas (CIEMAT), Madrid, Spain}

\author{S.~Avila}
\affiliation{Centro de Investigaciones Energ\'eticas, Medioambientales y Tecnol\'ogicas (CIEMAT), Madrid, Spain}

\author{A.~Porredon}
\affiliation{Centro de Investigaciones Energ\'eticas, Medioambientales y Tecnol\'ogicas (CIEMAT), Madrid, Spain}
\affiliation{Ruhr University Bochum, Faculty of Physics and Astronomy, Astronomical Institute, German Centre for Cosmological Lensing, 44780 Bochum, Germany}

\author{A.~Carnero~Rosell}
\affiliation{Instituto de Astrofisica de Canarias, E-38205 La Laguna, Tenerife, Spain}
\affiliation{Laborat\'orio Interinstitucional de e-Astronomia - LIneA, Av. Pastor Martin Luther King Jr, 126 Del Castilho, Nova Am\'erica Offices, Torre 3000/sala 817 CEP: 20765-000, Brazil}
\affiliation{Universidad de La Laguna, Dpto. Astrofísica, E-38206 La Laguna, Tenerife, Spain}

\author{A.~Drlica-Wagner}
\affiliation{Department of Astronomy and Astrophysics, University of Chicago, Chicago, IL 60637, USA}
\affiliation{Fermi National Accelerator Laboratory, P. O. Box 500, Batavia, IL 60510, USA}
\affiliation{Kavli Institute for Cosmological Physics, University of Chicago, Chicago, IL 60637, USA}

\author{D.~Anbajagane}
\affiliation{Kavli Institute for Cosmological Physics, University of Chicago, Chicago, IL 60637, USA}
\affiliation{Department of Astronomy and Astrophysics, University of Chicago, Chicago, IL 60637, USA}

\author{M.~R.~Becker}
\affiliation{Argonne National Laboratory, 9700 South Cass Avenue, Lemont, IL 60439, USA}

\author{K.~Bechtol}
\affiliation{Physics Department, 2320 Chamberlin Hall, University of Wisconsin-Madison, 1150 University Avenue Madison, WI  53706-1390}

\author{M.~Crocce}
\affiliation{Institut d'Estudis Espacials de Catalunya (IEEC), 08034 Barcelona, Spain}
\affiliation{Institute of Space Sciences (ICE, CSIC),  Campus UAB, Carrer de Can Magrans, s/n,  08193 Barcelona, Spain}

\author{A.~Fert{\'e}}
\affiliation{SLAC National Accelerator Laboratory, Menlo Park, CA, USA}

\author{M.~Gatti}
\affiliation{Kavli Institute for Cosmological Physics, University of Chicago, Chicago, IL 60637, USA}

\author{J.~Mena-Fern{\'a}ndez}
\affiliation{Universit\'e Grenoble Alpes, CNRS, LPSC-IN2P3, 38000 Grenoble, France}

\author{D.~Sanchez Cid}
\affiliation{Centro de Investigaciones Energ\'eticas, Medioambientales y Tecnol\'ogicas (CIEMAT), Madrid, Spain}
\affiliation{Physik-Institut, University of Zürich, Winterthurerstrasse 190, CH-8057 Zürich, Switzerland}

\author{M.~Yamamoto}
\affiliation{Department of Astrophysical Sciences, Princeton University, Peyton Hall, Princeton, NJ 08544, USA}
\affiliation{Department of Physics, Duke University Durham, NC 27708, USA}

\author{M.~Aguena}
\affiliation{INAF-Osservatorio Astronomico di Trieste, via G. B. Tiepolo 11, I-34143 Trieste, Italy}
\affiliation{Laborat\'orio Interinstitucional de e-Astronomia - LIneA, Av. Pastor Martin Luther King Jr, 126 Del Castilho, Nova Am\'erica Offices, Torre 3000/sala 817 CEP: 20765-000, Brazil}

\author{S.~S.~Allam}
\affiliation{Fermi National Accelerator Laboratory, P. O. Box 500, Batavia, IL 60510, USA}

\author{O.~Alves}
\affiliation{Department of Physics, University of Michigan, Ann Arbor, MI 48109, USA}

\author{F.~Andrade-Oliveira}
\affiliation{Physik-Institut, University of Zürich, Winterthurerstrasse 190, CH-8057 Zürich, Switzerland}

\author{D.~Bacon}
\affiliation{Institute of Cosmology and Gravitation, University of Portsmouth, Portsmouth, PO1 3FX, UK}

\author{J.~Blazek}
\affiliation{Department of Physics, Northeastern University, Boston, MA 02115, USA}

\author{S.~Bocquet}
\affiliation{University Observatory, Faculty of Physics, Ludwig-Maximilians-Universit\"at, Scheinerstr. 1, 81679 Munich, Germany}

\author{D.~Brooks}
\affiliation{Department of Physics \& Astronomy, University College London, Gower Street, London, WC1E 6BT, UK}

\author{R.~Camilleri}
\affiliation{School of Mathematics and Physics, University of Queensland,  Brisbane, QLD 4072, Australia}

\author{J.~Carretero}
\affiliation{Institut de F\'{\i}sica d'Altes Energies (IFAE), The Barcelona Institute of Science and Technology, Campus UAB, 08193 Bellaterra (Barcelona) Spain}

\author{R.~Cawthon}
\affiliation{Physics Department, William Jewell College, Liberty, MO, 64068}

\author{C.~Chang}
\affiliation{Department of Astronomy and Astrophysics, University of Chicago, Chicago, IL 60637, USA}
\affiliation{Kavli Institute for Cosmological Physics, University of Chicago, Chicago, IL 60637, USA}

\author{L.~N.~da Costa}
\affiliation{Laborat\'orio Interinstitucional de e-Astronomia - LIneA, Av. Pastor Martin Luther King Jr, 126 Del Castilho, Nova Am\'erica Offices, Torre 3000/sala 817 CEP: 20765-000, Brazil}

\author{M.~E.~da Silva Pereira}
\affiliation{Hamburger Sternwarte, Universit\"{a}t Hamburg, Gojenbergsweg 112, 21029 Hamburg, Germany}

\author{S.~Desai}
\affiliation{Department of Physics, IIT Hyderabad, Kandi, Telangana 502285, India}

\author{H.~T.~Diehl}
\affiliation{Fermi National Accelerator Laboratory, P. O. Box 500, Batavia, IL 60510, USA}

\author{P.~Doel}
\affiliation{Department of Physics \& Astronomy, University College London, Gower Street, London, WC1E 6BT, UK}

\author{C.~Doux}
\affiliation{Department of Physics and Astronomy, University of Pennsylvania, Philadelphia, PA 19104, USA}
\affiliation{Universit\'e Grenoble Alpes, CNRS, LPSC-IN2P3, 38000 Grenoble, France}

\author{S.~Everett}
\affiliation{California Institute of Technology, 1200 East California Blvd, MC 249-17, Pasadena, CA 91125, USA}

\author{B.~Flaugher}
\affiliation{Fermi National Accelerator Laboratory, P. O. Box 500, Batavia, IL 60510, USA}

\author{P.~Fosalba}
\affiliation{Institute of Space Sciences (ICE, CSIC), Campus UAB, Carrer de Can Magrans, s/n, 08193 Barcelona, Spain}
\affiliation{Institut d’Estudis Espacials de Catalunya (IEEC), Edifici RDIT, Campus UPC, 08860 Castelldefels (Barcelona), Spain}

\author{J.~Frieman}
\affiliation{Fermi National Accelerator Laboratory, P. O. Box 500, Batavia, IL 60510, USA}
\affiliation{Kavli Institute for Cosmological Physics, University of Chicago, Chicago, IL 60637, USA}

\author{J.~Garc\'ia-Bellido}
\affiliation{Instituto de Fisica Teorica UAM/CSIC, Universidad Autonoma de Madrid, 28049 Madrid, Spain}

\author{R.~A.~Gruendl}
\affiliation{Center for Astrophysical Surveys, National Center for Supercomputing Applications, 1205 West Clark St., Urbana, IL 61801, USA}
\affiliation{Department of Astronomy, University of Illinois at Urbana-Champaign, 1002 W. Green Street, Urbana, IL 61801, USA}

\author{K.~Herner}
\affiliation{Fermi National Accelerator Laboratory, P. O. Box 500, Batavia, IL 60510, USA}

\author{S.~R.~Hinton}
\affiliation{School of Mathematics and Physics, University of Queensland,  Brisbane, QLD 4072, Australia}

\author{D.~L.~Hollowood}
\affiliation{Santa Cruz Institute for Particle Physics, Santa Cruz, CA 95064, USA}

\author{K.~Honscheid}
\affiliation{Center for Cosmology and Astro-Particle Physics, The Ohio State University, Columbus, OH 43210, USA}
\affiliation{Department of Physics, The Ohio State University, Columbus, OH 43210, USA}

\author{D.~Huterer}
\affiliation{Department of Physics, University of Michigan, Ann Arbor, MI 48109, USA}

\author{D.~J.~James}
\affiliation{Center for Astrophysics $\vert$ Harvard \& Smithsonian, 60 Garden Street, Cambridge, MA 02138, USA}

\author{K.~Kuehn}
\affiliation{Australian Astronomical Optics, Macquarie University, North Ryde, NSW 2113, Australia}
\affiliation{Lowell Observatory, 1400 Mars Hill Rd, Flagstaff, AZ 86001, USA}

\author{O.~Lahav}
\affiliation{Department of Physics \& Astronomy, University College London, Gower Street, London, WC1E 6BT, UK}

\author{S.~Lee}
\affiliation{Department of Physics and Astronomy, Ohio University, Clippinger Labs, Athens, OH 45701}
\affiliation{Jet Propulsion Laboratory, California Institute of Technology, 4800 Oak Grove Dr., Pasadena, CA 91109, USA}
\affiliation{California Institute of Technology, 1200 East California Blvd, MC 249-17, Pasadena, CA 91125, USA}

\author{J.~L.~Marshall}
\affiliation{George P. and Cynthia Woods Mitchell Institute for Fundamental Physics and Astronomy, and Department of Physics and Astronomy, Texas A\&M University, College Station, TX 77843,  USA}

\author{R.~Miquel}
\affiliation{Instituci\'o Catalana de Recerca i Estudis Avan\c{c}ats, E-08010 Barcelona, Spain}
\affiliation{Institut de F\'{\i}sica d'Altes Energies (IFAE), The Barcelona Institute of Science and Technology, Campus UAB, 08193 Bellaterra (Barcelona) Spain}

\author{J.~Muir}
\affiliation{Department of Physics, University of Cincinnati, Cincinnati, Ohio 45221, USA}
\affiliation{Perimeter Institute for Theoretical Physics, 31 Caroline St. North, Waterloo, ON N2L 2Y5, Canada}

\author{J.~Myles}
\affiliation{Department of Astrophysical Sciences, Princeton University, Peyton Hall, Princeton, NJ 08544, USA}

\author{R.~L.~C.~Ogando}
\affiliation{Centro de Tecnologia da Informa\c{c}\~ao Renato Archer, Campinas, SP, Brazil - 13069-901 Observat\'orio Nacional, Rio de Janeiro, RJ, Brazil - 20921-400}

\author{A.~A.~Plazas~Malag\'on}
\affiliation{Kavli Institute for Particle Astrophysics \& Cosmology, P. O. Box 2450, Stanford University, Stanford, CA 94305, USA}
\affiliation{SLAC National Accelerator Laboratory, Menlo Park, CA 94025, USA}

\author{J.~Prat}
\affiliation{Department of Astronomy and Astrophysics, University of Chicago, Chicago, IL 60637, USA}
\affiliation{Nordita, KTH Royal Institute of Technology and Stockholm University, Hannes Alfv\'ens v\"ag 12, SE-10691 Stockholm, Sweden}

\author{E.~Sanchez}
\affiliation{Centro de Investigaciones Energ\'eticas, Medioambientales y Tecnol\'ogicas (CIEMAT), Madrid, Spain}

\author{T.~Schutt}
\affiliation{Department of Physics, Stanford University, 382 Via Pueblo Mall, Stanford, CA 94305, USA}
\affiliation{Kavli Institute for Particle Astrophysics \& Cosmology, P. O. Box 2450, Stanford University, Stanford, CA 94305, USA}
\affiliation{SLAC National Accelerator Laboratory, Menlo Park, CA 94025, USA}

\author{M.~Smith}
\affiliation{Physics Department, Lancaster University, Lancaster, LA1 4YB, UK}

\author{E.~Suchyta}
\affiliation{Computer Science and Mathematics Division, Oak Ridge National Laboratory, Oak Ridge, TN 37831}

\author{M.~E.~C.~Swanson}
\affiliation{Center for Astrophysical Surveys, National Center for Supercomputing Applications, 1205 West Clark St., Urbana, IL 61801, USA}

\author{C.~To}
\affiliation{Department of Astronomy and Astrophysics, University of Chicago, Chicago, IL 60637, USA}

\author{M.~A.~Troxel}
\affiliation{Department of Physics, Duke University Durham, NC 27708, USA}

\author{D.~L.~Tucker}
\affiliation{Fermi National Accelerator Laboratory, P. O. Box 500, Batavia, IL 60510, USA}

\author{V.~Vikram}
\affiliation{Central University of Kerala, Kasaragod, Kerala, India}

\collaboration{DES Collaboration}

\date{\today}

\begin{abstract}
As photometric surveys reach unprecedented statistical precision, systematic uncertainties increasingly dominate large-scale structure probes relying on galaxy number density. Defining the final survey footprint is critical, as it excludes regions affected by artefacts or suboptimal observing conditions. For galaxy clustering, spatially varying observational systematics, such as seeing, are a leading source of bias. Template maps of contaminants are used to derive spatially dependent corrections, but extreme values may fall outside the applicability range of mitigation methods, compromising correction reliability. The complexity and accuracy of systematics modelling depend on footprint conservativeness, with aggressive masking enabling simpler, robust mitigation. We present a unified approach to define the DES Y6 \textsc{joint} footprint, integrating observational systematics templates and artefact indicators that degrade mitigation performance. This removes extreme values from an initial \textsc{seed} footprint, leading to the final \textsc{joint} footprint. By evaluating the DES Y6 lens sample \textsc{MagLim++} on this footprint, we enhance the Iterative Systematics Decontamination (\texttt{ISD}) method, detecting non-linear systematic contamination and improving correction accuracy. While the mask’s impact on clustering is less significant than systematics decontamination, it remains non-negligible, comparable to statistical uncertainties in certain $w(\theta)$ scales and redshift bins. Supporting coherent analyses of galaxy clustering and cosmic shear, the final footprint spans $4,031.04 \rm \, deg^2$, setting the basis for DES Y6 1$\times$2pt, 2$\times$2pt, and 3$\times$2pt analyses. This work highlights how targeted masking strategies optimise the balance between statistical power and systematic control in Stage-III and -IV surveys.
\end{abstract}

\maketitle


\section{Introduction }\label{sec:intro}
The advent of large galaxy surveys over the past few decades has enabled us to study the evolution and dynamics of the Universe through its large-scale structure (LSS). Comparing low-redshift observations of the LSS (\citet{y3_3x2}) and geometric probes (\citet{y3clusters, y5sn, y6bao, desi_bao}) with early-Universe data from cosmic microwave background (CMB) experiments (\citet{Planck:2018vyg,atc}) enables tight constraints on the Universe's expansion history and the growth of cosmic structures. The combination of these observations allows even tighter constraints on the nature of cosmic acceleration and powerful tests of general relativity (\citet{desmultiprobes, y3extensions, y6baosn, desi_extended}).\\
\\
A primary reason for the increased precision in recent cosmological analyses based on LSS measurements is the vast amount of data collected by Stage-III and IV galaxy surveys, which cover wider areas while reaching greater depths. In this context, spectroscopic surveys, such as DESI (\citet{Dey_2019}), obtain three-dimensional information from millions of objects with exquisite precision, whereas photometric surveys like the Dark Energy Survey (\citet{DES:2005dhi}), the Kilo-Degree Survey (KiDS, \citet{2013Msngr.154...44D}), or the Hyper SuprimeCam Subaru Strategic Program (HSC-SSP, \citet{2018PASJ...70S...4A}) rely on the more challenging estimates of photometric redshifts. In turn, photometric surveys measure galaxy positions and shapes simultaneously, providing measurements of photometric galaxy clustering (\citet{y3weights}), cosmic shear (\citet{y3shear_1, y3shear_2}), and galaxy-galaxy lensing (\citet{y3gglensing}) from the same dataset. These measurements can be combined into the \twoxtwo (galaxy clustering + galaxy-galaxy lensing) and \threextwo (galaxy clustering + galaxy-galaxy lensing + cosmic shear) probes to set tighter constraints on cosmological parameters while breaking degeneracies and self-calibrating systematic effects (as demonstrated by DES: \citet{y3_3x2, y3_2x2_maglim, y3_2x2_redmagic}). Photometric galaxy clustering can also measure the transverse baryon acoustic oscillation (BAO) scale. DES obtained the most precise BAO measurement from a photometric survey, with an uncertainty comparable to that of spectroscopic surveys (\citet{y6bao}). Moreover, ongoing and  upcoming surveys —such as Euclid (\citet{2025A&A...697A...1E}), the Vera C. Rubin Observatory's Legacy Survey of Space and Time (LSST) (\citet{LSST:2008ijt}) and the Nancy Grace Roman Space Telescope (\citet{2015arXiv150303757S})— promise even more precise measurements, further constraining the $\Lambda$CDM model or potentially revealing new physics beyond it. \\
\\ 
As statistical uncertainties decrease, systematic effects become increasingly dominant in galaxy surveys. To mitigate their impact, the primary mitigation strategy involves defining a footprint that maximises the observed area while avoiding problematic regions of the sky, such as the Galactic plane. This footprint also accounts for the survey's observing strategy, for example, the depth reached with each photometric band (see, e.g., \citet{2015arXiv150900870R, 2019A&A...625A...2K, y1gold, y3gold, y6gold}). The next stage involves the definition of an angular mask to remove image artefacts (e.g., dead pixels or satellite streaks), astrophysical foregrounds (e.g., bright stars or globular clusters) and other observational effects. Additionally, the properties of the specific galaxy sample may impose further cuts on the observed area. For example, the angular mask applied to the galaxy samples in the DES Year 3 (Y3) analysis, namely \maglim (\citet{y3maglim}) and \redmagic (\citet{DES:2015pcw}), imposed further cuts to account for specific data selections (see \citet{y3weights}). \\
\\
The impact of these effects on the measurement of different cosmological probes, such as the photometric galaxy clustering angular correlation function, $w(\theta)$, and their removal via an angular mask have been thoroughly studied (see e.g., \citet{2011MNRAS.417.1350R}) and proven to be critical to obtaining unbiased results. \\
\\
However, Stage-IV surveys demand more flexible and adaptive strategies. DESI, for example, masks primarily regions with clearly compromised imaging or extreme values in the tails of imaging systematics maps (e.g., \citet{Kitanidis:2019rzi, desi_lss_catalogue, desi_target_selection, desi_sample}). Because DESI is optimised for BAO measurements–-a large-scale feature relatively robust to observational systematics–-their approach prioritises retaining survey area rather than the more aggressive masking strategy adopted here. Furthermore as a spectroscopic survey, DESI has precision measurements of both angular and radial modes with the latter providing additional information that dilutes the impact of the (largely angular) observational systematics that we address here. KiDS and HSC instead use static masking tailored to their depths and primary science cases, removing image artefacts and regions around bright stars \citet{Wright_2025, 2018PASJ...70S...4A}, but they do not flag regions based on extremity across a range of survey properties. Because both surveys have been used primarily for cosmic shear, where spurious density variations modulate only the uncertainty of the signal rather than biasing it (as they do for galaxy clustering), uniformity of galaxy selection across the footprint is a lower priority than for clustering samples. This is reflected in the much larger variation of organised-random densities (and thus weights) for the KiDS-Legacy sample compared with \maglimpp{} \citet{Johnston:2020pmu, Yan_2025}. While HSC's deeper imaging and brighter sample magnitude limit yield smaller selection inhomogeneity \citet{Li_2022}, this still must be corrected for when used for galaxy clustering, such as in \citet{hsc_clustering, y1weights} via linear mode de-projection, supplemented by manual removal of survey-property regions whose relation with galaxy density appeared visibly non-linear.
Designed for \threextwo{} analyses, the unified footprint presented here ensures consistency across probes. We introduce an adaptive masking approach that combines dynamic outlier rejection in both the 1-dimensional and multi-dimensional spaces of observational systematics, improving the detection of complex contamination patterns that may be overlooked by linear methods (e.g., DES Y3), as discussed later.

In the context of galaxy clustering, a dominant source of systematic contamination at current statistical precision are the so-called spatially varying observational systematics, or ``observational systematics'' for short. These systematic effects alter the galaxy selection function by modifying galaxy brightness or by obscuring them, causing the observed galaxy number density to differ from its true value. Observational systematics encompass a wide variety of effects with different natures, from those related with the observing conditions (e.g., average seeing per sky region), to survey-strategy effects (e.g., exposure time), and astrophysical foregrounds, such as stellar density or dust extinction. \\
\\
Observational systematics are typically characterised by pixel maps that track spatial variations in survey imaging conditions. These pixel maps serve as contamination templates for decontamination methods, which evaluate their spatial correlation with the galaxy density (see, e.g., \citet{2011MNRAS.417.1350R, Leistedt:2014wia, 2020MNRAS.495.1613R, Johnston:2020pmu, 2023A&A...675A.202V, y1weights, Weaverdyck_2021, y3weights}). Each method uses specific metrics to determine the impact of these templates, generally fitting the relation between their values and the observed galaxy number densities to infer a correction (e.g., by defining corrective weights as the inverses of those fits). In this context, extreme values in template maps can degrade the decontamination performance by reducing fit quality and correction reliability. Such values often correspond to unmasked image artefacts or other residual observational effects that persist after footprint definition and should ideally be addressed during earlier data-processing stages. To mitigate this, the DES Y1 analysis masked regions with galaxy number density fluctuations exceeding 20\% (\citet{y1weights}). For DES Y3, decontamination quality was ensured by evaluating the distributions of $\chi^2$ from fits of the observed galaxy number density to the observing conditions (\citet{y3weights}). \\
\\
This paper presents the Year 6 (Y6, which contains all DES data, from Y1 to its final observations) approach, which defines a new procedure to evaluate and mask extreme contamination values in observational systematic template maps. This novel approach allows us to upgrade the \code{Iterative Systematics Decontamination} (\isd) method relative to the Y3 implementation by incorporating up to third-order polynomial fits, thereby improving contamination detection and mitigation. In this work, we present how extreme values affect the performance of such cubic fits, while in the companion paper \citet*{y6weights} we study the improvement on the detection of some contaminants thanks to the usage of the cubic fits, which in turn are only reliable thanks to the application of the \jointmask footprint. \\
\\
In addition to addressing extreme observational systematic values that might degrade the decontamination performance, the final DES Y6 \jointmask footprint presented in this work is designed to incorporate cuts specific to both lens and source (\citet{y6shearcat}) galaxy samples used, facilitating their combined use. This \jointmask footprint is used for the Y6 \onextwo (\citet{y6shear}), \twoxtwo (\citet{y6_2x2}) and \threextwo (\citet{y6_3x2}) analyses. \\
\\
This paper is organised as follows: Section \ref{sec:inputs} introduces the individual footprints and masks, alongside the template maps characterising observational systematic effects, which are used to construct the final \jointmask footprint. Section \ref{sec:mask_creation} describes how the intermediate footprints and masks, namely the \seed footprint and the \baselinemask and \systmask masks, are defined and combined to form the \jointmask footprint. Section~\ref{sec:mask_impact} evaluates the impact of varying masking levels on the DES Y6 decontamination methods, demonstrating how our masking approach improves contaminant detection. This section also quantifies the effect of masking on the galaxy clustering correlation function, showing it to be subdominant compared to the correction applied via weights (\citet*{y6weights}). Finally, Section \ref{sec:conclusions} presents our conclusions. 

\section{Inputs for the creation of the DES Y6 \jointmask footprint }\label{sec:inputs}
In this section, we present the components used to construct the DES Y6 \jointmask footprint. These components address potential sources of systematic effects across the full \threextwo analysis. The first component is the footprint associated with the DES Y6 Gold catalogue (\citet{y6gold}) from which we remove some regions identified as problematic \emph{a posteriori}. Secondly, we incorporate an intermediate footprint derived as a by-product of the DES Y6 shape catalogue analysis (\citet{y6shearcat}), which is built on top of the DES Y6 Gold footprint. The third component comprises a set of contamination template maps that we use to identify and mask extreme observational systematic values in this work, as well as to define corrective weights for galaxy clustering in \citet*{y6weights}. We employ and combine these components at various stages of our analysis, as described in Section \ref{sec:mask_creation}.

\subsection{Terminology used in this work: \emph{footprints}, \emph{masks} and \emph{maps} }\label{sec:terms}
Before describing the inputs used to construct the final DES Y6 \jointmask footprint, we clarify the terminology employed to distinguish between the terms \emph{footprint}, \emph{mask} and \emph{map}. In all three cases, we work with pixelised objects defined using the \healpix\footnote{\url{https://healpix.sourceforge.io/}} (\citet{healpix}) scheme at a given \nside pixel resolution. We use the term \emph{footprint} when referring to an area of the sky defined by a set of pixels. Pixels are removed by means of a \emph{mask}, a full-sky binary object indicating whether a pixel at a given \nside should be retained or excluded. Our criterion is that a pixel is retained in the footprint if the corresponding \emph{mask} value is \code{False} and it is excluded if the \emph{mask} value is \code{True}. In this context, a \emph{footprint} has an associated \emph{mask} —its full-sky boolean representation— where pixels are \code{False} if contained within the footprint and \code{True} otherwise. Thus, we may also state that, for example, we apply the ``\seed mask'' to another object, meaning that we use the mask associated to the \seed footprint. A second object associated with a \emph{footprint}, which is the pixel fractional coverage or \fracdet. This quantity ranges from 0 to 1 and represents the observed fraction of each pixel at a given \nside. Fully observed pixels have a \fracdet of 1, whereas unobserved pixels have a \fracdet of 0. This quantity is important to properly degrade the values of a given map to lower \nside resolutions, as this is computed as the average of the sub-pixels weighted by their corresponding \fracdet (see Section \ref{sec:fracdet} and Appendix \ref{app:degrading} for more details). It is also important for correctly weighting the pixels in the computation of the correlation function. Finally, we use the term \emph{map} to refer to objects defined in pixel format on the sky, such as the contamination templates described in Section \ref{sec:spmaps}. By applying a selection criterion, such as excluding pixels with observational systematic values outside a given range, we can define new regions of the observed sky (i.e., within a \emph{footprint}) to be masked out. In other words, we can use a \emph{map} to define \emph{masks} based on various selection criteria, as well as apply different \emph{masks} to a \emph{map}. In this sense, a \emph{mask} is a boolean map, while the \fracdet is a map representing the fractional coverage of the pixels within the corresponding \emph{footprint}.  
 
\subsection{\goldfoot footprint}\label{sec:footprint}
This footprint, generated using \decasu \footnote{\url{https://github.com/erykoff/decasu}} as part of the construction of the DES Y6 Gold catalogue (\citet{y6gold}), includes all \nside = 16384 \healpix pixels in the initial footprint with information on \griz bands depth according to \decasu, and the condition of NOT being in the Y6 Gold foreground mask. In this case, pixels in the Y6 Gold foreground mask are flagged for removal because they represent known objects that obscure or degrade the images. This includes bright stars and associated artifacts from the brightest ones, nearby galaxies, globular clusters, and regions around the Magellanic Clouds. The diffraction spikes and satellite trails are removed or flagged at the image level. 
These cuts are applied by setting \texttt{FLAGS\_FOREGROUND}=0, with the other main contributors to the mask being \texttt{NUM\_IMAGE}>=2, and $f_{griz}^{4096}>0.5$. For further details on the construction of this footprint and the Y6 Gold foreground mask, we refer the reader to \citet{y6gold}. Later, using the information contained in this footprint, we apply depth cuts based on the specifics of the galaxy sample we work with, particularly its limiting magnitudes. This is detailed in Section \ref{sec:baseline_mask}. \\
\\
In addition, during the DES Y6 Gold processing, there were some issues that resulted in the removal of three tiles, corresponding to $\sim 2 \rm \, deg^2$. These three tiles were corrected in the public release of the DES Y6 Gold catalogue. However, the \maglimpp sample used for the DES Y6 \threextwo, \twoxtwo and galaxy clustering analyses was selected from a previous version of the Gold catalogue, which retained those problematic tiles. To address this, we applied an additional \nside = 16384 mask for the three faulty tiles to the Y6 Gold footprint. As with the foreground mask described in the previous subsection, the pixels included in the Gold faulty processing regions mask are removed, whereas those outside this mask are retained. Further details are provided in \citet{y6gold}.

\subsection{\seed footprint }\label{sec:shear_mask}
In addition to the \goldfoot footprint from the previous subsection, we consider an additional footprint obtained from the DES Y6 analysis to define the weak lensing shape catalogue (\citet{y6shearcat}): the \emph{shear} footprint. This footprint is created at a high resolution of \nside = 131072. It already accounts for some of the cuts from the \goldfoot footprint introduced in subsection \ref{sec:footprint} along with additional shear-specific cuts. In particular, the \emph{shear} footprint is composed of several object-level and pixel-level masks. Object masks include the footprint and foreground mask from \citet{y6gold}, Gaia stars (\citet{GaiaDR2}), DES stars from \citet{y6gold}, and HyperLEDA objects (\citet{hyperleda}). The pixel-level masks are the regions that are masked due to artifacts such as bad columns and cosmic rays in single-exposure images, and further include the apodized regions around object masks and edges of the coadd images. In total, a $4421.52 \, \rm deg^2$ footprint was left for the Y6 shear catalogue. For more details on the creation of the \emph{shear} footprint, we refer the reader to Section 4.2 of \citet{y6shearcat}. \\
\\
Next, we incorporate cuts provided by the galaxy clustering analysis. In this regard, the Y6 lens sample uses the redshift-bin optimised colour-based star-galaxy cut described in \citet{stargalsep} to remove residual stellar contamination in the sample. This uses \texttt{unWISE} NIR photometry from \decals DR9 (\citet{Dey_2019}). We therefore apply an additional \decals-based LRG mask as described in \citet{DESI:2022gle}, which uses a larger radius around bright stars and also removes star wings, ensuring high-quality W1 NIR photometry. This mask is defined at \nside=16384. A small number of objects at the southern edge of the footprint (RA, DEC) $\approx (-66^{\circ}, 47^{\circ})$ were not removed by this mask, but lacked \decals photometry, and these were removed by dropping $\sim$1 deg$^2$ of \nside = 512 pixels containing them. \\
\\
We refer to the combination \emph{shear} footprint, \texttt{unWISE} mask and \goldfoot footprint as \seed footprint. As we later explain in Section \ref{sec:mask_creation}, we use this footprint as a basis on top of which we apply the cuts that lead to the final \jointmask footprint and to obtain lower resolution versions of other intermediate masks. 

\subsection{DES Y6 template maps of contamination}\label{sec:spmaps}
We present here the template maps used to construct the \systmask mask, as detailed in Section \ref{sec:syst_mask}. Our map selection includes the maps used to compute corrective weights in  \citet*{y6weights}. The DES Y6 methods—\isd and \texttt{ENet}—are regression techniques that employ template maps as explanatory variables to identify observational systematic signatures in the galaxy number density field. While including more maps increases model flexibility to fit potential contamination, this approach risks overcorrection, particularly when template maps inadvertently contain real LSS introduced during their construction, which can be difficult to characterise (see \citet*{y6weights} for details). Additionally, highly correlated maps introduce redundant information without necessarily improving systematic detection. Thus, we must balance comprehensive systematic coverage against the risk of overcorrection and suppressing true LSS modes due to spurious correlations or inadvertent LSS embedding in the model space. \\
\\
However, even though it is preferable to limit the list of template maps used for weights estimation (note that the number we use is still significantly larger than e.g., that typically used in DESI, see: \citet{Kitanidis:2019rzi}), we can nonetheless employ the broader set of templates to identify and mask extreme pixels within the footprint. This ensures footprint uniformity across a broader range of metrics, without the risk of LSS mode suppression that comes from using them to estimate weights. \\
\\
The following subsections briefly describe each template map employed in this work. For detailed information about these maps' characteristics, see \citet{y6gold}. Table \ref{tab:spmaps_table} summarises their key properties: native pixel resolution, photometric bands used for measurements, and their role in systematic weights computation.
\begin{table*}
	\centering
	\begin{tabular}{l|lcccc} 
	\hline
	Template map & Description & Units & \nside & Bands & Weights \\
	\hline
    \texttt{AIRMASS\_WMEAN} & Secant of the zenith angle & - & 16384 & \griz & Yes \\ 
    \texttt{FWHM\_WMEAN} & Average FWHM of 2D elliptical Moffat function fit & pixels & 16384 & \griz & Yes \\ 
    \texttt{MAGLIM\_WMEAN} & Magnitude limit estimated from the weight maps & mag & 16384 & \griz & Yes \\ 
    \texttt{SKYSIGMA\_WMEAN} & Square root of sky variance & e$^{-}$/CCD pix. & 16384 & \griz & Yes \\  
    \texttt{SKYSIGMA\_WMEAN\_SCALED} & \texttt{SKYSIGMA\_WMEAN} scaled by coadds' zero-points & e$^{-}$/CCD pix. & 16384 & \griz & No \\ 
    \texttt{SKYBRIGHTNESS\_WMEAN} & Sky brightness from the sky background model & e$^{-}$/CCD pix. & 16384 & \griz & No \\  
    \texttt{SKYBRIGHTNESS\_WMEAN\_SCALED} & \texttt{SKYBRIGHTNESS\_WMEAN} scaled by coadds' zero-points & e$^{-}$/CCD pix. & 16384 & \griz & No \\  
    \texttt{DCR\_DDEC\_WMEAN} & Diff. chromatic refraction on positions (rel. shifts) & - & 16384 & \griz & No \\ 
    \texttt{DCR\_DRA\_WMEAN} & Diff. chromatic refraction on positions (rel. shifts) & - & 16384 & \griz & No \\ 
    \texttt{DCR\_E1\_WMEAN} & Diff. chromatic refraction on ellipticity (rel. shifts) & - & 16384 & \griz & No \\ 
    \texttt{DCR\_E2\_WMEAN} & Diff. chromatic refraction on ellipticity (rel. shifts) & - & 16384 & \griz & No \\ 
    \texttt{BDF\_DEPTH} & Mag. limit (de-reddened) & mag & 4096 & \griz & No \\
    \hline
    \texttt{GAIA\_DR3} & Stellar density defined from Gaia DR3 & stars/$\rm deg^{-2}$ & 512 & \emph{G} & Yes \\
    \texttt{EBV\_SFD98} & $E(B-V)$ interstellar extinction from dust IR emission & mag & 4096 & - & No \\
    \texttt{EBV\_CSFD23} & \texttt{EBV\_SFD98} corrected from cosmic infrared background & mag & 2048 & - & Yes \\
    \hline
    \texttt{BFD\_BACKGROUND\_OFFSET} & Position dependent avg. residual bkgrd. of galaxy images & counts\footnote{Zero-point calibrated counts (originally electrons, rescaled to a fixed zero point of 30). For more information about this map, see \citet*{y6bfdshearcat}.} & 512 & - & No \\ 
    \hline
    \texttt{CIRRUS\_NEB\_MEAN} & Mean prob. of nebular emission from \maximask & - & 4096 & \griz & No \\ 
    \texttt{CIRRUS\_SB\_MEAN} & Mean surf. brightness in image in \healpix (relative) & - & 4096 & \griz & No \\
    \hline
    \end{tabular}
    \caption{List of template maps for observational systematics considered in DES Y6. We indicate the native \nside resolution at which these maps were created, the photometric bands in which they are defined (if any) and whether they are used to obtain systematic weights from with our decontamination methods (see \citet*{y6weights}). Each block on this table corresponds to one of the categories introduced in Section \ref{sec:inputs}. }
    \label{tab:spmaps_table}
\end{table*}

\subsubsection{Survey properties and observing conditions }
Among these effects, we consider the survey properties, which are effects related to the observing strategy, such as exposure time, airmass or magnitude limit, and the observing conditions, such as seeing (characterised by the full width at half maximum (FWHM) of the point spread function, PSF) or the sky brightness. In the context of this analysis, we refer to this set of contamination templates as \emph{survey property} (SP) maps. As noted in Table \ref{tab:spmaps_table}, the first four SP maps are also used to compute correcting weights that are applied to the DES Y6 \maglimpp sample. The maps in the first block of this table are in \healsparse format, while the rest of them are in \healpix format. As it can be seen, \healsparse maps come in a much higher \nside resolution, which would allow to test systematic effects at smaller scales. However, in order to reduce noise when computing the outliers and since we are limited by the lower resolutions from other maps, we degrade them to \nside = 4096 later on the analysis (see Appendix \ref{app:degrading}).   

\subsubsection{Astrophysical foregrounds}\label{sec:foregrounds}
The DES footprint is designed to avoid the Galactic plane, since that area of the sky contains stars and dust that can contaminate the data. However, this does not completely avoid the presence of stars, stellar clusters and interstellar dust within our region of interest. These foregrounds affect the galaxy selection function through multiple mechanisms: for instance, high stellar density can complicate star–galaxy separation when selecting a galaxy sample. Interstellar dust results in extinction and reddening, affecting both sample selection and photometric redshift estimation. Additionally, regions near bright stars may exhibit obscuration effects, reducing the observed galaxy counts.\footnote{Stellar contamination and obscuration affect the observed density in opposite ways, complicating its treatment. For example, a high density of stars in the field increases the likelihood of misclassifying them as galaxies, which would alter the clustering amplitude, while also potentially obscuring background galaxies, such that we want to remove regions of the sky where the contamination significantly impacts our ability to measure the true number density. We apply a new and more optimized star-galaxy separation algorithm for the \maglimpp sample, rendering most of the residual star-related systematics to be obscuration. See \citet*{y6weights}.} \\ 
\\
To account for these effects, we use three maps external to DES: regarding the effects related with stars, both stellar obscuration and contamination should trace stellar density. Thus, we use the density of Gaia DR3 (\citet{GaiaDR3}) objects that satisfy the following star-galaxy separation cuts as a template map:
\begin{itemize}
    \item 
    $\begin{aligned}[t]
        \log_{10}(\texttt{gaia\_astrometric\_excess\_noise}) &< \\
        \max(0.3,\ 0.3(G_{\rm Gaia} - 18.2) + 0.15)
    \end{aligned}$
    \item $G_{\rm Gaia} < 20$
\end{itemize}
For masking extreme values, we start with the full-sky map and count the number of stars within pixels of \nside = 512 resolution with no masks applied. We select this \nside to ensure that pixels are signal-dominated, with an average of $\approx 35$ stars per pixel. After this, we convert the star counts to stellar density by dividing each pixel's count by the area corresponding to \nside = 512. Once the stellar density map is defined, we upgrade it to \nside = 4096. We detail the degrading/upgrading process in Appendix \ref{app:degrading}. \\ 
\\
To account for dust extinction, we use two maps: (1) The $E(B-V)$ interstellar extinction map from \citet{sfd98} (\texttt{EBV\_SFD98}), which is used to correct fluxes in the DES Y6 Gold catalogue (\citet{y6gold}), but includes cosmic infrared background (CIB) contamination. (2) An updated $E(B-V)$ map from \citet{csfd23} (\texttt{EBV\_CSFD23}), which mitigates the CIB contamination present in \texttt{EBV\_SFD98}. We use both \texttt{EBV\_SFD98} and \texttt{EBV\_CSFD23} maps for masking extreme-valued pixels, adopting the latter as our primary extinction template for corrective weights. Note that one might naively want to use both the \texttt{EBV\_SFD98} and \texttt{EBV\_CSFD23} maps as templates for weights, or indeed the difference between them as this would track ``false de-reddening'' more closely. However, since the difference is dominated by CIB, which traces LSS, it would be dangerous to do this for weights estimation (see \citet*{y6weights} for a discussion). We work with \nside = 4096 versions of these two maps.\footnote{The \nside = 4096 \texttt{EBV\_SFD98} map we work with comes from upgrading an \nside = 512 \healpix distribution of this map (\url{https://lambda.gsfc.nasa.gov/product/foreground/fg_sfd_get.html}). Similarly, the original resolution of the \texttt{EBV\_CSFD23} map is \nside = 2048, which we then upgrade to \nside = 4096. For this reason, the \nside indicated on Table \ref{tab:spmaps_table} for these maps does not represent their native resolution.} 

\subsubsection{BFD background offset }
An offset background map is generated as a by-product of the Bayesian Fourier Domain (BFD; \citet{Bernstein:2015gya}) shear measurement pipeline (see \citet*{y6bfdshearcat}). This map represents the average residual background of galaxy images as a function of survey position. For each galaxy in the catalogue, the residual background is estimated for every individual exposure and then averaged across exposures. To construct the offset background map, the mean residual background per image is computed for all galaxies falling within the same pixel. \\
\\
The BFD pipeline estimates the residual background in analogue-to-digital units (ADU) for each single-epoch image used in shear estimation. Each single-epoch image (typically $40\times40$ pixels) defines an external frame from the five outermost pixels, and the mean ADU value in this frame is taken as the background estimate. Pixels flagged for masking or containing neighbouring objects (according to the segmentation map) are excluded. To ensure a conservative estimate, both the bad-pixel and segmentation masks are expanded by up to five pixels in all directions, through iterative axis shifts and mask combinations. If fewer than 50 valid pixels remain, the single-epoch image is excluded from the calculation. Individual background estimates from different bands and epochs are then averaged using the same per-band, per-exposure weights as in the shear measurement pipeline. Notably, the offset background map strongly correlates with stellar density.

\subsubsection{Galactic cirrus / nebulosity estimate maps}\label{sec:cirrus_maps}
The faint diffuse light from Galactic cirrus can lead to problems in object detection, inducing galaxy number density fluctuations of systematic origin. Galactic cirrus tends to correlate with maps of Galactic dust extinction, such as \texttt{EBV\_SFD98}. However, dust extinction does not completely trace cirrus, because prominent cirrus emission requires the presence of both dust and a source of light illuminating it. \\
\\
To account for these diffuse features, we use two maps that estimate the presence of Galactic cirrus over the DES Y6 footprint: the mean nebular-emission probability, \texttt{CIRRUS\_NEB\_MEAN}, calculated with \maximask (\citet{2020A&A...634A..48P}) and the mean surface brightness in image from \healpix pixels, \texttt{CIRRUS\_SB\_MEAN}. For more information about these maps, we refer the reader to \citet{y6gold}, Appendix C.1. 

\section{Creation of the \systmask mask and the DES Y6 \jointmask footprint}\label{sec:mask_creation}
\begin{figure*}
    \centering
    \includegraphics[width=0.8\linewidth]{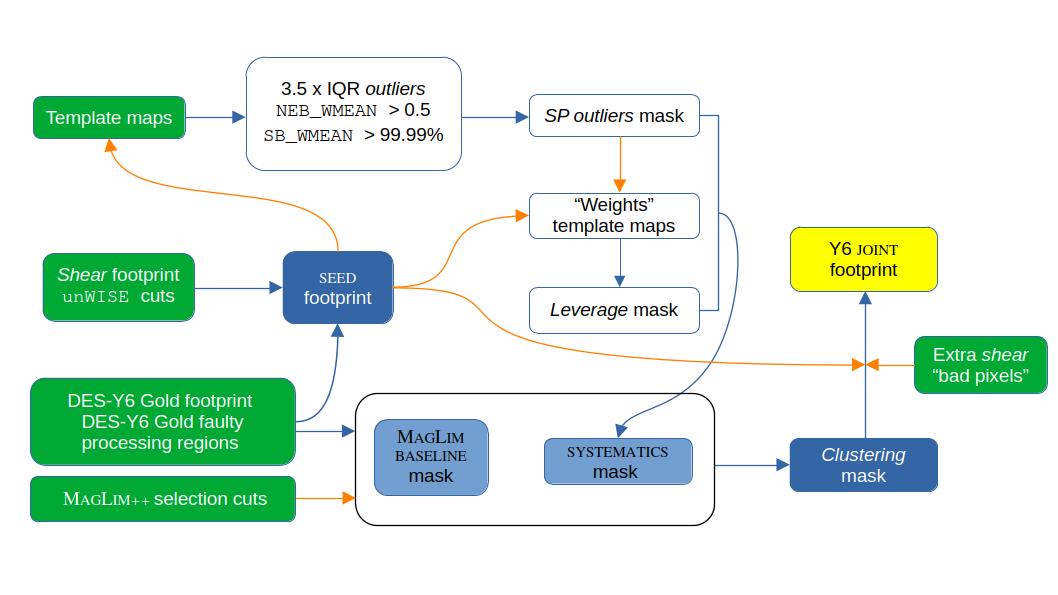}
    \caption{Scheme of the process to obtain the \jointmask mask from the individual masks. Green boxes correspond to the initial components presented in Section \ref{sec:inputs} and white boxes represent intermediate masks defined along the process. Orange arrows starting at any mask mean that that mask is applied to a given element of the process. Note how in this sense, the \seed mask is used to provide the \fracdet necessary to degrade to lower resolutions when required. Finally, blue arrows indicate that the combination of certain masks defines a new object, either a footprint or a mask, such as in the case of the combination of the outlier masks from the different $3.5\times$IQR cuts, the \texttt{CIRRUS\_NEB\_MEAN} > 0.5 and the \texttt{CIRRUS\_SB\_MEAN} > $99.99\%$ cuts, giving rise to the \emph{SP outliers mask}.}
    \label{fig:mask_process}
\end{figure*}

The creation of the \jointmask footprint for DES Y6 requires all footprints and maps presented in the previous section, as well as several sub-masks derived from them. Here, we describe the process to assemble these elements into a single joint footprint. Figure \ref{fig:mask_process} depicts schematically the process of creating this footprint from its individual components. 

\subsection{Fractional coverage from the \seed footprint }\label{sec:fracdet}
We use the \seed footprint presented in Section \ref{sec:shear_mask} as the foundation upon which we progressively remove area using different masks or cuts. In addition to setting the initial area, we also use this footprint to define the \fracdet (as defined in Section \ref{sec:terms}) of the pixels at different resolutions. To obtain this \fracdet, we start from the original \nside = 131072 version of the \seed footprint. At such high resolution (pixels at this \nside correspond to an angular scale of $\sim 4.5\cdot10^{-4} \rm \, deg$ or $\sim 1.6 \rm \, arcsec$), we assume that all pixels are fully observed, so we can assume that their \fracdet is 1. Then, we degrade this high-resolution \fracdet to \nside = 16384, 4096, 2048, and 512, corresponding to the native \nside values at which the template maps presented in Section \ref{sec:spmaps} (see Table \ref{tab:spmaps_table}) are defined. As detailed in the following subsections, in some cases we need to degrade some of the aforementioned template maps. To do this for a given map, we first apply the \seed mask to it at the same \nside and then we do a \fracdet-weighted average of pixel values. Appendix \ref{app:degrading} provides further details on the \fracdet computation and template map degradation. 

\subsection{DES Y6 \baselinemask mask }\label{sec:baseline_mask}
As mentioned in Section \ref{sec:footprint}, we need to apply specific cuts to the \goldfoot footprint based on the particular selection properties of the galaxy sample being used. In the case of the \maglimpp sample, it has an upper magnitude limit at $i < 22.2$ (see \citet{y3maglim} and \citet*{y6weights}). In this sense, we ensure high SNR in the selection by removing pixels with $i_{\rm depth} < 22.2$ at $10\sigma$ level (\texttt{BDF\_DEPTH} photometry; see \citet{y6gold}) from the \goldfoot footprint. We further remove the small number of pixels that lie in the DES Y6 Gold faulty processing regions presented in Section \ref{sec:footprint}. Combined, these cuts form the \baselinemask mask at \nside = 16384. 

\subsection{\systmask mask }\label{sec:syst_mask}
The \systmask mask gathers information about potentially problematic pixels coming from the different templates of contamination that we consider. In addition, this allows us to account for the impact that those bad pixels may have on the performance of the decontamination methods used to obtain corrective weights. In the following subsections we describe the process to put together all systematic contributions into a single mask. 

\subsubsection{SP outliers mask}
The presence of extreme-valued pixels in contamination templates poses a challenge for decontamination, as such pixels introduce non-representative variations that differ from the footprint's dominant systematic modes, which reduces the robustness of the decontamination methods. This is important, because the different decontamination methods work by fitting the relation between the observed \ngal and the templates of contamination, and assuming that the true cosmological signal corresponds to the residual of those fits. Therefore, extreme values can affect these fits, as we later show in Section \ref{sec:mask_impact}, which can translate into an important impact on the recovered signal. \\ 
\\
For most of the template maps, we identify extreme values using a version of Tukey's fences (\citet{tukey1977exploratory}), which is a non-parametric approach that uses the central half of the distribution as a characteristic scale and flags pixels that lie far from the bulk, relative to this scale. In contrast, the \texttt{CIRRUS\_NEB\_MEAN} and \texttt{CIRRUS\_SB\_MEAN} maps were produced for exploratory analysis, and are highly skewed or likely contain LSS signal, respectively. We therefore visually inspect representative pixels at different values of these maps to identify upper limits above which we mask. The combination of the different cuts from the template maps that yields the \emph{SP outliers mask} is described below. 

\subsubsection*{Galactic cirrus outliers }
\begin{figure}
    \centering
    \includegraphics[width=0.8\linewidth]{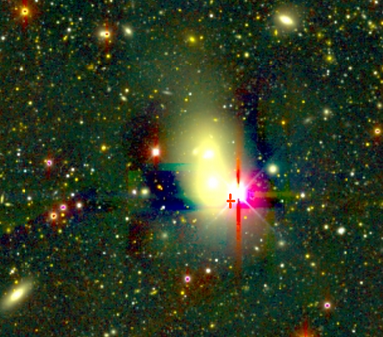}
    \includegraphics[width=0.8\linewidth]{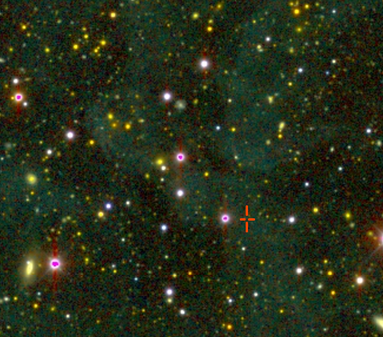}
    \caption{\emph{Top:} sky region presenting image artefacts found at sky positions where the \texttt{CIRRUS\_SB\_MEAN} estimate maps have extreme values belonging the 0.9999 quantile of the distribution. \emph{Bottom:} sky region showcasing Galactic cirrus, identified with \texttt{CIRRUS\_NEB\_MEAN} values greater than 0.5. }
    \label{fig:cirrus}
\end{figure}

During the DES Y6 analysis to obtain corrective weights for the lens galaxy sample, \citet*{y6weights} found that the \texttt{CIRRUS\_SB\_MEAN} maps likely contain actual LSS (this may be, e.g., from undetected sources). This leaked LSS would be identified as a systematic correlation with the data and therefore removed, which would result in over-correction. For this reason, we discard these maps to obtain corrective weights from. 
This is motivated by the initial suspicions during DES Y3 that a leakage of LSS into some of our template maps might be responsible for the so-called X lens effect that appeared in the case of the \redmagic sample (see \cite{y3_3x2} and \cite{y3weights} for a discussion on this). We found that the Galactic cirrus maps were capable of tracing some of that potentially leakable LSS, so we decided to use them as tracers of this problem rather than for weights. \\
\\
However, they are still useful for masking, since when inspecting them we find that while they do not seem to track visual cirrus very effectively, their extreme values were consistent indicators of image artefacts in the coadds. We therefore use a custom threshold determined by visual inspection, masking the 0.01\% of \nside = 4096 pixels with the highest \texttt{CIRRUS\_SB\_MEAN} values. The top panel of Figure \ref{fig:cirrus} shows an example artefact removed by this cut. To define this threshold, we inspected two pixels each at \texttt{CIRRUS\_SB\_MEAN} values closest to the [0.1, 0.5, 0.9, 0.95, 0.99, 0.999, 0.9999]’th quantiles of the map (two pixels each at the same quantiles). Each pixel at the 99.99\% and above appeared very near a bright object or clear visible artifact, while each random pixel sampled below the threshold did not. The sampling is admittedly small and the threshold imperfect but it helps remove a small number of the most egregious areas. \\
\\
The \texttt{CIRRUS\_NEB\_MEAN} maps were not used either for systematic weights, as these maps were an experimental effort and are produced by applying a convolutional neural net rather far outside of its training space. It nominally corresponds to the probability that a given pixel contains cirrus, though this should be interpreted with significant caution for the aforementioned reason. The maps range from  $[0, 1]$ and are extremely skewed, with most of the footprint $\sim0$, rendering them ill-suited as templates. However we remove \nside = 4096 pixels with $\texttt{CIRRUS\_NEB\_MEAN} > 0.5$, as we find these serve as good indicators of visual cirrus (see bottom panel of Figure \ref{fig:cirrus}). This threshold is defined by inspecting the \texttt{CIRRUS\_NEB\_MEAN} maps the same way as the surface brightness ones, i.e., checking two pixels at nebulosity values closest to the same quantiles as described before. We found that those at the 99th percentile and above consistently showed visible cirrus, while those below showed none. Since the 99th percentile corresponded to a nebulosity probability value of 0.505, we picked 0.5 as a reasonable threshold. A more careful study could likely refine this value. \\
\\
We note that, given that at the early stages of this analysis the \emph{shear} footprint was not finalised yet, a preliminary version of the \baselinemask mask for \maglimpp introduced in Section \ref{sec:baseline_mask} is applied to the \texttt{CIRRUS\_NEB\_MEAN} and \texttt{CIRRUS\_SB\_MEAN} maps. \\
\\
Taken separately, the cuts from \texttt{CIRRUS\_NEB\_MEAN}$>0.5$ remove $77.54 \rm \, deg^2$ from the \seed footprint, while the cuts from \texttt{CIRRUS\_SB\_MEAN}$>99.99\%$ remove only $1.69 \cdot 10^{-3} \rm \, deg^2$. We note that for both estimates of Galactic cirrus we compute their outliers in the four photometric bands considered and then combine them into single \texttt{CIRRUS\_NEB\_MEAN} and \texttt{CIRRUS\_SB\_MEAN} masks. From these two estimates of cirrus, \texttt{CIRRUS\_NEB\_MEAN} yields most area removal. We verify that the area removal from these cirrus estimate maps comes from two blocks: one constituted by the \emph{gri}-bands and another one for the \emph{z}-band. We find that, for the \texttt{CIRRUS\_NEB\_MEAN} maps in \emph{gri}-bands the outlier pixels are present over the whole initial \seed footprint, although they tend to appear closer to the edges, while in the case of the \emph{z}-band the outliers are more evenly distributed. Appendix \ref{app:outliers} provides further details about this. \\
\\
One issue that could arise from masking extreme values from maps that might contain actual LSS is that this can bias the estimated clustering signal, even if such maps are not used for corrective weights. We verify that this is not the case for the \texttt{CIRRUS} maps in Appendix \ref{app:cirrus_mask}.

\subsubsection*{$\rm 3.5\times IQR$ outliers from SP maps }\label{sec:iqrcuts}
\begin{figure}
    \centering
    \includegraphics[width=\linewidth]{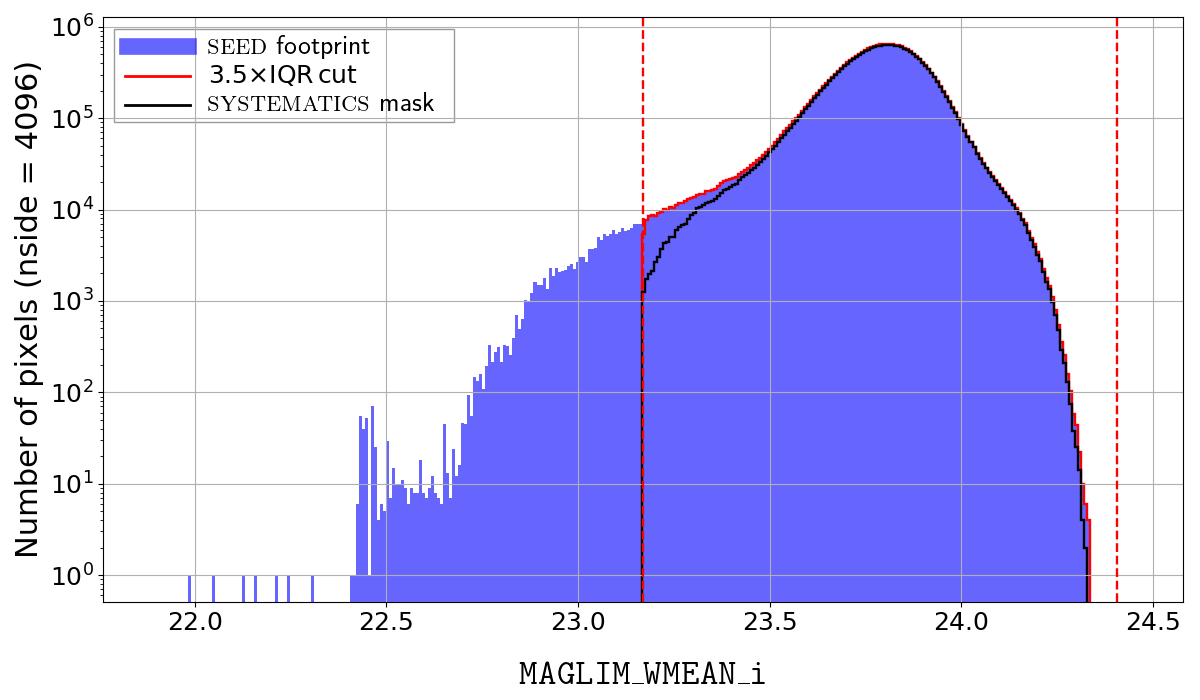}
    \caption{Values distribution of the \texttt{MAGLIM\_WMEAN\_i} map. The filled histogram corresponds to the full range of values when we just apply the \seed mask; the red-outlined histogram represents the same distribution after identifying the $3.5\times \rm IQR$ outliers of this map (represented by the red dashed lines) and excluding values beyond these limits; the black-outlined histogram shows the distribution of values of this map after applying to it the total \systmask mask. }
    \label{fig:outliers}
\end{figure}

\begin{figure*}
    \centering
    \includegraphics[width=1.0\linewidth]{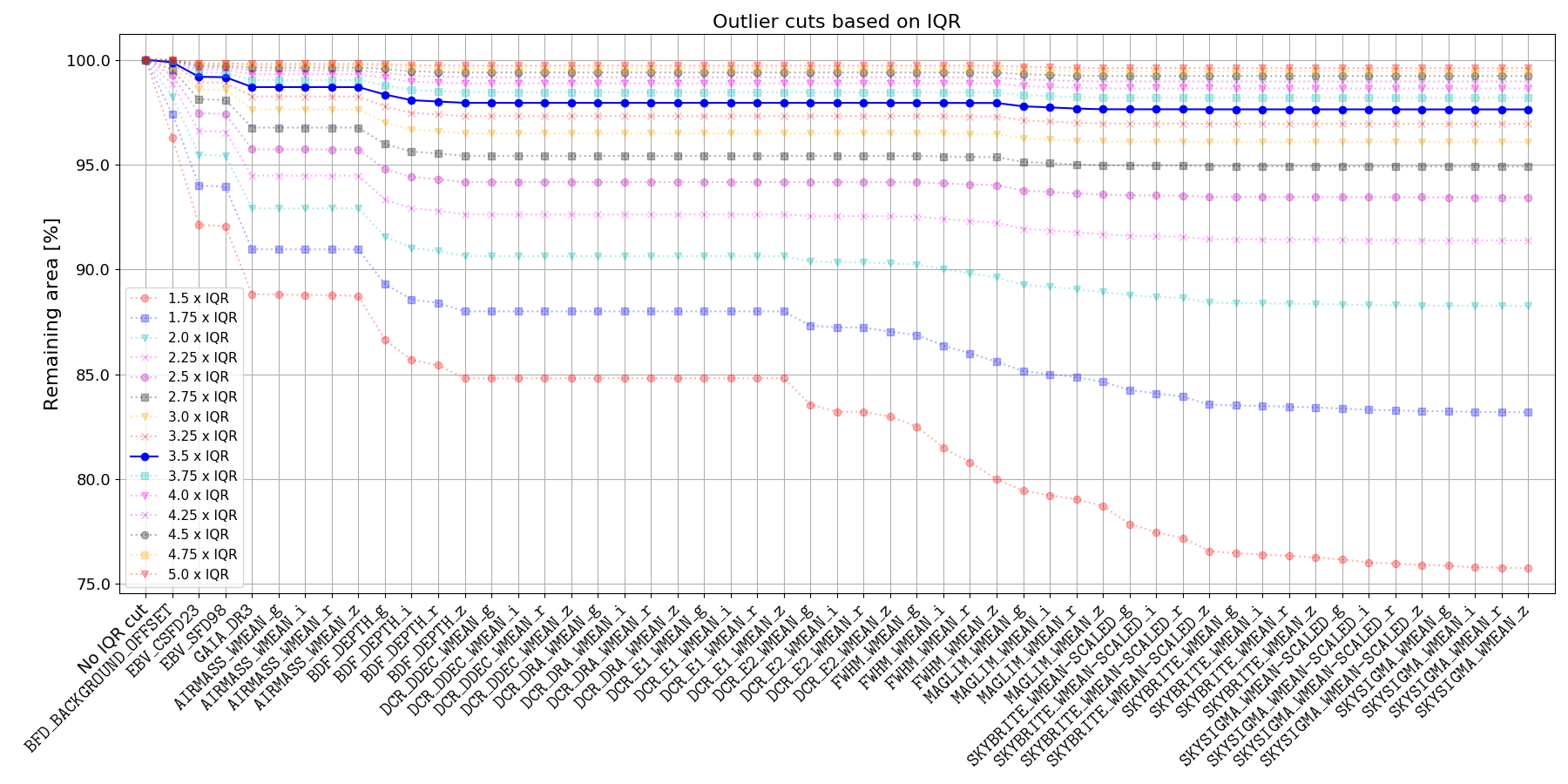}
    \caption{Remaining area (in percentage) with respect to the area given by the \seed footprint (i.e. with no IQR cuts applied) as a function of progressive $a \times \rm IQR$ cuts from the different template maps of contamination (with the exception of the cirrus tracers maps). Each line corresponds to different values of $a$. The solid blue line corresponds to our chosen threshold of $a=3.5$. }
    \label{fig:iqr_mask}
\end{figure*}

For all other SP maps, we define ``outlier'' pixels as those with pixel values outside the range $[\mathrm{Q_1} - a \times \mathrm{IQR}, \mathrm{Q_3} + a \times \mathrm{IQR}]$, where IQR is the interquartile range, defined as 
\begin{equation}\label{eqn:iqr}
    \rm IQR = Q_3 - Q_1 \, , 
\end{equation}
with $\rm Q_1$ and $\rm Q_3$ the first and the third quartiles of the distribution, and $a$ a distance factor. To avoid being biased by pixels that were already removed by the \seed mask and to make this cut more stringent, we evaluate the value distributions after applying the \seed mask. In addition, to reduce noisiness on the value distributions and to make them more comparable from one map to another, we degrade the maps with \nside = 16384 to 4096 as described in Appendix \ref{app:degrading}. In the case of the maps whose native resolution is 4096 or lower (see Table \ref{tab:spmaps_table}), we simply apply the \seed mask pre-degraded to that \nside and evaluate their value distributions in the same way as the other maps. In both cases, we also assign the \fracdet at the corresponding resolution to keep track of the area loss as a function of the IQR cut applied. In Figure \ref{fig:outliers} we show an example of the distribution of values for the template map \texttt{MAGLIM\_WMEAN\_i}, where the filled histogram corresponds to the values when only the \seed mask is applied, while the red and black-outlined histograms depict the distribution of values after applying the $3.5 \times \rm IQR$ outlier cut (indicated by the vertical dashed red lines) from this map and the total \systmask mask, respectively. In total, we compute these $3.5 \times \rm IQR$ cuts on 52 SP maps. \\
\\
In order to calibrate the value of $a$, we calculate the cumulative area loss for a range of $a$'s that covers the two extreme cases in which we reject points very close and very far from the IQR limits of each template map. To do this, for each SP map separately we find the pixels with values out of the range given by $a \, \times$ IQR, producing an individual rejection mask for each map at each $a$ value. Then, for that same $a$, the individual masks are combined into a total ``outlier mask'' that contains the intersection of all pixels that have survived the individual cuts. In Figure \ref{fig:iqr_mask} we show the evolution of the area loss (in percentage with respect the area from the \seed footprint) as we progressively remove pixels beyond $a \, \times$ IQR for several values of $a$. As shown in the figure, $a$ values ranging from 2.75 to 5 result in a loss of area from $\sim 0\% - 5\%$. For $a<2.75$ the area loss increases rapidly. We choose a threshold $a = 3.5$ (solid blue line in the plot) as a reasonable compromise that removes the most extreme pixels without significantly decreasing the area of the footprint. This choice for the cut is somewhat arbitrary within the range of $a$ values that do not cause a rapid decline of area, $a \in [3, 4]$. Nevertheless, we verify that perturbing $a$ and recomputing \enet weights does not significantly affect our $w(\theta)$ measurement, obtaining a $\Delta \chi^2 (w(\theta)) < 1$ for a $\Delta a = 0.25$ with respect to a = 3.5 for each redshift bin. \\
\\
Combining the cuts from the 52 SP maps considered, the $3.5 \times \rm IQR$ outliers cut removes $100.45 \rm \, deg^2$ from the \seed footprint. In Appendix \ref{app:outliers} we show the amount of area removed from the \seed footprint by the $3.5 \times \rm IQR$ cut of each individual SP map. As shown in Figure \ref{fig:iqr_individual} that the SP maps that cause most of the area loss given our definition of outliers are those related with the depth of the survey, namely \texttt{BDF\_DEPTH} and \texttt{MAGLIM\_WMEAN} and with the astrophysical foregrounds, i.e., \texttt{EBV\_CSFD23}, \texttt{EBV\_SFD98} and \texttt{GAIA\_DR3}. On the top and bottom panels of Figure \ref{fig:outlier_sky_maps} we show the sky projections of the \texttt{EBV\_CSFD23} and \texttt{BDF\_DEPTH\_i} maps, respectively, together with the positions of the \nside = 4096 pixels that are removed according to the $3.5 \times \rm IQR$ criterion. It can be seen how the outlier pixels from the \texttt{EBV\_CSFD23} map mainly distributed close to Galactic plane, while in the case of the \texttt{BDF\_DEPTH\_i} map the outliers concentrate on the borders of the initial \seed footprint. \\
\\
We refer to the mask resulting from the combination of the cuts from the \texttt{CIRRUS\_SB\_MEAN} and \texttt{CIRRUS\_NEB\_MEAN} maps with the $3.5 \times \rm IQR$ cuts from all considered SP maps as \emph{SP outlier} mask.

\subsubsection{Leverage mask } \label{sec:leverage}
\begin{figure}
    \centering
    \includegraphics[width=1\linewidth]{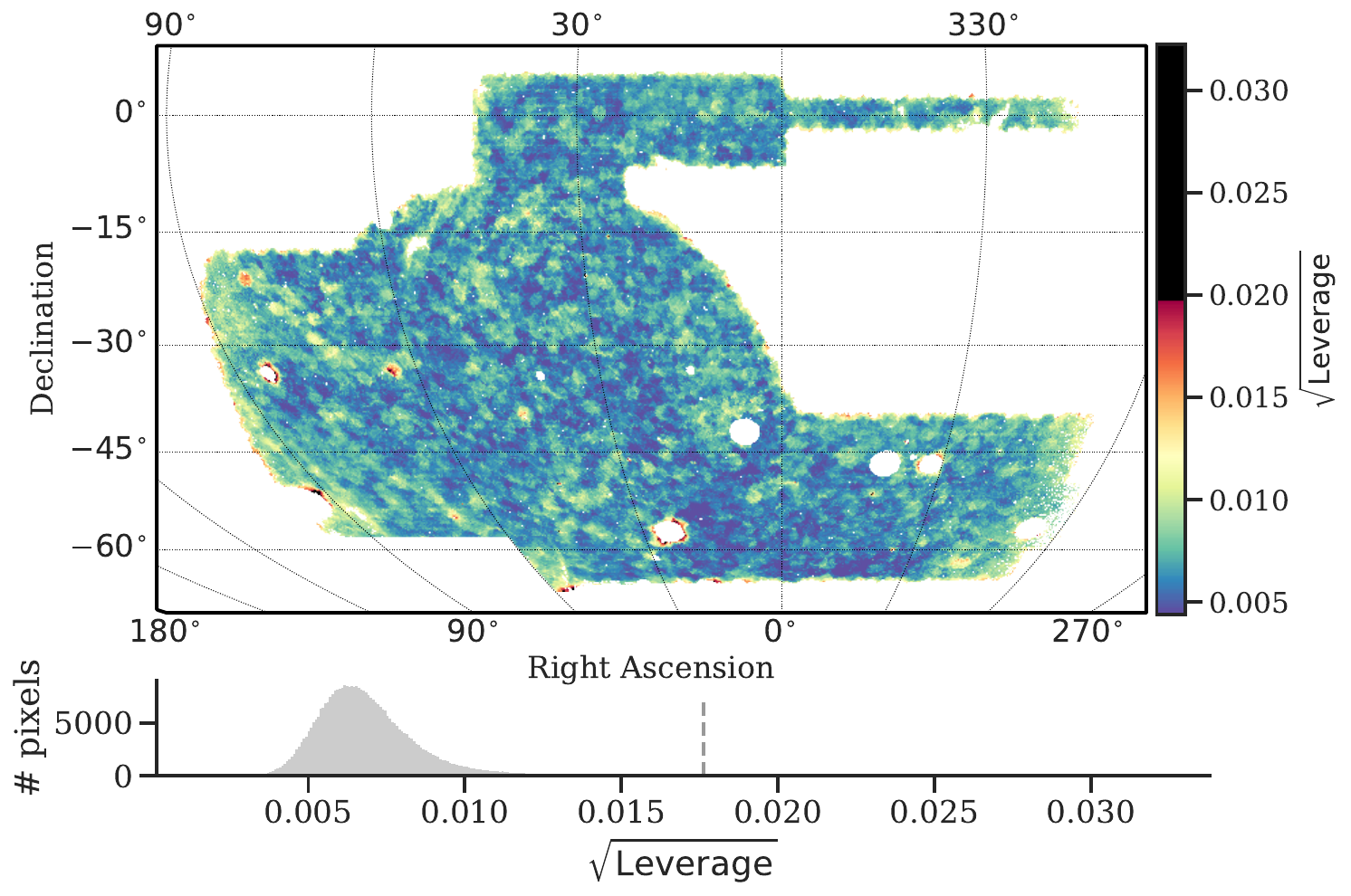}
    \caption{\textit{Top:} Square root of the pixel leverage across the DES footprint, a measure of extremity in the $N_{\rm tpl}$ dimensional template space (\nside = 512). Pixels with high leverage are least likely to satisfy the assumptions of perturbative contamination models, exert proportionally greater influence on weights estimation, and are likely to have greater map-level error on the estimated over-density. The highest 0.1\% of pixels are removed for the \emph{leverage} mask, indicated in black. \textit{Bottom:} The distribution of leverage values for pixels in the footprint, with the leverage cut indicated by the dashed line. See Section \ref{sec:leverage}.}
    \label{fig:leverage}
\end{figure}

We also compute and mask pixels that are extreme in the full 19-dimensional template space that is used for computing galaxy weights. This is done using the so-called \textit{leverage} statistic proposed in \citet{Weaverdyck_2021}, which is defined as 
\begin{equation}
    h_{i} = \left[T\left(T^\dagger T\right)^{-1}T^\dagger\right]_{ii}
\end{equation}
for pixel $i$, where $T_{ij}$ is the value of template map $j$ at pixel $i$. The resultant (square root) leverage map is a measure of how far each pixel is from the centre of mass of all pixels in the 19-dimensional space, after accounting for the covariance between SP maps. This allows us to capture highly unusual \textit{combinations} of observing conditions, as opposed to the unidimensional  outlier cuts described in the previous section.\\
\\
As noted in \citet{Weaverdyck_2021}, the leverage map is useful not only for identifying atypical observations, but can be further interpreted as encoding the relative influence each pixel is likely to have when determining galaxy weights, as the model fits are most sensitive to observations far from the centre of mass.\footnote{As a toy example, consider a simple linear fit to predict $y$ from $x$ using a scatter of observations $(x_i,y_i)$ centred around (0,0). Intuitively it is clear that points near the edge impact the model fit parameters more, that is those points with $|x|\gg0$, in this case the leverage captures this as it reduces to $h_i = x_i^2/\lVert x\rVert^2$.  Here $x$ is analogous to a standard template map, $y$ the observed number density, and $y_{\rm pred}\sim f(x)$ the contribution to the observed density that can be explained by systematics from the SP map, which is used to define the galaxy weights.} In the case of linear regression this connection is direct, with the leverage map encoding the direct response of the (inverse) weight to the observed over-density in that pixel: 
\begin{equation}
h_i = \frac{\partial(1/w_i)}{\partial\delta_i}.    
\end{equation}
This means that the inferred weights are particularly sensitive to pixels with high leverage, and the uncertainty of the weights in these pixels will be higher than average. Furthermore, these pixels are least likely to satisfy the assumptions of perturbative contamination models, which fall far from the mean. Note that the leverage measure comes only from the template maps themselves and incorporates no information from the observed density field, as it is purely a function of the distribution and correlation of points in the maps. We therefore use the leverage statistic as an additional indicator of (joint) survey property extremity, computing it for the 19 template maps used to derive weights (see Table \ref{tab:spmaps_table}) and after applying the \emph{SP outlier} mask described in the previous section. We mask the top 0.1\% of pixels, which removes negligible area but addresses the most offending pixels in the tail of the distribution. \\
\\
The square root leverage is shown in Figure \ref{fig:leverage}, with masked values indicated in red and the number of pixels at each value given by the inset histogram. We show the square root leverage as this is roughly how the RMS map error scales, assuming the templates span the contaminating systematics (\citet{Weaverdyck_2021}). As to be expected, the \emph{leverage} mask removes pixels at the edges of the footprint, largely adjacent to regions already removed by other masks, but offers further protection by accounting for the covariance between template maps when defining outliers. 

\subsubsection{Combination of SP outliers and leverage masks } 
\begin{figure}
    \centering
    \includegraphics[width=\linewidth]{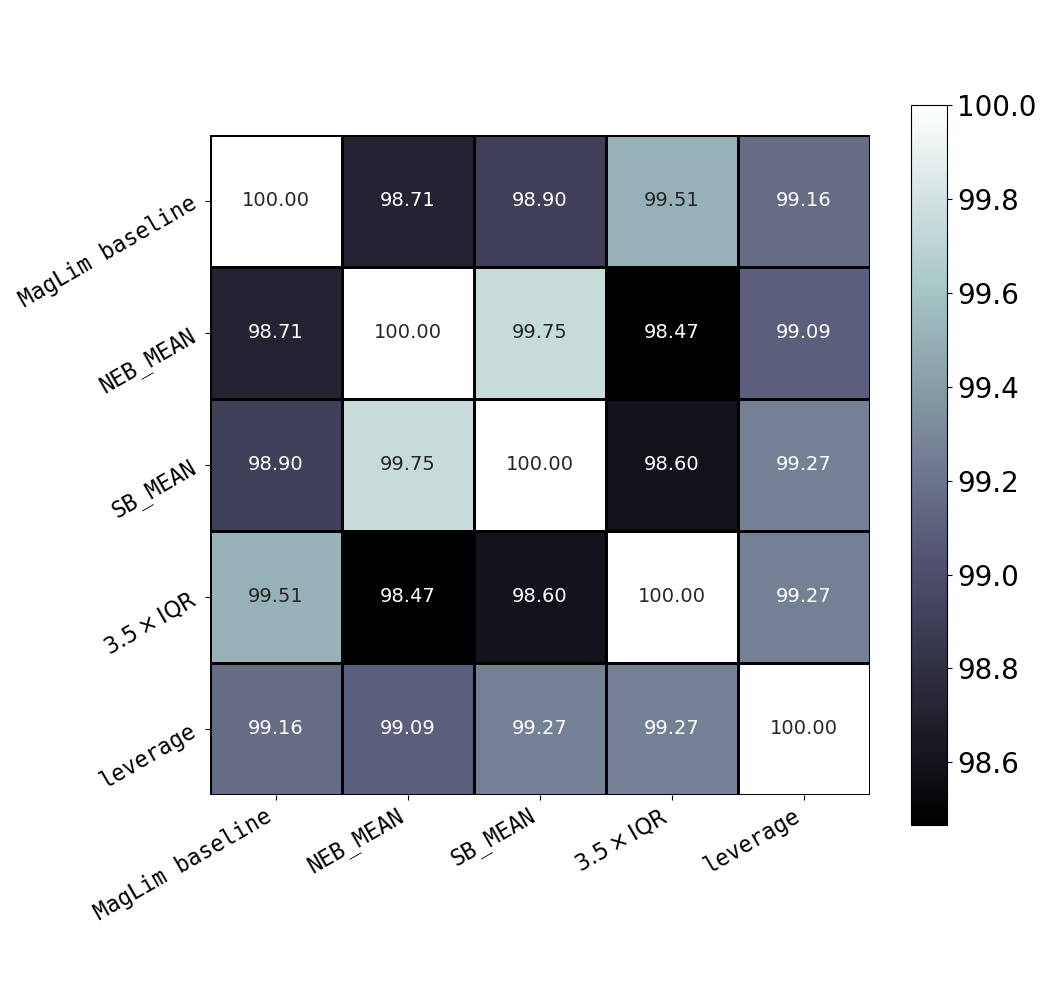}
    \vspace{-0.5cm}
    \caption{Matrix of overlap coefficients between the resulting sub-footprints after applying progressive masks to the initial \seed footprint, namely the \baselinemask mask and the components of the \systmask mask.}
    \label{fig:overlap_matrix}
\end{figure}

\begin{figure*}
    \centering
    \includegraphics[width=0.4\linewidth]{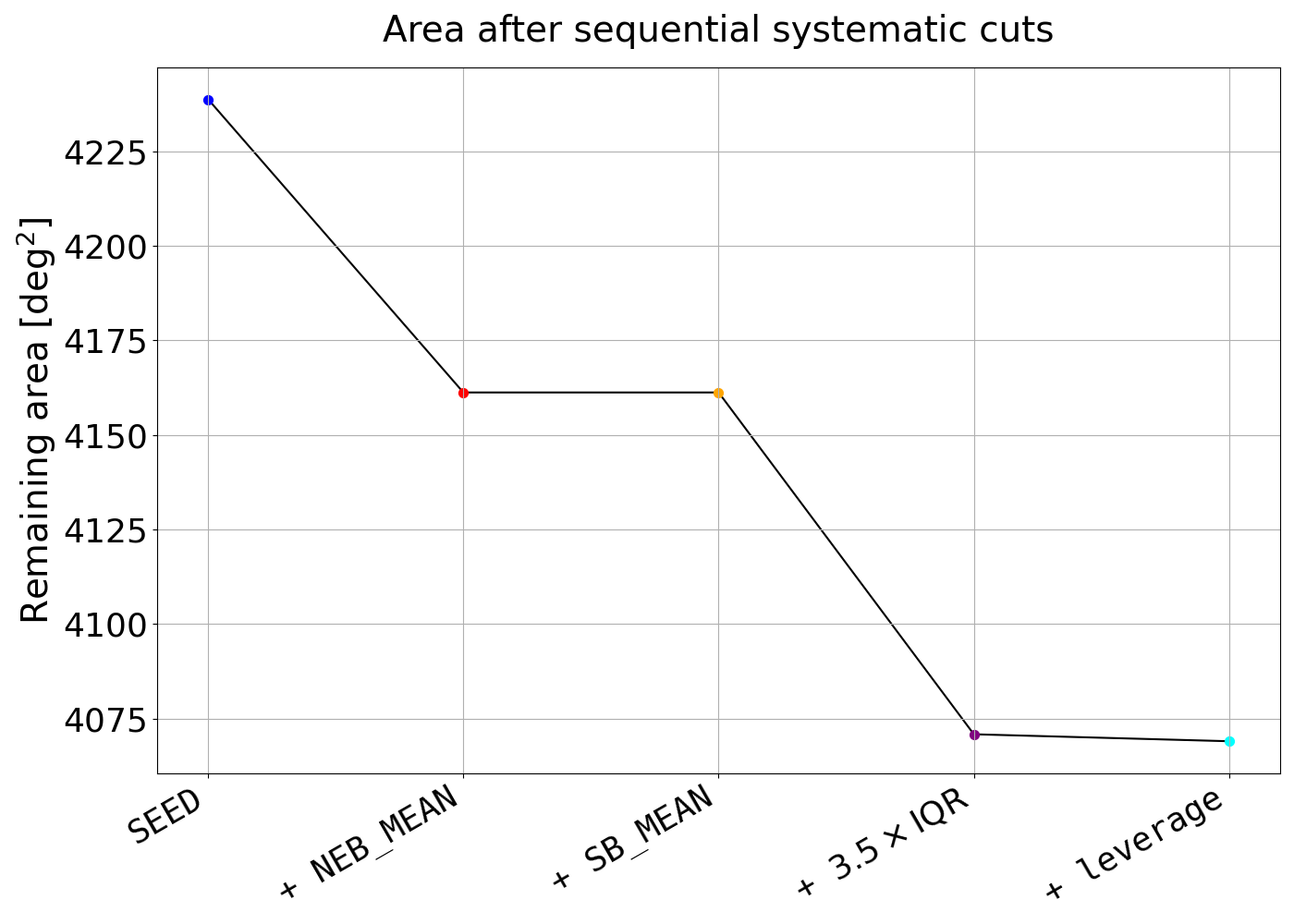}
    \includegraphics[width=0.55\linewidth]{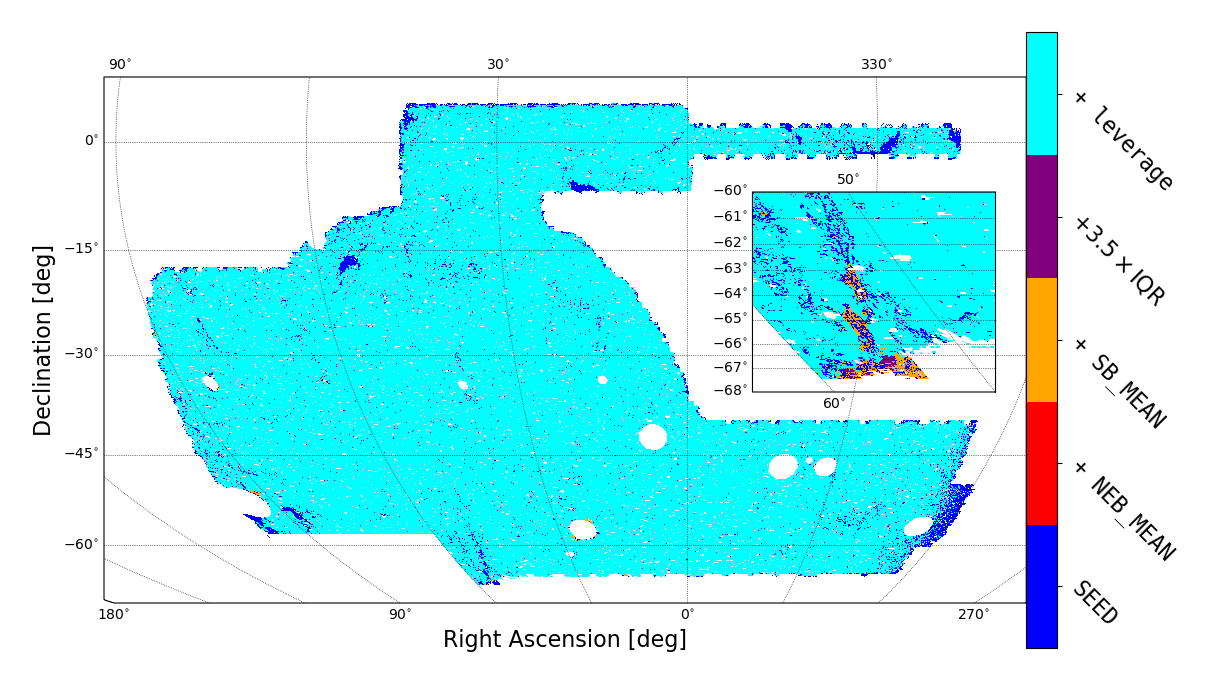}
    \includegraphics[width=0.9\linewidth]{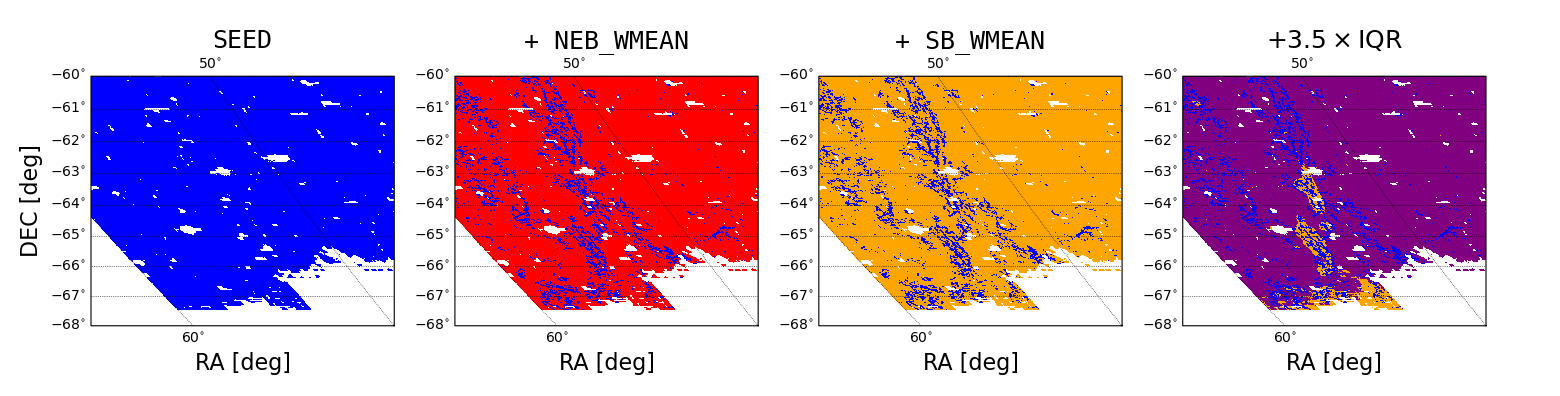}
    \caption{\emph{Top left:} Evolution of the remaining area (in square degrees) with respect to the area given by the \seed footprint after progressively applying the different cuts contributing to the \systmask mask. \emph{Top right:} Evolution on the sky of the \textbf{remaining footprint} after applying progressively the individual cuts contributing to the \systmask mask following the order from the plot on the top left panel. The dark blue area corresponding to the initial \seed footprint shrinks to the cyan area given by the \systmask mask after applying all individual systematic cuts. \emph{Bottom:} Detail of the footprint evolution on a zoomed-in region. \textbf{Each colour represents the surviving area after applying a new mask on top of the previous ones.} One should imagine five different footprints, one for each colour, lying on top of each other. Under this interpretation, we can see the area covered by a footprint at step $i$ through the holes created by the next mask when applied to it, which gives footprint $i+1$. For this reason, the dark blue regions from the \seed footprint are only visible through the holes left by the progressive masking. The same applies to the intermediate colours, i.e. we see them on the zones where the next mask removes them. This is depicted more in detail in the zoomed-in region: from the dark blue area, we pass to the orange one, then to the purple one and finally to the cyan one. We do not see the red area corresponding to the \texttt{CIRRUS\_NEB\_MEAN} cut because the following cut, \texttt{CIRRUS\_SB\_MEAN}, removes very little area with respect to the former (see plot on the left). The same applies to the purple ($3.5 \times \rm IQR$) and the cyan (\emph{leverage}) areas.}
    \label{fig:syst_mask}
\end{figure*}

Once the \emph{leverage} mask is created, we combine it with the previous \emph{SP outlier} mask, giving rise to the \systmask mask, which contains information about extreme values that may impact the reliability of the systematic corrections obtained from the two decontamination methods that we consider for Y6. We note that the overlap between the four sub-masks that form the systematics mask is large. Thus, when we apply them sequentially, some outlier pixels have already been removed in the previous masking steps. This is due to the existing spatial correlations between the different template maps considered. To quantify this, we compute the matrix of overlap between masks, shown in Figure \ref{fig:overlap_matrix}. To obtain this matrix, we define the overlap coefficient (percentage) between two masks on the full sky at \nside~=~4096 as the number of pixels where both masks are equal in a boolean sense divided by the total number of pixels. This way, a complete overlap yields a coefficient of 100\% and completely non-overlapping masks would result in a coefficient of 0\%. We confirm on the matrix that the level of overlap between masks is very high, presenting the highest values for the overlap between the two sub-masks from the cirrus estimate maps and between the $3.5 \times$IQR and \emph{leverage} masks. This is expected, since in the case of the Galactic cirrus maps, they are tracing the same quantity with different methods or statistics, while in the case of the $3.5 \times$IQR and \emph{leverage} it is also expected, as their dealing with the outliers from the same SP maps just in different ways. Nevertheless, all sub-masks present high levels of overlap,
indicating that an important part of the area removed by each of them is already removed by other sub-mask(s).  \\
\\
On the top left panel of Figure \ref{fig:syst_mask} we show the evolution of the remaining area after applying the different systematic cuts described in this section. The cuts coming from masking the outliers of both cirrus estimator maps remove $\sim 77.54 \, \rm deg^2$, and then the combined masking of outliers from the rest of template maps from both the $3.5\times \rm IQR$ cuts and the \emph{leverage} mask removes another $\sim 92.26 \, \rm deg^2$. Note how this area is smaller than the $100.45 \rm \, deg^2$ reported before for the $3.5\times \rm IQR$ cuts alone. This is again a consequence of the existing overlap with the masks from the cirrus estimate maps. From the point of view of area removed from the initial \seed mask, the overlaps between masks are clear on the same panel of Figure \ref{fig:syst_mask} as well: for the case of the cirrus maps, it can be seen how the outliers from the \texttt{CIRRUS\_SB\_MEAN} maps have a negligible removal of additional area with respect to what is removed by the outliers from the \texttt{CIRRUS\_NEB\_MEAN} maps. In the case of the rest of the template maps, we see how the additional area removal from the \emph{leverage} mask is also negligible with respect to the $3.5 \times$IQR outliers. This is so because the \emph{leverage} mask is obtained only from the main SP maps, i.e. a sub-set of the full list of template maps available. Even if we obtained the \emph{leverage} mask from the full list of templates, we do not expect both masks to differ significantly, since they are the result of identifying outliers in two different ways on the same template space. \\ 
\\
In total, after accounting for the overlap between its constituents, the \systmask mask removes $169.80 \, \rm deg^2$ from the initial \seed footprint. This makes the \systmask mask the element that removes the biggest area in this analysis. The top right panel of Figure \ref{fig:syst_mask} depicts the evolution of the surviving pixels at an \nside = 4096. Each colour in this figure represents the footprint resulting from sequentially applying the different systematic sub-masks presented in this section. For clarity, on the bottom panel of Figure \ref{fig:syst_mask} we depict the evolution of the footprint on a zoomed-in region. The final step, corresponding to the application of the \emph{leverage} mask, is shown within the plot on the top right panel in this same figure. 

\subsection{DES Y6 \jointmask footprint } 
\begin{figure*}
    \centering
    \includegraphics[width=\linewidth]{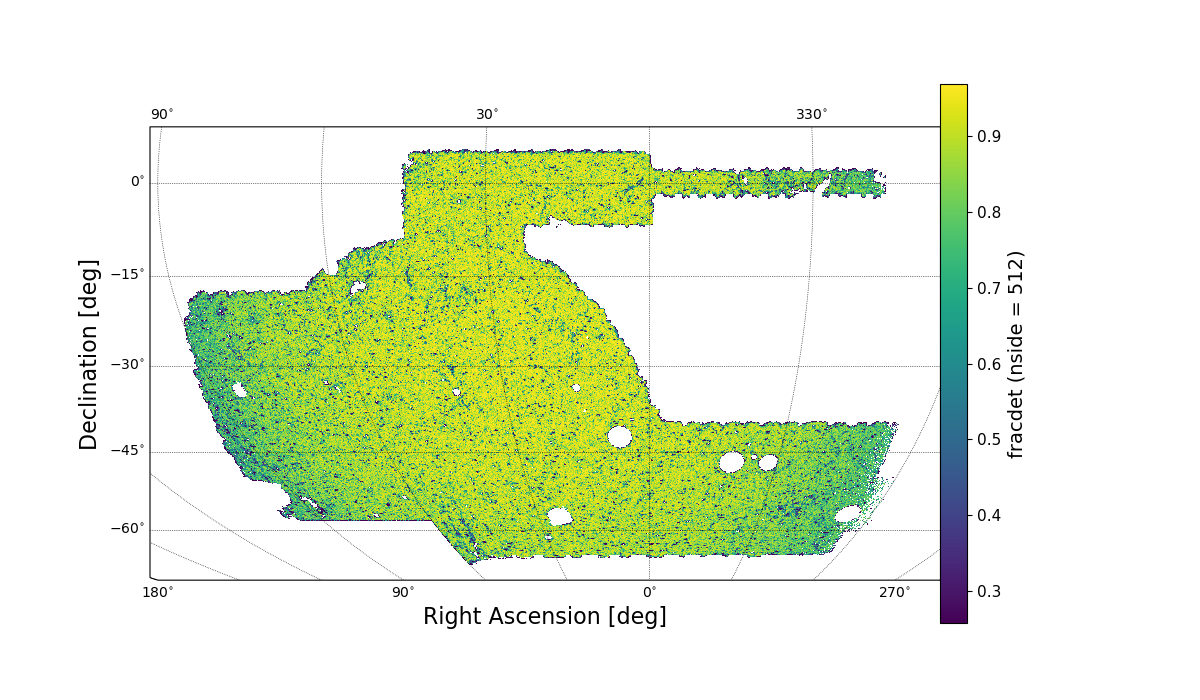}
    \caption{Final DES Y6 \jointmask footprint. It results from applying the \baselinemask and the \systmask masks to the initial \seed footprint. The final area, after removing the few extra pixels identified as problematic by the shear analysis team, is $4031.04 \rm \deg^2$. The colour scale corresponds to the fractional coverage or \fracdet of the pixels. To make the image less noisy, we show the final \jointmask footprint at \nside = 512. }
    \label{fig:joint_mask}
\end{figure*}

Finally, we combine the two masks produced in the previous steps, namely the \baselinemask and the \systmask masks presented in Sections \ref{sec:baseline_mask} and \ref{sec:syst_mask}, respectively, with the \seed footprint from Section \ref{sec:shear_mask} to obtain the \jointmask footprint. We note that this \jointmask footprint has some extra pixels removed, which were identified as problematic by the shear analysis team when this footprint was already defined. Therefore, we could not remove them from the beginning. Nevertheless, these few extra pixels were not taken into account in the production of the systematic weights in \citet*{y6weights}, as their removal was purely motivated by quality cuts affecting the shear part of the \threextwo analysis together with the fact that they only represent $\rm 9.49 \cdot 10^{-3} \, deg^2$ of additional masking. \\
\\
The final DES Y6 \jointmask footprint provides an area of $4031.04 \, \rm deg^2$ over which the main Y6 cosmology analyses are conducted. In Figure \ref{fig:joint_mask} we show the final DES Y6 \jointmask footprint at \nside = 512 (for visualisation purposes only) along the \fracdet per pixel. In Table \ref{tab:area_evol} we provide a synthesis of the area evolution after progressively applying the \baselinemask and the \systmask masks to the \seed footprint. The values in the last column also account for the few \emph{extra pixels} from the shear analysis mentioned above. We also note how the cumulative masked area after applying both these two masks (last row, fourth column) is smaller than the sum of areas masked by them individually (second row on the table). This occurs because there is overlap between these masks, as can be seen in the overlap matrix in Figure \ref{fig:overlap_matrix}. \\
\\
\begin{table}
    \centering
    \begin{tabular}{c c c c c}
        \multicolumn{5}{c}{\textbf{Area evolution}}\\ \hline
        \hline
        Incremental & & + \maglim & & \\
        masking & \textsc{seed} & \textsc{baseline} & + \textsc{syst.} & \textsc{joint} \\ \hline
        \multirow{1}*{Remaining area} & & & & \\ 
        $[\rm deg^2]$ & 4238.76 & 4198.21 & 4031.05 & 4031.04 \\ \hline
        \multirow{1}*{Masked} & & & \\
        area $[\rm deg^2]$ & 0 & 40.55 & 169.80 & 207.72 \\ \hline
        \multirow{1}{*}{Cumulative masked} & & & \\
        area $[\rm deg^2]$ & 0 & 40.55 & 207.71 & 207.72 \\ \hline
        \hline
    \end{tabular}
    \caption{Evolution of the remaining area as we progressively combine the three main levels of masking considered, namely from the initial \seed mask, the \baselinemask mask and the \systmask mask. We consider the area given by the \seed mask as the starting point from which we remove further pixels. The values on the last column correspond to the combination of all sub-masks into the final \jointmask mask, which also account for the $\rm 9.49 \cdot 10^{-3} \, deg^2$ mask from the few extra pixels identified a posteriori during the DES Y6 shear analysis. }
    \label{tab:area_evol}
\end{table}

\section{Impact of masking on observational systematics decontamination for DES Y6 \maglimpp lens galaxy sample} \label{sec:mask_impact}
Here we show how the elimination of extreme SP map values not only helps mitigate contamination in those particular pixels, but also enhances the flexibility and accuracy of systematic corrections via galaxy weights on the rest of the footprint. We further demonstrate that the \isd metrics improve after applying the \jointmask mask to the \maglimpp sample. We note that in the context of this section, as explained in Section \ref{sec:terms}, ``\jointmask mask'' refers the boolean object associated with the \jointmask footprint. When needed for degrading template maps, we use the \fracdet associated with this footprint as well. \\
\\
We focus on \isd because its one-dimensional metric makes it easy to visualize the impact of masking on the estimated contamination as compared to $N$-dimensional parameter space in which \enet operates. The following subsections present these results and assess the effect of masking on the two-point angular correlation function.

\subsection{Impact on the performance of \isd }\label{sec:impact_1ds}
Briefly, the \isd method works as follows (c.f. \citet{y1weights}, \citet{y3weights} and \citet*{y6weights} for more details): 
\begin{itemize}
    \item compute the relation between the observed galaxy number density, $n_{\rm obs}$, and a given template map, $s$, over the observed area. This is usually referred to as ``1D relation''; 
    \item compute $\chi^2_{\rm null}$ from the fit of the 1D relation to the null test line, i.e. to 1, which corresponds to no systematic effect from $s$ on the data (when compared with the average $n_{\rm obs}$ calculated on the full footprint);
    \item compute $\chi^2_{\rm model}$ from the fit of the 1D relation to a given function of the systematic template, $f(s)$; 
    \item define the systematic impact from $s$ on the data as $\Delta \chi^2_{\rm data} = \chi^2_{\rm null} - \chi^2_{\rm model}$; 
    \item evaluate the significance of $\Delta \chi^2_{\rm data}$ by computing the same quantity on a set of mocks with no contamination and compare against the distribution of $\Delta \chi^2_{\rm mocks}$. 
\end{itemize}
The 1D relations are constructed by binning template map values into 10 equally-spaced bins and evaluating $n_{\rm obs}$ within the corresponding sky regions for each bin. To reduce noise, we compute these relations using the galaxy sample and template maps degraded to \nside = 512. For a given template map $s$, its 1D coordinates, $s^b$, are computed as the average value of the pixels belonging to that bin weighted by their \fracdet at \nside = 512, that is 
\begin{equation}\label{eqn:sysmap_1d_bin}
    s^b = \frac{\sum_{p \in b}s_p \cdot f_p}{\sum_{p \in b} f_p} \, ,
\end{equation}
where the index $p$ runs over all pixels within the 1D bin $b$. Similarly, the average value of the observed galaxy number density at each 1D bin is computed as 
\begin{equation}\label{eqn:ngal_1d_bin}
    n_{\rm obs}^b = \frac{\sum_{p \in b} N_{{\rm obs},p}}{\sum_{p \in b} f_p} \, ,
\end{equation}
where $N_{{\rm obs},p}$ is the number of galaxies in each pixel $p$. Usually, the 1D relations are referred to the average observed galaxy number density over the full footprint, $\bar{n}_{\rm obs}$, which is computed as $n_{\rm obs}^b$, but running over all pixels on the footprint. This way, the null test line for no systematic impact is set at 1. \\
\\
Regarding the normalising factor $f_p$ in Equations \ref{eqn:sysmap_1d_bin} and \ref{eqn:ngal_1d_bin}, it corresponds to the effective observed area of pixel $p$ at \nside = 512, i.e., the \fracdet at that resolution. These two normalised quantities are unbiased under the following assumptions: a) the observed sub-area of a partially covered pixel is representative of the whole pixel; b) the missing area does not correlate with galaxy density or with the template value beyond what the mask already encodes (keeping in mind that the \nside = 512 mask is obtained from higher resolutions); c) that the counts scale approximately linearly with observed area; and d) that the low-\fracdet tail does not destabilise the binned estimates. We note that we do not impose any cut to remove low-\fracdet pixels at any point of this analysis, being the lowest value $1.53 \cdot 10^{-5}$ for \nside = 512. In Appendix \ref{app:degrading} we provide more information about the effects of this choice. \\
\\
With these considerations, when $\Delta \chi^2_{\rm data}$ exceeds a predetermined significance threshold, we correct for the impact of systematic $s$ by generating a map of correcting weights. This map is created by evaluating $1/f(s)$ at each pixel of $s$, where $f(s)$ is the function fitted to the binned 1D relation.\\
\\
The DES photometry is exceptionally uniform, and we are in the regime where systematic contamination of the lens galaxies (which are high SNR) is a small fraction of the average density. Thus contamination can be well-approximated via Taylor expansion, where we test both first and third order contamination models as the functional form of $f(s)$. This helps avoid misidentifying cosmological signal as systematic contamination, which would lead to overcorrection, while allowing us to use a relatively large number of template maps (19), ensuring we do not place strong priors on exactly the spatial pattern that systematic contamination can take.\\
\\
For the DES Y3 galaxy clustering analysis, we restricted ourselves to linear $f(s)$ dependencies for all template maps. However, for Y6 we decide to increase the flexibility of these dependencies by using a polynomial $f(s)$. This upgrade is motivated by \isd's contamination metric, $\Delta \chi^2$, as this quantity determines contaminant significance in both data and mocks. Low $\Delta \chi^2$ values can arise from two scenarios: minimal systematic impact where both $\chi^2_{\rm model}$ and $\chi^2_{\rm null}$ are small (with $f(s) \sim 1 \, \forall s$), or significant systematic impact where both values are large—indicating that while contamination exists (large $\chi^2_{\rm null}$), the inflexible $f(s)$ cannot adequately capture the 1D relation (resulting in large $\chi^2_{\rm model}$). This means that the sensitivity to different forms of contamination is crucial for \isd performance. To address this limitation, we enhanced \isd's flexibility by allowing $f(s)$ to be a polynomial up to third order. This choice accommodates the varying even or odd behaviours observed in different template map relations, with the cubic term coefficient free to vanish when unnecessary, while maintaining sufficient simplicity to avoid overcorrection. In this same regard, we did not consider higher order polynomials due to the limited number of 1D bins used (10).  \\ 
\\
In Y3, we ensured good performance of \isd by examining the $\chi^2_{\rm model}$ and $\chi^2_{\rm null}$ distributions at the start (i.e. with no weights) and at the end (i.e. with the final weights applied) of the process, respectively, finding neither evidence of deviations from linearity on the starting 1D relations nor hints of uncorrected systematic signal. However, for Y6, we decide to implement and test this upgraded version of the 1D fits from the very beginning of the analysis, so we avoid the aforementioned pathological $\Delta \chi^2$ values, should they arise. This allows us to optimise the decontamination process by detecting problems with any SP map earlier. Nevertheless, we keep the $\chi^2_{\rm model}$ and $\chi^2_{\rm null}$ tests in our battery of validation tests. \\
\\ 
Given these considerations, masking systematic template map outliers proves crucial at two stages of the decontamination process. First, outliers can degrade the quality of 1D fits (both linear and cubic), even when using binned data, potentially compromising the identification of contaminating templates. Second, extreme systematic values produce extreme correcting weights when evaluating the fitted $f(s)$ function, which can destabilize the correction procedure.

\subsubsection{Detection of significant systematic contamination}
\begin{figure*}
    \centering
    \includegraphics[width=\linewidth]{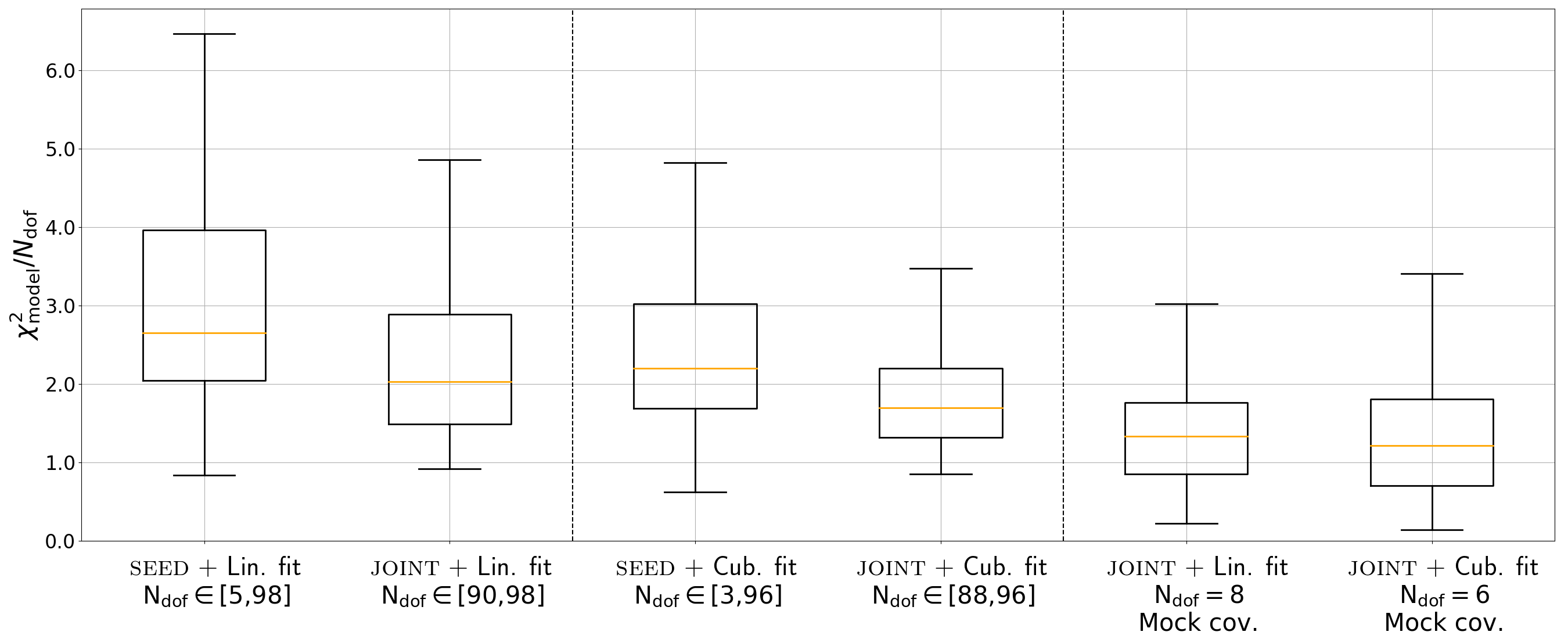}
    \caption{Median value and IQR of the reduced $\chi^2_{\rm model}$ values obtained from fitting the 1D relation between the observed \ngal in \maglimpp at each of its redshift bins and the 19 template maps used to compute correcting weights with a linear and a cubic function in combination with the \seed or the \jointmask masks. Each block groups together the reduced $\chi^2_{\rm model}$ values from all \maglimpp redshift bins. The lower and upper whiskers indicate the limits given by $\rm Q_1 - 1.5\,IQR$ and $\rm Q_3 + 1.5 \, IQR$, respectively. It can be seen how the lowest median values and the narrowest IQR are those from applying the \jointmask mask and using cubic fits. The values for the first four blocks are obtained without accounting for the covariance between 1D bins, so these are qualitative estimations. The last two blocks correspond to fiducial \isd runs with linear and cubic fits, respectively, where the 1D covariance is taken into account. }
    \label{fig:chi2_boxplot}
\end{figure*}

\begin{figure*}
    \centering
    \includegraphics[width=0.495\linewidth]{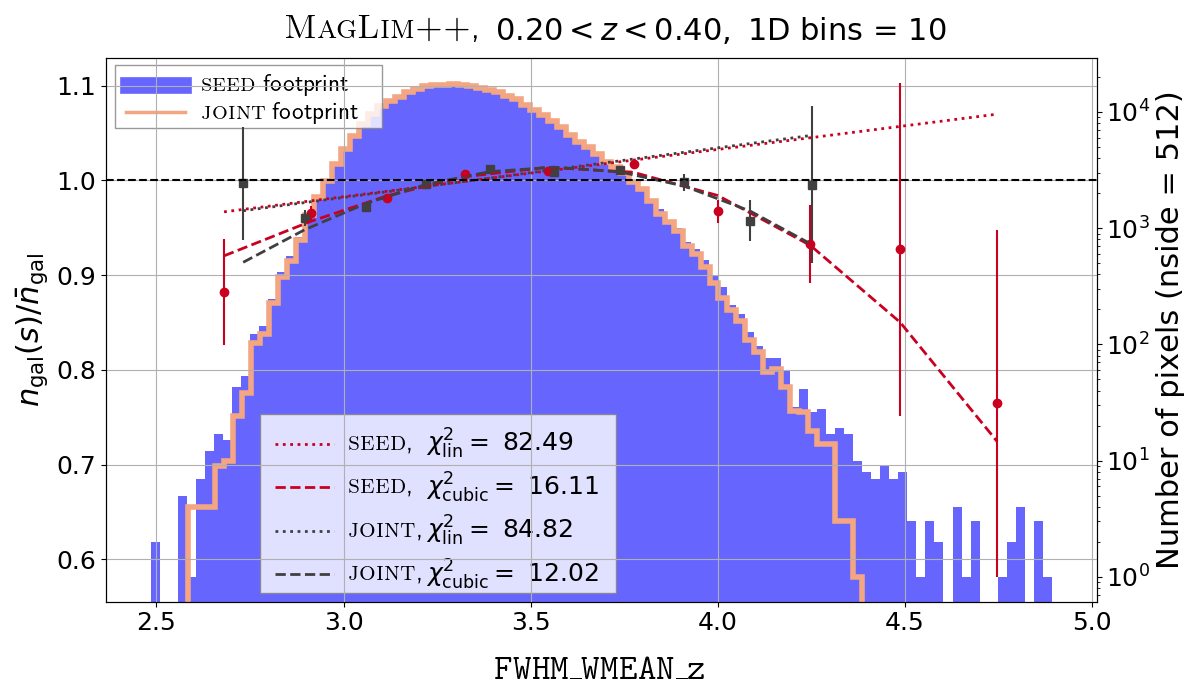}
    \includegraphics[width=0.495\linewidth]{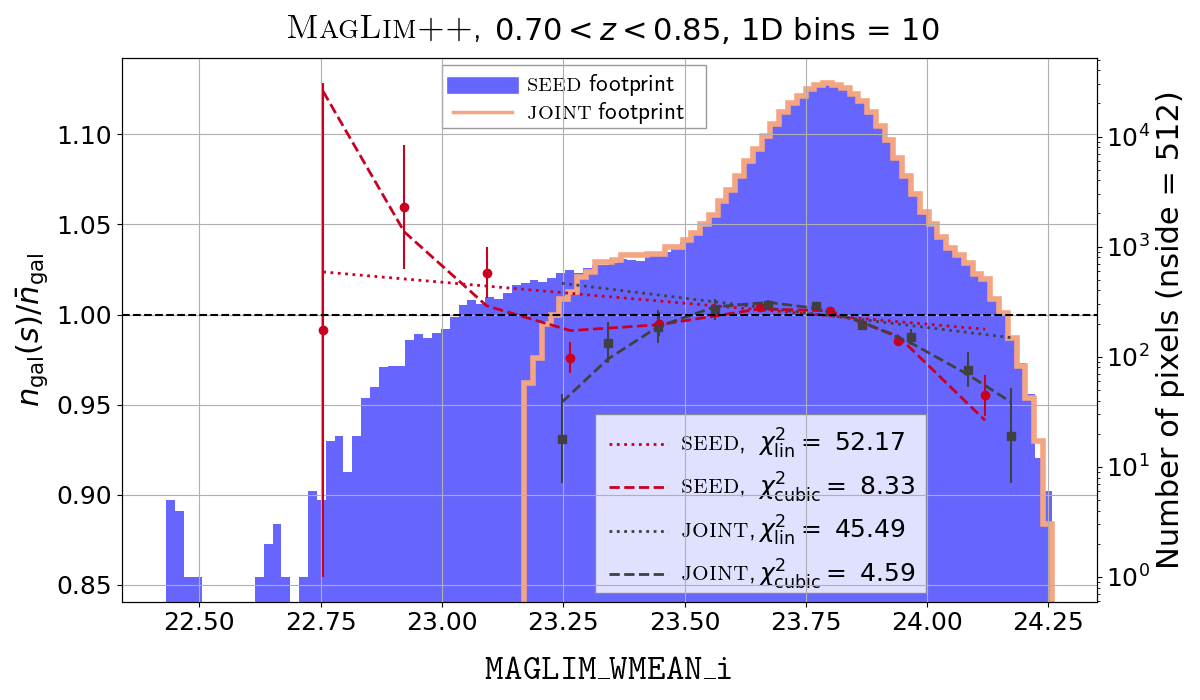}
    \includegraphics[width=0.49\linewidth]{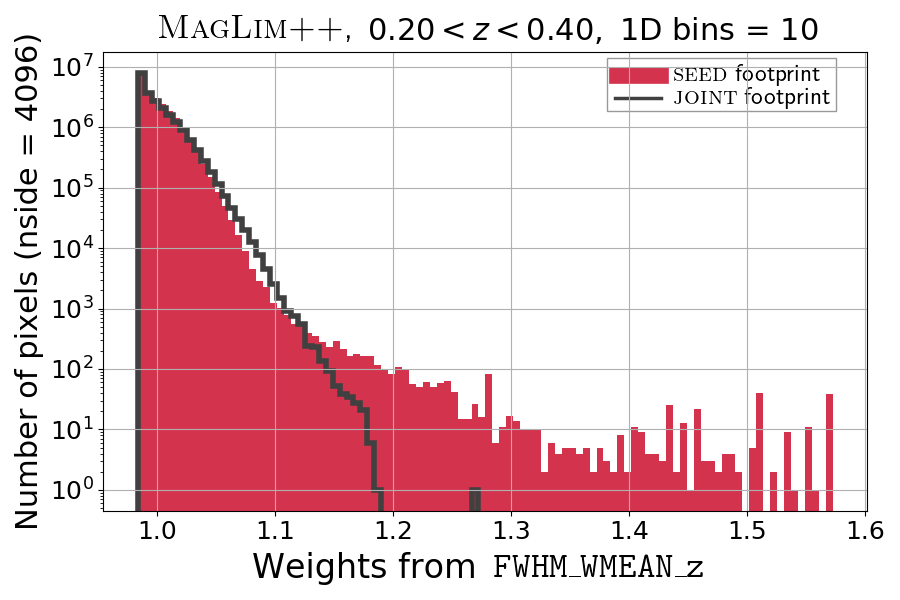}
    \includegraphics[width=0.49\linewidth]{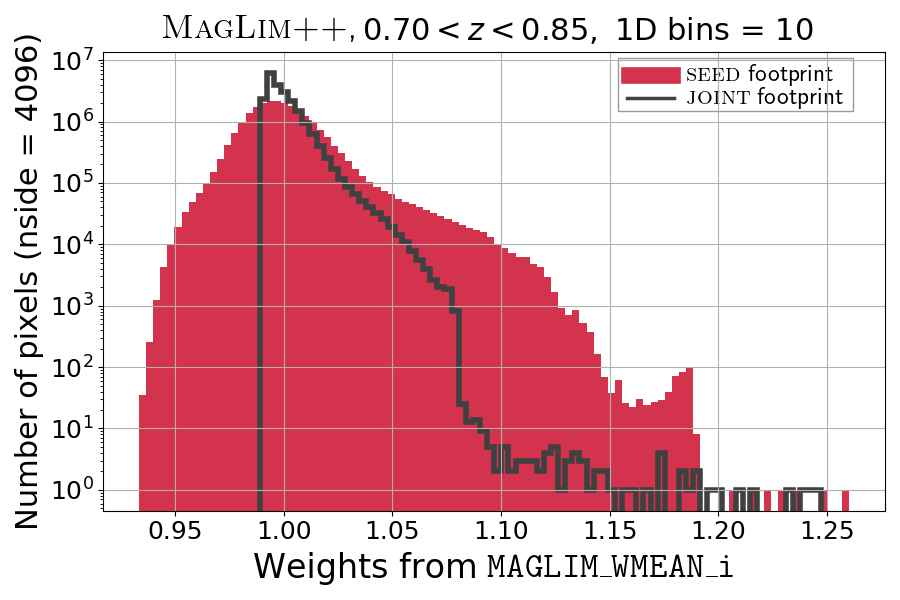}
    \includegraphics[width=0.49\linewidth]{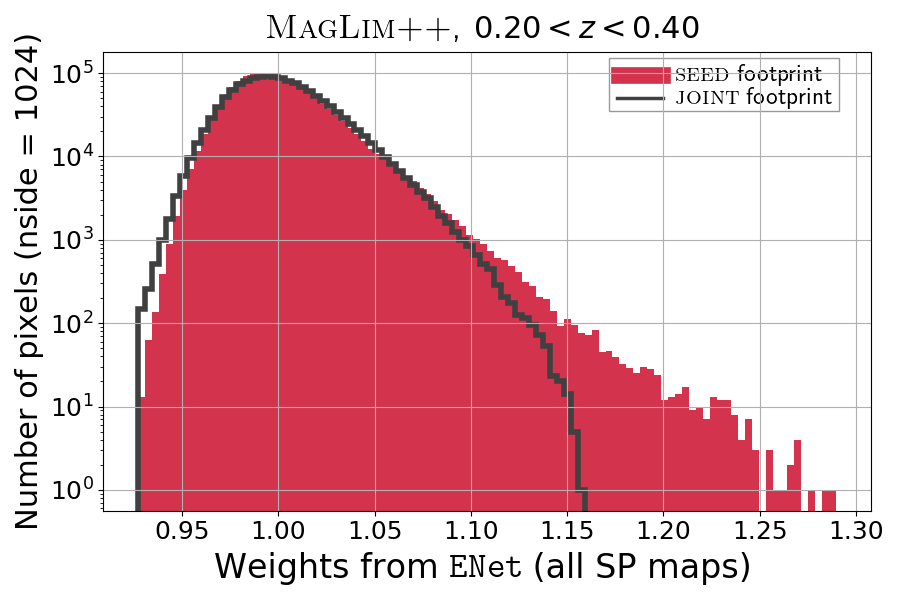}
    \includegraphics[width=0.49\linewidth]{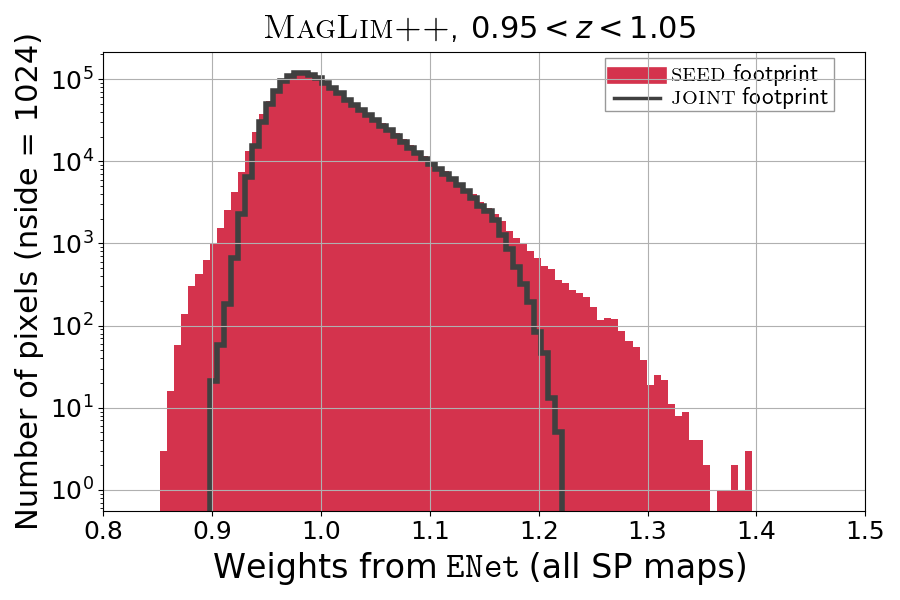}
    \caption{\emph{Top:} 1D relations for the \texttt{FWHM\_WMEAN\_z} and \texttt{MAGLIM\_WMEAN\_i} template maps with the \maglimpp sample evaluated at its first and fourth redshift bins with the \seed mask (red points) or the \jointmask mask (dark grey points) applied. To define these 1D relations, we use 10 equal width bins spanning the full range of template map values in both masking scenarios. 
    These value ranges are shown by the filled histogram (\seed mask) and the orange-outlined histogram (\jointmask mask). The $\chi^2_{\rm model}$ value for the cubic fit improves drastically after applying the \jointmask mask to the sample. \emph{Middle:} weight distributions obtained from the same template maps at \nside = 4096. It can be seen how the weights obtained from the \seed mask case reach greater values, which means that more extreme corrections would be applied to the data at those pixels. Note, that because these are 1D marginal fits at only a single step of the \isd{} process, we do not expect the curves to perfectly trace the observed density fluctuations, as other (non-independent) SP maps also trace systematics. \emph{Bottom:} weight distributions obtained with \enet. We note that in this case the weight maps come from all SP maps together, as \enet performs a simultaneous fit to all of them. The resolution of these weight maps is lower as the ones obtained with \isd, with an \nside = 1024. As in the previous case, we observe how the weights obtained with the \seed mask reach greater values as a result of the presence of extreme SP map values, which are not present in the case of the \jointmask mask. We show the two redshift bins where we find that the effect is more evident. }
    \label{fig:mask_1ds}
\end{figure*}

As previously discussed, the \isd method identifies contaminating templates by fitting the binned 1D relation of each map against the data and evaluating the significance of $\Delta \chi^2$ by comparing against the same quantity from mocks. 
Increasing the flexibility of the fitting function, $f(s)$ allows more freedom to capture systematic signal, but the presence of outliers can yield extreme 1D bins that cannot be properly fitted even with this greater flexibility and which may violate the perturbative assumptions made about contamination. To evaluate the effect of masking on the quality of the 1D fits, and thus on the detectability of contaminants, we look at the $\chi^2_{\rm model}$ values obtained from four different cases: linear / cubic fits with only the \seed mask applied and linear / cubic fits with the \jointmask mask applied to the \maglimpp sample at each redshift bin separately and to each SP map. We use the same set of 19 baseline template maps as used for the fiducial analysis. \\
\\
Ideally, to test changes on the performance of \isd, we should evaluate each set $\{$template map, mask, sample$\}$ as if they were going to be used to obtain correcting weights with \isd, that is, with the same configuration settings for the 1D relations. For DES Y6, the main configuration that we use consists of 10 1D bins with equal width. However, the presence of outliers when applying only the \seed mask poses some problems in this regard, so we need to compute the $\chi^2_{\rm model}$ values in a slightly different way from what is done to obtain the final weights. First, outliers lead in some cases to empty 1D bins. If the outliers are very extreme, the number of non-empty bins can be less than the number of parameters we want to fit. This problem arises especially in the cubic case. To overcome this, we start binning the distribution of each template map in 100 equal width bins and compute the $\chi^2_{\rm model}$ from a linear and a cubic fit on the non-empty 1D bins. This means that the number of degrees of freedom of each fit will vary from one map to other, so a direct comparison with a $\chi^2$ distribution is not possible. Instead, we consider the reduced $\chi^2_{\rm model}$ from each map, that is $\chi^2_{\rm model} / N_{\rm dof}$. \\
\\
The second difference is related with the uncertainties used in the fits: when computing correcting weights, the different $\chi^2$ values are obtained using errors from log-normal mocks and taking into account the covariance between 1D bins (see \citet*{y6weights}). This is done by obtaining the 1D relations from the mocks with the exact same configuration as in the data in terms of 1D binning and masking, so any change in the setup would lead to re-evaluating on the set of 1000 mocks. Furthermore, the mocks are created on a preliminary version of the final \jointmask footprint, so they are not defined in the full area given by the \seed footprint. However, since this is intended only for testing purposes, we calculate the $\chi^2_{\rm model}$ values for this test computing the errors as $\sigma_{y^i} = \sigma_{n_{\rm gal}^i}/\sqrt{N_{\rm pix}^i}$, i.e., not considering the covariance between 1D bins. Since this covariance is non-negligible, its contribution to the reduced $\chi^2_{\rm model}$ cannot be neglected, and therefore \textbf{we cannot simply compare the obtained distributions with the expected value of 1 for a good fit.} \\
\\
Considering these differences with respect to the main configuration, we show in Figure \ref{fig:chi2_boxplot} the box-and-whisker plot for the four cases considered. Each box groups together the reduced $\chi^2_{\rm model}$ values from the 19 1D fits at the 6 redshift bins of the unweighted \maglimpp sample. The orange lines correspond to the median value, the vertical range of the boxes represents the IQR (defined as in Equation \ref{eqn:iqr}), and the lower and upper whiskers show the values at $\rm Q_1-1.5\,IQR$ and $\rm Q_3+1.5\,IQR$ (the value of 1.5 is just the usual choice in this kind of representation), respectively. On the left half of the plot we show the box-and-whiskers for the reduced $\chi^2_{\rm model}$ distributions obtained when the 1D relations are fitted with a linear function. Similarly, on the right half of the plot we show the same but the cubic fits. \\
\\
It can be seen how the median reduced $\chi^2_{\rm model}$ gets lower in the two cases where we apply the \jointmask mask. If we compare the linear and cubic fits, the cubic ones yield better median reduced $\chi^2_{\rm model}$ values than the linear fits in both masking cases, since the model gains flexibility. We need to keep in mind that, \textbf{since we are not accounting for the 1D covariance obtained from the log-normal mocks, these comparisons are only qualitative, and these are not exact $\chi^2$ values.} Furthermore, each SP map only explains some (or none) of the density variations; unexplained density variations are then due to both noise/cosmic variance and systematics due to other SP maps, so we do not expect a single SP map 1D fit to explain all the variation. We also observe that the IQRs become narrower when using cubic fits, with the narrowest IQR corresponding to the \jointmask mask scenario. This indicates that the reduced $\chi^2_{\rm model}$ values are less scattered in this case. We note that the IQR for the \jointmask mask + linear fits configuration is slightly wider than that for the \seed mask + cubic fits. This is not unexpected, as in the presence of moderated 1D outliers a cubic fit could still provide a better description of the 1D relation, i.e. better reduced $\chi^2_{\rm model}$ values. Nevertheless, this can result in issues when computing correcting weights, as we discuss in the next subsection. In the labels on the x-axis of the plot we include the range of $N_{\rm dof}$, i.e. non-empty 1D bins minus 2 or 4 for the linear and cubic fits, respectively, obtained for each masking case. Note how these ranges are much shorter when we apply the \jointmask mask, meaning that there are much fewer empty bins as a result of the removal of SP outliers. \\
\\
In order to compare with true $\chi^2$ values, on the right part of Figure \ref{fig:chi2_boxplot} we show the $\chi^2_{\rm model}$ distributions from a fiducial run of \isd, that is using 10 1D bins, accounting for the covariance between them and with the \jointmask mask applied. We compute these values for both the linear and the cubic fits. We see how the median $\chi^2_{\rm model}$ further lowers, being mildly lowest for the cubic case. The outliers for this case have been identified, being \texttt{MAGLIM\_WMEAN\_i} evaluated at the first redshift bin the only map that is beyond the $\rm Q_3 + 1.5IQR$ upper limit. As it can be seen in Figure \ref{fig:1d_maglim_i} in Appendix \ref{app:chi2_model_vals}, this high $\chi^2_{\rm model}$ value is driven by the third 1D bin. We also show the 1D relation after having weighted for this map, where it can be seen how that is the only 1D bin that is not consistent with 1. Nevertheless, we confirm that it is not necessary to weigh for this map for a significance threshold of $T_{1D}  = 2$, which is our fiducial configuration to obtain corrective weights (see \citet*{y6weights}). We also verify that none of the SP maps with the highest $\chi^2_{\rm model}$ has to be corrected for. In Table \ref{tab:spmaps_weighted} we present the set of maps that \isd finds it has to correct for.  \\
\\ 
These results show that the removal of anomalous values from the template maps leads to an improvement on the characterisation of the 1D relations. In fact, this facilitates the usage of cubic fits, which enhances the ability of \isd to identify systematic contaminants to which a linear fit could be insensitive to. \\
\\
To further exemplify the importance of this improvement, we check the effect of the two masks considered in this section on the 1D fits of two template maps using 10 1D bins, the same as to compute the official weights. We still do not use the covariance matrix from the mocks mentioned before, but this represents a case closer to what we do to obtain such systematic corrections. In the upper panel of Figure \ref{fig:mask_1ds} we show the impact on the quality of both linear and cubic fits to the 1D relations of the \texttt{FWHM\_WMEAN\_z} and \texttt{MAGLIM\_WMEAN\_i} maps at \maglimpp's first and fourth redshift bins, respectively, when evaluated only with the \seed mask and with the \jointmask one applied. Focusing on the former masking scenario, it can be seen in the blue filled histograms and in the red dots of both maps how the range of values that they cover is wider, as expected. For the linear fits (dotted red lines) in both maps, we find very high $\chi^2_{\rm model}$ values ($N_{\rm dof} = 8$). On the other hand, the cubic fits (dashed red lines) show better $\chi^2_{\rm model}$ values ($N_{\rm dof} = 6$), but this is influenced by the big error bars from the most extreme 1D bins, where either the galaxy number density can be very low as in the case of the highest \texttt{FWHM\_WMEAN\_z} values, or the number of available pixels is very small, as happens for the shallowest \texttt{MAGLIM\_WMEAN\_i} values. \\
\\
If we now look at the case with the \jointmask mask applied, we can see how the $\chi^2_{\rm model}$ for the linear fit (dark grey dotted line) remains very high (for \texttt{FWHM\_WMEAN\_z} it is in fact higher than in the previous masking case) given the non-linear contamination behaviour of these maps, but the $\chi^2_{\rm model}$ for the cubic fits (dark grey dashed line) experiences an improvement. \\
\\
We have chosen to illustrate the impact of masking on the 1D fits with the \texttt{FWHM\_WMEAN\_z} and \texttt{MAGLIM\_WMEAN\_i} maps because, as it is shown in the companion paper for DES Y6 galaxy clustering (\citet*{y6weights}), they are an example of how two contaminants that have an effect on the observed \ngal of $\sim 5\% - 10\%$ (as showcased by the dark grey dots in Figure \ref{fig:mask_1ds}) may go unflagged as significant (for the statistical power of DES Y6) or less significant with linear fits, but do trigger \isd's contamination metric when cubic fits are applied. Furthermore, in Table \ref{tab:spmaps_weighted} we show a comparison of the maps flagged as significant when using cubic and linear fits. As it can be seen in the table, at some redshift bins it is necessary to weigh for more maps in the linear case than in the cubic one. This happens because the quality of the linear fits is worse, so \isd needs more iterations to achieve convergence, while the cubic fits are more efficient at detecting and correcting contamination. \\ 
\\
These results demonstrate the importance of allowing the decontamination methods to be more flexible and how the removal of pixels containing extreme contamination contributes to achieving this. Thus, combining the joint mask with cubic 1D fits improves the characterisation of dependence between the observed \ngal and observational systematics. 

\subsubsection{Computation of correcting weights}
Upgrading \isd's contamination metric to a cubic polynomial fit enhances its sensitivity to non-linear contamination modes. However, the usage of such cubic function can raise a problem when obtaining correcting weights with it. In general, for a given template map $s_i$ the corresponding weights are defined as $1/f(s_i)$, where the form of $f(s_i)$ and how it is obtained depends on the correction method. In the case of \isd, it corresponds to the cubic function mentioned before and how it is obtained is described below. \\
\\
From a general perspective, we must consider the effect of the fitted coefficients, since the quadratic and cubic terms in $f(s_i)$, used to fit the observed $n_{\rm gal}$, can be very sensitive to extreme values of a given template map, $s_i$. Also, the presence of such extreme values decreases the validity of the perturbative description that we use for the contamination, as we later explain. This affects the accuracy of the corrections, since it depends on the uncertainty of the estimated coefficients of the fit. As a consequence, the corrections at pixels with such outlier values would not only be large, but would also have large associated uncertainties. \\ 
\\
It is at this point where the masking of extreme values plays a major role as well, because by removing such values we are not only improving the fitting (and therefore, the detection) of contaminants, but we are also making the corrections more reliable by avoiding pixels where the contamination cannot be characterised by a function with the flexibility limitations that we impose. Of course, corrections provided by linear functions, as in DES Y3, are less sensitive to extreme values. However, their description of the systematic contamination is limited when the statistical power increases, as with Y6, so the accuracy of those corrections decreases as well (compare for example the red dashed and dotted lines on the 1D relations of the upper right panel in Figure \ref{fig:mask_1ds}). For this reason, there must be a trade-off between flexibility of the correction and domain of contamination where it can be applied. \\ 
\\
On the other hand, looking at the specifics of how \isd computes correcting weights, the problems derived from template map outliers are related to two aspects of how this method works. For a given template map, $s_i$, \isd fits a cubic polynomial, $f(s_i)$, to its 1D relation, obtaining the parameters of this function taking into account the following details: \\
\begin{itemize}
    \item first, the 1D relation is obtained from the sample and the template map at \nside = 512, that is $s_i^{512}$. As mentioned before, this is done to reduce the noise on the fits. This \nside is also limited by the resolution of the log-normal mocks needed to determine the significance of the contamination\footnote{In Y3 this limited us to \nside = 512, but in Y6 we have produced mocks at \nside = 1024 as well. While this allows us to look for systematic contamination at smaller scales, it also slows down substantially the performance of \isd. For this reason, after testing that we obtain similar results in terms of number and type of detected contaminants when using a subset of mocks at \nside = 512 and 1024, we decide to keep using the lower resolution ones. Nevertheless, this corresponds to an angular resolution of $\sim 6.87 \rm \, arcmin$, which is still below the scale cuts for our galaxy clustering measurements.}; 
    \item secondly, the 1D relation is computed by binning the distribution of $s_i^{512}$ values.\footnote{We find no significant difference in terms of number and type of detected contaminants when choosing different numbers of bins.}
\end{itemize}
Both aspects imply that high-resolution pixels containing outlier values can be absorbed into more average values either when degrading to lower \nside or when binning the data. If that is the case, the problem can arise if $s_i$ is flagged as contaminant, because the corresponding correcting weights are defined by evaluating $1/f(s_i)$ \textbf{on a higher resolution version of that map}. In our case, this is done at \nside = 4096. Therefore, if any outlier pixels were hidden either by the degrading or the binning, they will reappear at this point. As mentioned before, the cubic and quadratic terms in $f(s_i)$ can be very sensitive to extreme values of $s_i$, so when evaluating $f(s_i^{4096})$ on the reappeared outliers, they can generate large weights, and then introduce large corrections. For the reasons already exposed, we want to avoid this kind of situation, as the accuracy of such extreme corrections is harder to control. In addition, given the iterative nature of \isd, correcting for some map $s_i$ at a given iteration affects the 1D relations of this map (and of all maps that correlate with it) at the following iterations, so weighting for map values not properly described by $f(s_i)$ can prevent the iterative process from converging. \\
\\
Considering all this, the usage of the \jointmask mask allows us not only to improve the detectability of complex forms of contamination with \isd, but also to enhance the accuracy of the derived corrections. To illustrate this, in the middle panel of Figure \ref{fig:mask_1ds}, we show the distribution of weights obtained from \texttt{FWHM\_WMEAN\_z} and \texttt{MAGLIM\_WMEAN\_i} at \maglimpp's first and fourth redshift bins, respectively, when calculated with the \seed mask (red filled histograms) and the \jointmask mask (black-outlined histogram) applied. To compute these weights, we fit separately the cubic function $f(s_i^{512})$ to the corresponding 1D relation (red and black dashed lines on the upper panel of Figure \ref{fig:mask_1ds}) and then we evaluate $1/f(s_i^{4096})$ on the map with the corresponding mask applied. In this sense, if we consider a limit for the systematic fluctuations of 10\% on the 1D relations (i.e., $\Delta n_{\rm gal} / \bar{n}_{\rm gal}<0.1$), we can see how the tails of the weight distributions reduce considerably after using the \jointmask mask. For instance, the amount of pixels with values higher than 1.3 (i.e., pixels that require a correction of 30\%, where the perturbative assumption may break down) reduces substantially in the case of \texttt{FWHM\_WMEAN\_z}. In particular, the number of pixels with weight values greater than 1.3 reduces from 320 in the case of the the \seed footprint to 0 for the \jointmask footprint, the former corresponding to $190.0 \, \rm arcmin^2$ for an \nside = 4096. This indicates that the needed corrections are smaller, ensuring that we remain within the range of validity of the perturbative approach. \\
\\
We also account for the impact of masking systematics on \enet by applying the \emph{leverage} mask as well. As shown in the overlap matrix in Figure \ref{fig:overlap_matrix}, this mask greatly overlaps with the \emph{SP outlier} mask constituents (the other main element of the \systmask mask). Therefore, the effect of the \emph{leverage} mask in the performance of both \isd and \enet should be negligible with respect to what is already accounted for by the \emph{SP outlier} mask, but we still apply both to be more strict in terms of systematics control. Regarding the impact of the \jointmask mask on the quality of \enet weights, we also test their distributions when calculated on the \seed and the \jointmask footprints, similarly to what is done for \isd. As it can be seen in the bottom panel of Figure \ref{fig:mask_1ds}, the \enet weights obtained on the \seed footprint (red filled histograms) cover a wider range of values, while computing the weights on the \jointmask footprint (black-outlined histograms) results in narrower weight distributions, similar to the case of \isd. We note that, given the multi-dimensional fitting nature of \enet, the weight distributions shown in the figure do not correspond to individual SP maps, but to the full set of them used to compute systematic weights (see \citet*{y6weights}). These results demonstrate that the usage of the \jointmask footprint impacts the performance of our two methods for observational systematics decontamination by reducing the size of the corrections needed. 

\subsection{Impact on the measurement of $w(\theta)$}
\begin{figure*}
    \includegraphics[width=\linewidth]{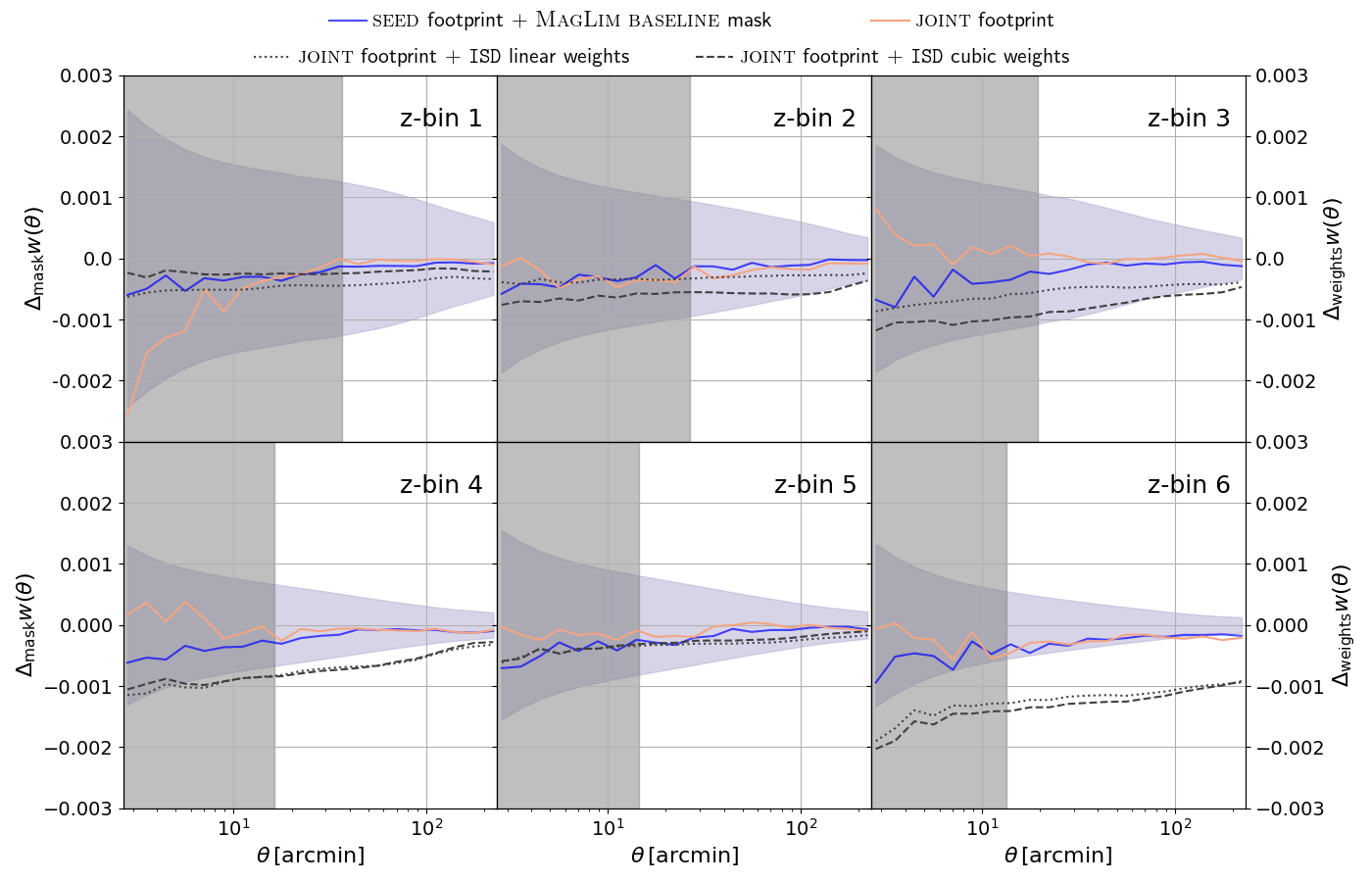}
    \caption{Impact of the different levels of masking on the measured angular correlation function of galaxy clustering, $w(\theta)$. We define $\Delta w_{\rm mask}(\theta) = w_{\rm mask}(\theta) - w_0(\theta)$, where $w_0(\theta)$ corresponds to the data with the \seed mask applied and $w_{\rm mask}(\theta)$ to further masks applied. In both cases, no corrective weights are applied. We consider the intermediate case of \seed footprint + \baselinemask mask (blue solid line) and the final case with the \jointmask footprint (orange solid line). Negative values of $\Delta w(\theta)$ indicate a reduction on the $w(\theta)$ amplitude as a result of masking. This behaviour is compatible with that observed when mitigating observational systematics with correcting weights (\citet*{y6weights}), although there is no evident trend. The grey dotted and dashed lines depict $\Delta_{\rm weights} = w_{\rm JM}^{\rm weights}(\theta) - w_{\rm JM}(\theta)$, i.e. the impact of the linear and cubic correcting systematic weights, respectively, on the data with the \jointmask mask applied with respect to $w(\theta)$ with this mask also applied, but no correcting weights. We observe at some redshift bins, e.g., the second one, that the cubic weights result in a $\sim 1\sigma$ reduction of the amplitude of $w(\theta)$ at large angular scales with respect to the linear weights. After studying the level of over-correction introduced by \isd's cubic fits in \citet*{y6weights}, we determine that this reduction on the $w(\theta)$ amplitude is a signature of a better identification and correction of systematic signal, enabled by the \jointmask mask.} 
    \label{fig:delta_wtheta}
\end{figure*} 

Until this point, we have evaluated the impact of different masks only on the 1D relations the systematic weights derived from them, i.e. only on 1-point statistics, since the observational systematics decontamination methods that we consider work at this level, either in 1D or $N$D. However, the effect of masking should be noticeable at the 2-point statistics level as well, especially at small angular scales, where the impact of removing high resolution pixels could be more important. Motivated by this, we test the sensitivity of the \maglimpp $w(\theta)$ measurements to variations in the applied mask. In order to check this, we define the impact of a given mask on $w(\theta)$ as 
\begin{equation}
    \Delta_{\rm mask}w(\theta) = w_{\rm mask}(\theta) - w_0(\theta) \, , 
\end{equation}
where $w_0(\theta)$ is the clustering correlation function measured on \maglimpp in our initial footprint, i.e. the \seed footprint, and with no corrective weights applied. We consider two levels of footprint to compare $w_0(\theta)$ against: an intermediate step consisting of the combination \seed footprint + \baselinemask masks, and the final \jointmask footprint, which contains all cuts presented in this work, including the intermediate step. \\ 
\\
To compute $w(\theta)$ we use \treecorr\footnote{\url{https://rmjarvis.github.io/TreeCorr/_build/html/index.html}} with the pixel-based Landy-Szalay estimator (\citet{1993ApJ...412...64L}). We use this option instead of the original, pair-count version of this estimator, for simplicity, since this is intended only for testing purposes and not for cosmology inference. Nevertheless, we compute $w(\theta)$ in each case using the masks and the \maglimpp sample at \nside = 4096, corresponding to an angular resolution of $\sim 0.86 \, \rm arcmin$, which is well below the scale cuts for galaxy clustering established by the Y6 modelling analysis (\citet*{y6modelling}). In addition, we also used this pixel-based approach in DES Y3, where we checked that there are no significant differences with respect to the pair-count estimator for our scales of interest. \\ 
\\
In Figure \ref{fig:delta_wtheta}, we showcase the $\Delta_{\rm mask} w(\theta)$ from the two levels of masking mentioned above compared with the statistical error at each redshift bin of the \maglimpp sample. These errors correspond to the diagonal of the galaxy clustering part of the DES Y6 covariance matrix (see \citet*{y6modelling}) evaluated at each redshift bin. According to its definition, negative values of $\Delta_{\rm mask} w(\theta)$ indicate that the amplitude of $w(\theta)$ reduces with respect to the \seed only case. Higher amplitudes would be expected in the presence of outlier pixels, since the observed number of objects is biased at them by the presence of extreme systematics, which imprints structure on the data. It can be seen how the impact of both the \baselinemask and the \jointmask masks is more remarkable at low angular scales, although within the ranges excluded by the scale cuts. For larger angles, the impacts are not significant compared with the statistical error, except for the last redshift bin, where we appreciate that both masks have an effect of $\sim 1 \sigma$. \\ 
\\
From these results, we find that after applying both levels of masking the trend is to mildly reduce the $w(\theta)$ amplitude for redshift bins 1, 2, 4, and 5. For the third bin, the \jointmask footprint seems to have the opposite effect, while for the sixth bin the effect of both the \seed + \baselinemask and \jointmask footprints is to reduce the clustering amplitude to $\sim 1\sigma$ level. This is not contradictory, since that curve corresponds to the data containing observational systematics contamination yet, and these systematic effects impact each redshift bin in different ways. This is only an indication that observational systematics have a different effect on the sample when evaluating them on the \jointmask footprint or the \seed footprint + \baselinemask mask. In this sense, we stress that the effect of the mask is not to completely remove the impact of observational systematics, but to make it easier to mitigate them. The negative values of $\Delta_{\rm mask} w(\theta)$ are equivalent to the effect that we see when we mitigate contamination with correcting weights, as we show in \citet*{y6weights}. All redshift bins show that the impact of masking is well within the error bars, except for the last one, where we find both masking levels to have a relevant impact, comparable to the statistical uncertainties at increasing angular scales of \maglimpp. \\
\\
In accordance with what is presented in the previous section in terms of 1-point statistics, we also evaluate the effect that the removal of observational systematic outliers has indirectly on the $w(\theta)$ once it is corrected from observational systematic effects. To illustrate this, on the right y-axis of Figure \ref{fig:delta_wtheta} we show the quantity 
\begin{equation}
    \Delta_{\rm weights}(\theta) = w_{\rm JM}^{\rm weights}(\theta) - w_{\rm JM}(\theta) \, , 
\end{equation}
which shows the effect of correcting observational systematics on \maglimpp with weights obtained with \isd \footnote{The systematic weights used for this test correspond to those obtained with \isd with a significance threshold $T_{\rm 1D} = 2$. Nevertheless, we note that the fiducial weights used for DES Y6 galaxy clustering and galaxy-galaxy lensing correspond to those from the \enet method. The difference between $w(\theta)$ measured on \maglimpp corrected with weights from both methods is taken as systematic contribution to the covariance, as we detail in \citet*{y6weights}. } after applying the \jointmask mask. We compute this correction from linear (dotted grey line) and cubic (dashed grey line) fits. We observe how observational systematics impact represents the dominant source of systematic signal in the correlation function with respect to the effect of removal of regions with extreme values of these contaminants. We also note that in all redshift bins but the first one the cubic weights yield values of $\Delta_{\rm weights} w(\theta)$ lower than or equal to those from the linear ones. Looking at the second redshift bin, the cubic fit yields a correction that reaching a $\sim 1\sigma$ difference relative to the unweighted case at large scales, while the linear fit results in a $\sim 0.5 \sigma$ difference. At the third bin we also note a relevant difference at intermediate angular scales. \\
\\
In the context of observational systematics decontamination, a lower $w(\theta)$ amplitude results from the detection and removal of additional contaminants. Of course, this can also arise from over-correction. To test this second possibility, in \citet*{y6weights} we run both \isd and \enet on a set of log-normal mocks with and without systematic contamination and determine the level of over-correction bias by comparing with reference mocks. The levels of over-correction found are negligible with respect to the statistical uncertainty, as it is detailed in Section V. B of the aforementioned companion paper. We refer the reader to that work for more information on the different variations of validation tests for the DES Y6 corrective weights. 
We conclude from those tests that the lower amplitudes from the cubic corrective weights seen in Figure \ref{fig:delta_wtheta} are due to them capturing more systematic signal and being able to produce weights that model and mitigate these effects better. As discussed in the previous section, masking extreme regions improves the detection of observational systematics and increases the accuracy of the corrections obtained from them. \\ 
\\

\section{Conclusions}\label{sec:conclusions}
In this work we present the procedure to create the DES Y6 \jointmask footprint, which introduces a novel approach by constructing it together hand-in-hand with the analysis to obtain corrective weights to mitigate observational systematic effects from the Y6 lens galaxy sample, \maglimpp (\citet*{y6weights}). We combine information from several inputs, including image artefacts, sample selection effects and most importantly spatially varying observational systematics for galaxy clustering, together with cuts from the definition of the DES Y6 weak lensing shape catalogue. This makes the DES Y6 \jointmask footprint jointly usable for both galaxy clustering and cosmic shear analyses. \\
\\
Our goal is to deliver a footprint that improves the detection and mitigation of contamination from observational systematic effects with our decontamination methods, namely \isd and \enet. In this regard, we place special emphasis on the presence of outlier values on all our template maps for systematic contamination, as a subset of 19 of these are the main inputs to both methods. We use two approaches to identify outliers on these maps: for the \texttt{CIRRUS\_SB\_MEAN} and \texttt{CIRRUS\_NEB\_MEAN} maps, we visually identify and correlate with image artefacts or Galactic cirrus, respectively; for the rest of the SP maps, we compute the inter-quantile range on their value distributions and remove pixels with values outside the range $[\rm Q_1 - a \times \mathrm{IQR}, \rm Q_3 + a \times \mathrm{IQR}]$, where $a = 3.5$. This value is chosen as a reasonable compromise that removes the most discrepant pixels while not decreasing the area of the footprint significantly. The combination of these cuts from each individual template map results in the \emph{SP outlier} mask. We identify as the main contributors to this mask, in terms of area loss with respect to the initial the \seed footprint, the \texttt{CIRRUS\_NEB\_MEAN} maps, the astrophysical foreground maps and the \texttt{BDF\_DEPTH} and \texttt{MAGLIM\_WMEAN} depth maps. The \emph{SP outlier} mask is then used to degrade our template maps to lower resolution versions on which we look for outlier values in the $N$-dimensional space formed by them (with $N = N_{\rm tpl} = 19$, corresponding to the main template maps used for weights). This yields the \emph{leverage} mask. \\
\\
Both the \emph{SP outlier} and \emph{leverage} masks are combined to yield the \systmask mask, which removes extreme observational systematic values defined at the 1-dimensional and 19-dimensional levels. This is done to address how such outlier values can impact the performance of \isd and \enet. The \systmask mask removes $169.80 \rm \, deg^2$ from the initial area of $4238.76 \rm \, deg^2$ given by the \seed footprint, therefore representing the main masking element of this analysis. \\ 
\\ 
In order to test the impact of the \jointmask mask (i.e. the mask associated with the \jointmask footprint) on the performance of our systematic decontamination methods, we focus on \isd, since the 1-dimensional metric of this method allows us for better evaluating the improvement on the detection of contaminants. We test the impact of applying the \jointmask mask to the \maglimpp sample by computing the reduced $\chi^2_{\rm model}$ values obtained from fitting the 1D relations of the subset of main template maps without (i.e. only with the \seed mask) and with this mask. In addition, we carry out these fits using a linear and a cubic form for the function describing the 1D-relation, $f(s)$. We find that the removal of outliers from the template maps results in an improvement in the characterisation of the 1D relations, with the cubic fits providing the best results. The consequence of upgrading the contamination metric of \isd to use cubic fits is to enhance its ability to identify more complex systematic contaminants, to which a linear function could be insensitive. In addition, we demonstrate that the \jointmask mask not only improves the detection of observational contaminants on the data, but also yields corrective weights obtained from the template maps that are less dispersed and more accurate. By restricting the range of systematic values, we ensure that we remain within the validity range of the perturbative form assumed for the contamination. \\
\\
Finally, we evaluate the impact on the measurement of the galaxy clustering angular correlation function from the \maglimpp sample. When evaluating this sample on two footprints with different levels of masking, namely the \seed footprint + \baselinemask mask and the \jointmask footprint, we find that the effect is to slightly reduce the $w(\theta)$ amplitude, except for the \jointmask case at the third bin, which is compatible with the effect of observational systematics being different from the \seed footprint + \baselinemask mask only case. We also note that for the last redshift bin we find that the effect of both masks is significant at $\sim 1\sigma$, especially at the largest angular scales. We determine that masking template map outliers has an indirect effect on $w(\theta)$ when correcting from observational systematic effects with weights, since removing such extreme values is precisely what allows to improve the detection of systematic contamination at the 1-point level. Therefore, enhancing the detection at this level results in a better recovery of the true galaxy clustering signal when correcting with weights. \\ 
\\
In summary, the DES Y6 \jointmask footprint and its associated mask, which accounts for different sources of systematic contamination on the \maglimpp sample, improve \texttt{ISD}’s flexibility, enhancing the detection and correction of observational systematics. In particular, from the point of view of \isd and given its characteristics and limitations, using the \jointmask mask makes the cubic fits usable, rendering this method and the results obtained with it more robust and reliable. We consider that the definition of masking strategies similar to the one presented in this work would also enhance the performance of other decontamination methods. Furthermore, using a common footprint for all DES Y6 analyses using simultaneously both lens and source galaxy samples facilitates the combination of data and the computation of covariances. \\
\\ 
Considering the fractional coverage of the high resolution pixels provided by the \seed footprint, the final \jointmask footprint that results from the combination of the \seed footprint with the \baselinemask and the \systmask masks covers an area of $4031.04 \, \rm deg^2$. This footprint is created at \nside = 16384 and later adapted to the resolutions required by the different Y6 analyses that make use of it. \\ 
\\
In the current context of machine learning-based methodologies, which offer far greater flexibility than classical techniques, we emphasise the critical need to monitor outliers caused by systematic effects, as such outliers can undermine both the performance and reliability of these advanced methods. This will be especially important for cosmological inference with upcoming LSS surveys, such as Euclid, LSST, and Roman, as shrinking statistical errors push us into a systematics-dominated regime.

\section*{Data Availability}
The main product of this work is the final footprint on which the different DES Y6 key analyses are conducted. This footprint is part of the data products that will be made available as part of the DES Y6 coordinated release at \url{https://des.ncsa.illinois.edu/releases} following publication of the DES Y6 Cosmology Results papers (\url{https://www.darkenergysurvey.org/des-y6-cosmology-results-papers/}). 

\section*{Acknowledgements}
\emph{Author's contributions:} We would like to acknowledge everyone who made this work possible. MRM developed the code to identify and cut observational systematic extreme values to create the \systmask mask, combined it with different footprints to produce the final \jointmask footprint, and tested the impact of these on \isd and $w(\theta)$. NW defined the cuts for the cirrus maps and the additional \texttt{DECaLS}-based LRG mask, and created the \emph{leverage mask}. MRM and NW co-led this research project. JEL created the log-normal mocks needed to run \isd and run tests on the masking impact. ISN, ACR and KB provided the \goldfoot footprint, applied the cuts from the faulty processing regions to this footprint and formatted the survey property maps. ADW and ACR provided essential information about the different footprints and masks used in this work and were internal reviewers. DA provided this project with information about leaked LSS on our template maps of contamination. AF contributed to the quality assessment of the initial \seed footprint. MG provided the \texttt{BFD\_BACKGROUND\_OFFSET} map used in this work and helped in paper writing. JMF contributed to the formatting of the SP maps used in this work. DSC performed validation tests on the survey property maps used to define the \systmask mask and provided the $w(\theta)$ covariance. MY provided the \emph{shear} footprint that was used to create the \seed footprint. AP, SA, MC and MRB coordinated, respectively, the Large-Scale Structure Working Group and the $3\times$2pt project, communicating with different teams to obtain the inputs needed for this work. The remaining authors have made contributions to this paper that include, but are not limited to, the construction of DECam and other aspects of collecting the data; data processing and calibration; developing broadly used methods, codes, and simulations; running the pipelines and validation tests; and promoting the science analysis.\\
\\
Funding for the DES Projects has been provided by the U.S. Department of Energy, the U.S. National Science Foundation, the Ministry of Science and Education of Spain, the Science and Technology Facilities Council of the United Kingdom, the Higher Education Funding Council for England, the National Center for Supercomputing Applications at the University of Illinois at Urbana-Champaign, the Kavli Institute of Cosmological Physics at the University of Chicago, the Center for Cosmology and Astro-Particle Physics at the Ohio State University,
the Mitchell Institute for Fundamental Physics and Astronomy at Texas A\&M University, Financiadora de Estudos e Projetos, Funda{\c c}{\~a}o Carlos Chagas Filho de Amparo {\`a} Pesquisa do Estado do Rio de Janeiro, Conselho Nacional de Desenvolvimento Cient{\'i}fico e Tecnol{\'o}gico and the Minist{\'e}rio da Ci{\^e}ncia, Tecnologia e Inova{\c c}{\~a}o, the Deutsche Forschungsgemeinschaft and the Collaborating Institutions in the Dark Energy Survey. \\
\\
The Collaborating Institutions are Argonne National Laboratory, the University of California at Santa Cruz, the University of Cambridge, Centro de Investigaciones Energ{\'e}ticas, Medioambientales y Tecnol{\'o}gicas-Madrid, the University of Chicago, University College London, the DES-Brazil Consortium, the University of Edinburgh, the Eidgen{\"o}ssische Technische Hochschule (ETH) Z{\"u}rich, Fermi National Accelerator Laboratory, the University of Illinois at Urbana-Champaign, the Institut de Ci{\`e}ncies de l'Espai (IEEC/CSIC), the Institut de F{\'i}sica d'Altes Energies, Lawrence Berkeley National Laboratory, the Ludwig-Maximilians Universit{\"a}t M{\"u}nchen and the associated Excellence Cluster Universe, the University of Michigan, NSF NOIRLab, the University of Nottingham, The Ohio State University, the University of Pennsylvania, the University of Portsmouth, SLAC National Accelerator Laboratory, Stanford University, the University of Sussex, Texas A\&M University, and the OzDES Membership Consortium.\\
\\
Based in part on observations at NSF Cerro Tololo Inter-American Observatory at NSF NOIRLab (NOIRLab Prop. ID 2012B-0001; PI: J. Frieman), which is managed by the Association of Universities for Research in Astronomy (AURA) under a cooperative agreement with the National Science Foundation.\\
\\
The DES data management system is supported by the National Science Foundation under Grant Numbers AST-1138766 and AST-1536171. Data access is enabled by Jetstream2 and OSN at Indiana University through allocation PHY240006: Dark Energy Survey from the Advanced Cyberinfrastructure Coordination Ecosystem: Services and Support (ACCESS) program, which is supported by U.S. National Science Foundation grants 2138259, 2138286, 2138307, 2137603, and 2138296.
The DES participants from Spanish institutions are partially supported by MICINN under grants PID2021-123012, PID2021-128989 PID2022-141079, SEV-2016-0588, CEX2020-001058-M and CEX2020-001007-S, some of which include ERDF funds from the European Union. IFAE is partially funded by the CERCA program of the Generalitat de Catalunya.\\
\\
We  acknowledge support from the Brazilian Instituto Nacional de Ci\^encia
e Tecnologia (INCT) do e-Universo (CNPq grant 465376/2014-2).\\
\\
This document was prepared by the DES Collaboration using the resources of the Fermi National Accelerator Laboratory (Fermilab), a U.S. Department of Energy, Office of Science, Office of High Energy Physics HEP User Facility. Fermilab is managed by Fermi Forward Discovery Group, LLC, acting under Contract No. 89243024CSC000002.

\bibliographystyle{mnras}
\bibliography{desy12,desy32,desy62,nondes2,software2}

\appendix

\section{Pixel fractional coverage and map degrading / upgrading}\label{app:degrading}
\begin{figure}
    \centering
    \includegraphics[width=\linewidth]{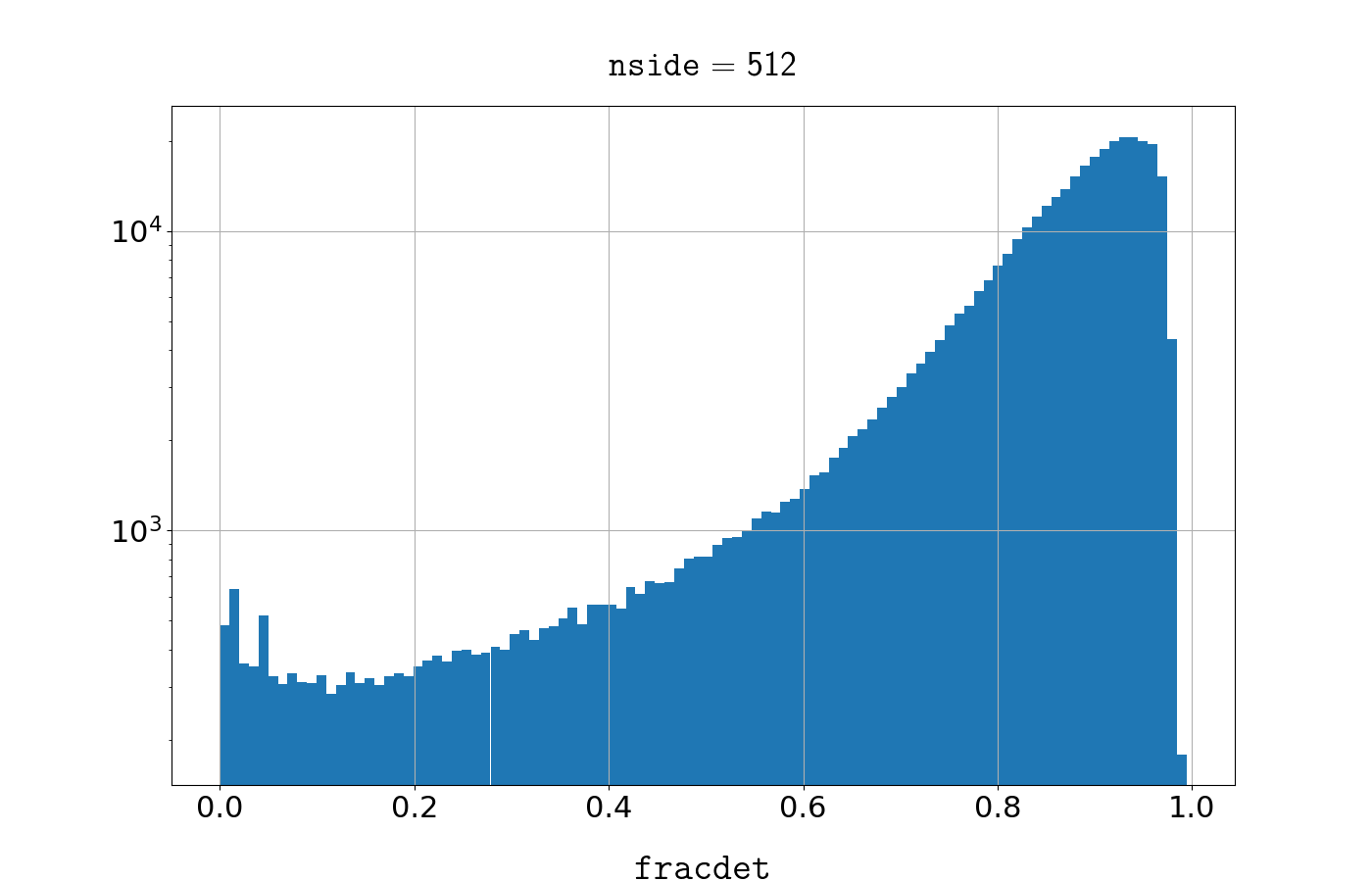}
    \caption{Distribution of \fracdet values at \nside = 512 for the \jointmask footprint. The area corresponding to pixels with values lower than 0.3 is $21.39 \rm \, deg^2$. }
    \label{fig:fracdet_dist}
\end{figure}

\begin{figure}
    \centering
    \includegraphics[width=\linewidth]{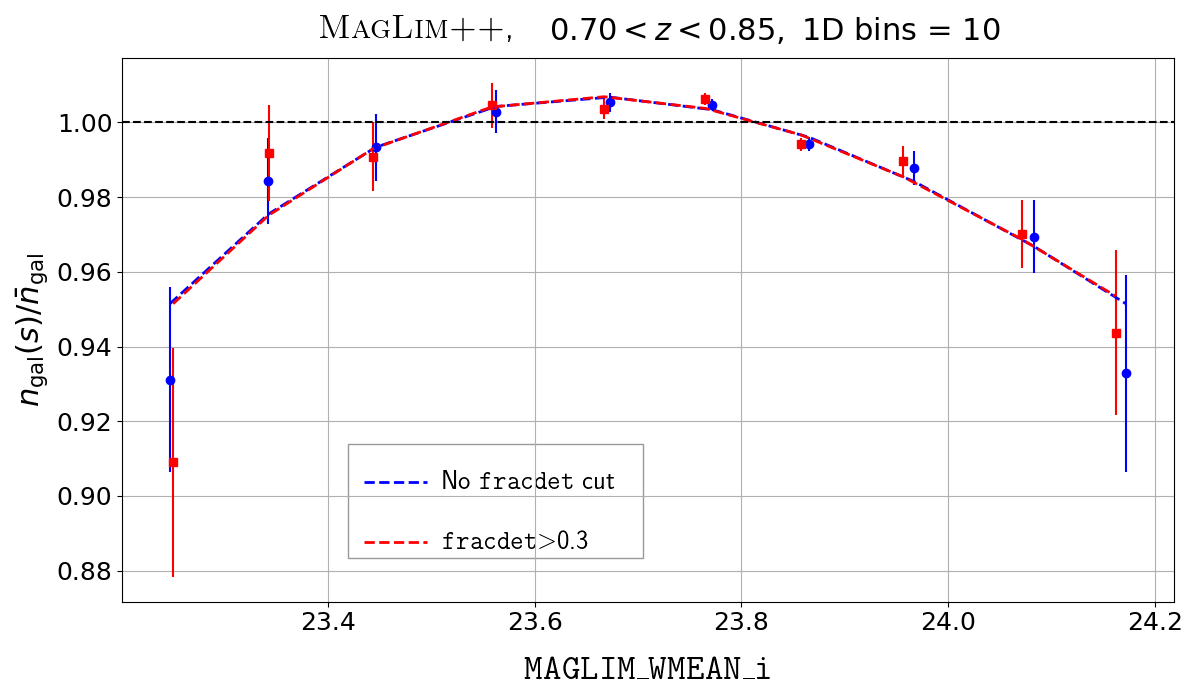}
    \caption{Impact of the \fracdet > 0.3 cut on the 1D relation of \texttt{MAGLIM\_WMEAN\_i}. The blue dots are obtained on the \jointmask footprint, while the red ones correspond to the 1D relation computed on the area with low fractional coverage pixels removed at \nside = 512. As it can be seen, the effect of the cut on the 1D coordinates is negligible.  }
    \label{fig:1d_fracdet_cut}
\end{figure}

\begin{figure*}
    \centering
    \includegraphics[width=\linewidth]{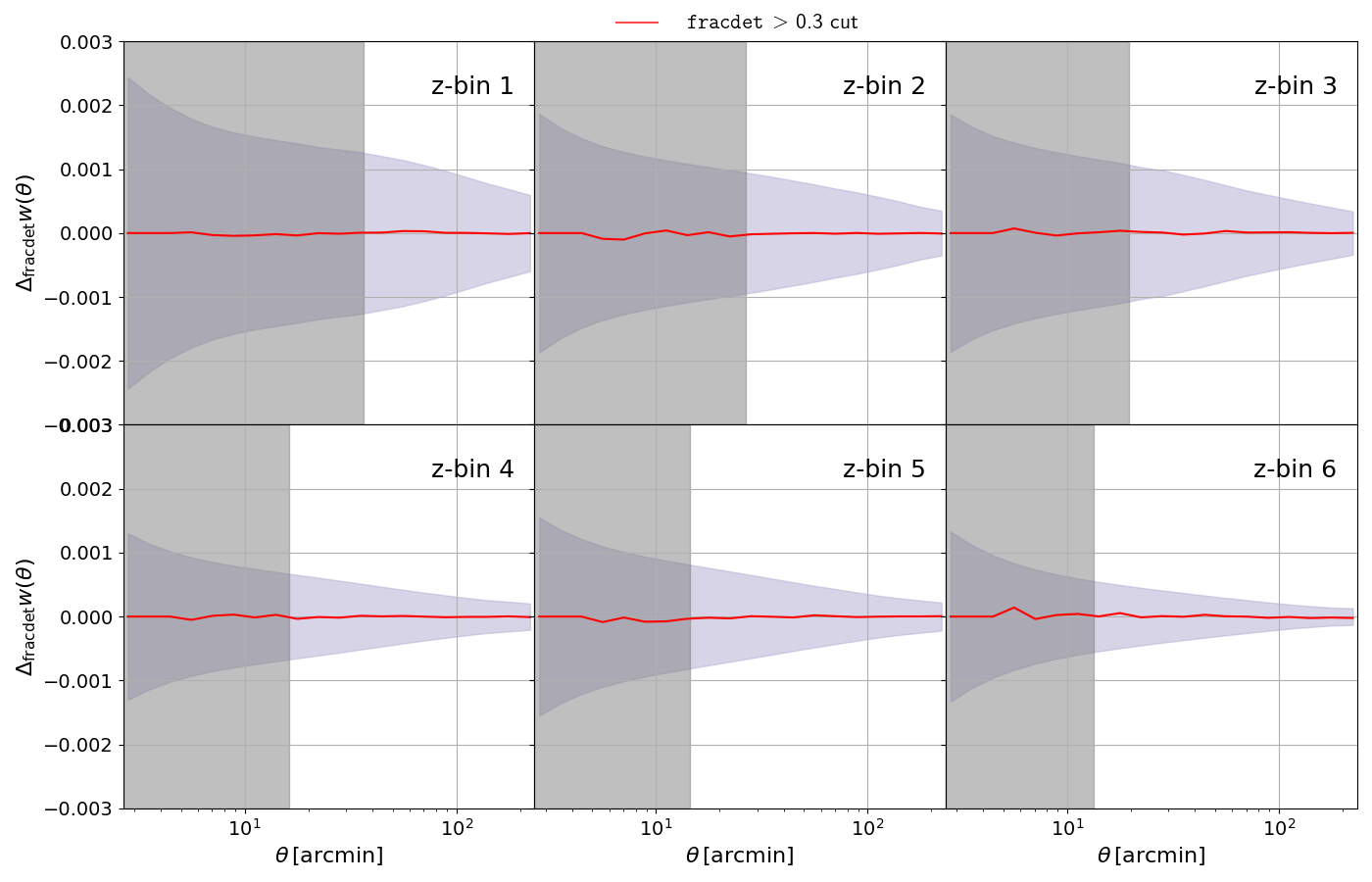}
    \caption{Difference between $w(\theta)$ measured on the \jointmask footprint and on the same footprint minus the pixels with \fracdet < 0.3. This correlation function is computed on the data with no systematic weights applied and at an \nside = 512. The blue shadowed region represents the errors of the diagonal from the galaxy clustering part of the DES Y6 covariance matrix. We determine that the quantity $\Delta_{\fracdet} w(\theta) = w_{\rm JM}(\theta) - w_{\fracdet>0.3}(\theta)$ is compatible with zero at all redshift bins within the statistical uncertainty. These results indicate that the fractional coverage is properly accounted for in the computation of the angular correlation function and that the effect of small \fracdet is negligible.}
    \label{fig:wtheta_fracdet_cut}
\end{figure*}

The fractional coverage or \fracdet of a pixel at a given \nside resolution is a value between 0 and 1 that corresponds to the fraction of that pixel that has been covered or observed by the survey. This quantity is important because it makes it possible to weigh or rescale the content values of the pixels, which is very relevant when computing statistics from them, such as two-point correlation functions. Ideally, working at very high \nside results in a finer covering of the footprint, but handling maps at such high resolutions poses computational problems, while also rendering such maps noisier. In addition, in order to avoid non-linearities and other modelling systematics, we apply a series of scale cuts for each correlation function of the \threextwo analysis that exclude angular scales smaller than those limits. The smallest scale cut for the \threextwo{} analysis corresponds to $\sim 3.7 \, \rm arcmin$, while for an \nside = 16384 the angular separation between pixel centres is $\sim 0.22 \, \rm arcmin$. We are therefore most concerned with scales larger than this. \\
\\ 
To obtain the \fracdet at a lower resolution, we start from a very high \nside pixelation, $N_{\rm high}$, of the observed sky region. At such high resolution, we consider that all pixels are fully observed, so we assume that their \fracdet is 1. Nevertheless, we need to note that this is an approximation, since the mask making procedure does not include image-level masks from bright stars, satellite streaks, cosmic rays, etc, which can result in lower fractional coverages, even at very high resolutions. Then, considering this approximation, the \fracdet of a lower resolution pixel, $f_i^{\rm low}$, is obtained by degrading from $N_{\rm high}$ as 
\begin{equation}\label{eqn:degrade_fracdet}
    f_i^{\rm low} = \frac{1}{(N_{\rm high}/N_{\rm low})^2} \sum_{j \in i} f_j^{\rm high} \, , 
\end{equation}
where $j$ runs over the high resolution sub-pixels contained within the $N_{\rm low}$ pixel $i$. In our case, we define our \fracdet starting from an \nside = 131072 version of the \seed footprint, which we then degrade to the native \nside resolutions of the template maps presented in Table \ref{tab:spmaps_table}, namely 16384, 4096, 2048 and 512. \\
\\
Then, in order to degrade a template map to lower resolutions, we need to assign the \fracdet at its native resolution to the corresponding pixels and then we do a weighted average as  
\begin{equation}\label{eqn:degrade_map}
    m_i^{\rm low} = \frac{\sum_{j \in i} m_j^{\rm high} \cdot f_j^{\rm high}}{(N_{\rm high}/N_{\rm low})^2 \cdot f_i^{\rm low}} \, . 
\end{equation}
\\
On the other hand, in some cases we need to upgrade some maps from a lower resolution, e.g., the Gaia stellar density map (see subsection \ref{sec:foregrounds}). In those cases, for each pixel $i$ of \nside = $n_{\rm low}$ with value $m_i^{\rm low}$, we assigned to all $(N_{\rm high}/N_{\rm low})^2$ sub-pixels $j$ at \nside = $n_{\rm high}$ contained within it the same value as at $n_{\rm low}$, that is $m_i^{\rm low}$. Since all sub-pixels have the same value, it is straightforward to verify that if we degrade this higher resolution map applying Equation \ref{eqn:degrade_map} to any \nside lower than it up to the initial $n_{\rm low}$, the resulting lower resolution pixel will have value $m_i^{\rm low}$. \\
\\
Regarding the \fracdet values, as mentioned in Section \ref{sec:impact_1ds}, we do not consider any cut for the fractional coverage at any \nside resolution. As stated in that section, the minimum value for \fracdet at \nside = 512 is $1.53 \cdot 10^{-5}$. The distribution of values can be seen in Figure \ref{fig:fracdet_dist}. In order to test the impact of this choice of not removing pixels based on their \fracdet value, we evaluate the 1D relations of different SP maps computed on the \jointmask footprint at \nside = 512 and on a version of it with a \fracdet cut that removes pixels with values lower than 0.3. This cut removes $21.39 \rm \, deg^2$ of area. As shown in Figure \ref{fig:1d_fracdet_cut}, the impact of this cut is negligible and it does not affect the shape of the 1D relation (we do not compute the $\chi^2$ values for the \fracdet$>0.3$ case because that would entail to apply the same cut to all log-normal mocks and recompute the $\Delta \chi^2(68)$ from all of them). In addition, we also evaluate the effect of this \fracdet cut on the computation of $w(\theta)$. In Figure \ref{fig:wtheta_fracdet_cut} we depict the quantity $\Delta_{\fracdet} w(\theta) = w_{\rm JM}(\theta) - w_{\fracdet>0.3}(\theta)$, computed on data with no systematic weights applied (\citet*{y6weights}) and at an \nside resolution of 512. As it can be seen, the effect of this cut is completely negligible compared with the statistical error from the covariance matrix (\citet*{y6modelling}).

\section{Outlier cuts from Galactic cirrus estimates and SP maps}\label{app:outliers}
\begin{figure}
    \centering
    \includegraphics[width=0.9\linewidth]{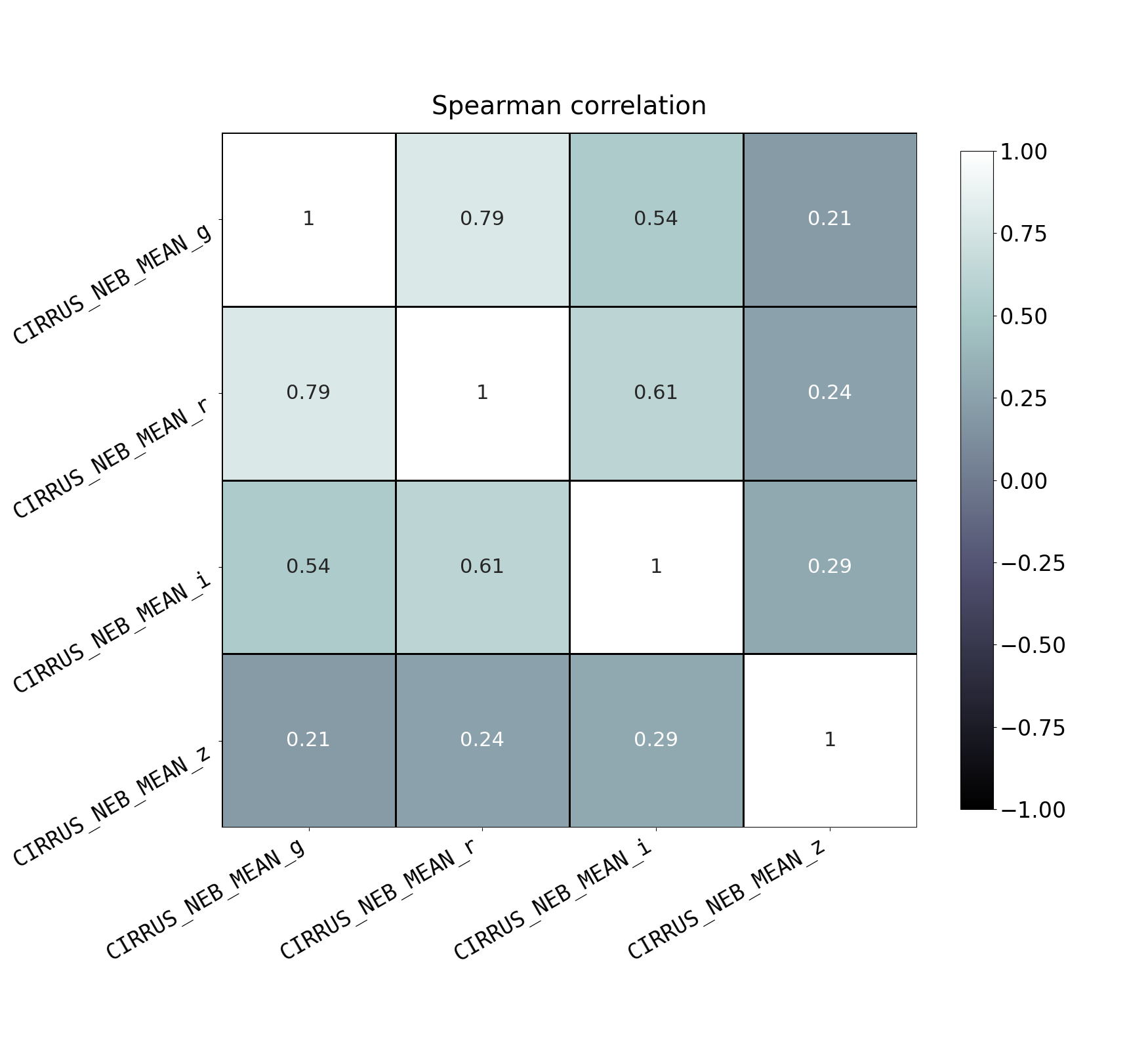}
    \vspace{-0.5cm}
    \caption{Spearman's correlation coefficient matrix for the \texttt{CIRRUS\_NEB\_MEAN} maps in the four DES Y6 photometric bands, \emph{griz}. Two blocks of correlations can be identified: one for the maps in \emph{gri}-bands and one for the \emph{z} band alone. }
    \label{fig:spearman_corr_cirrus}
\end{figure}

\begin{figure*}
    \centering
    \includegraphics[width=0.52\linewidth]{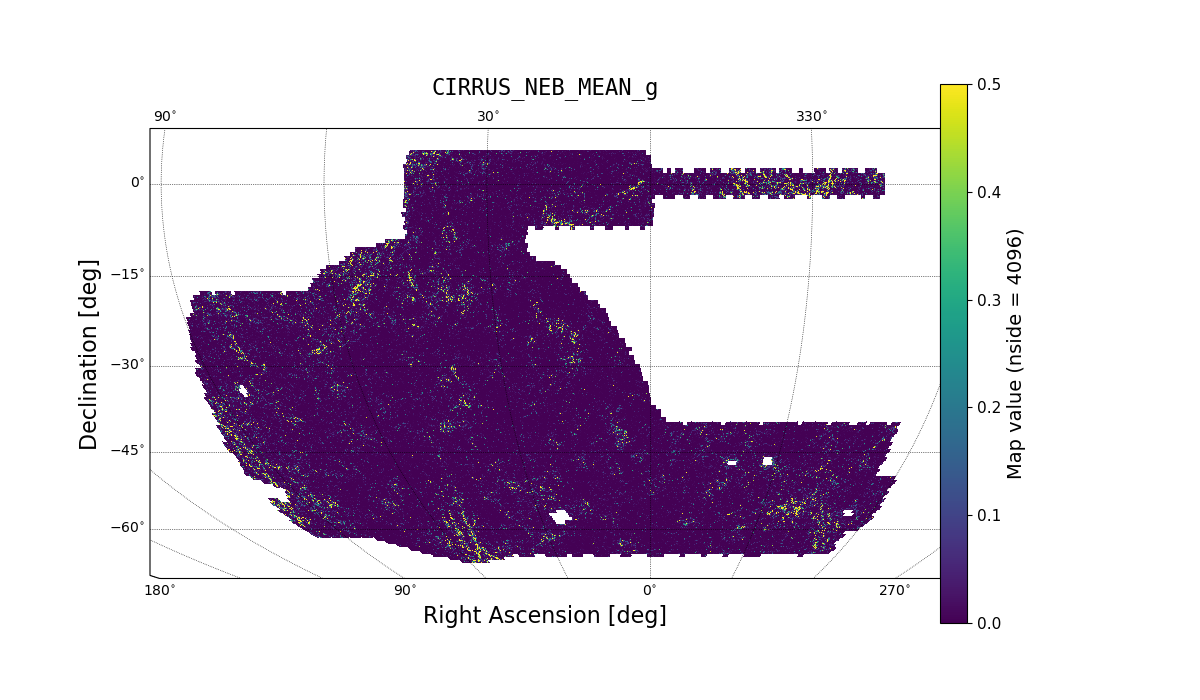}
    \hspace{-1.5cm}
    \includegraphics[width=0.52\linewidth]{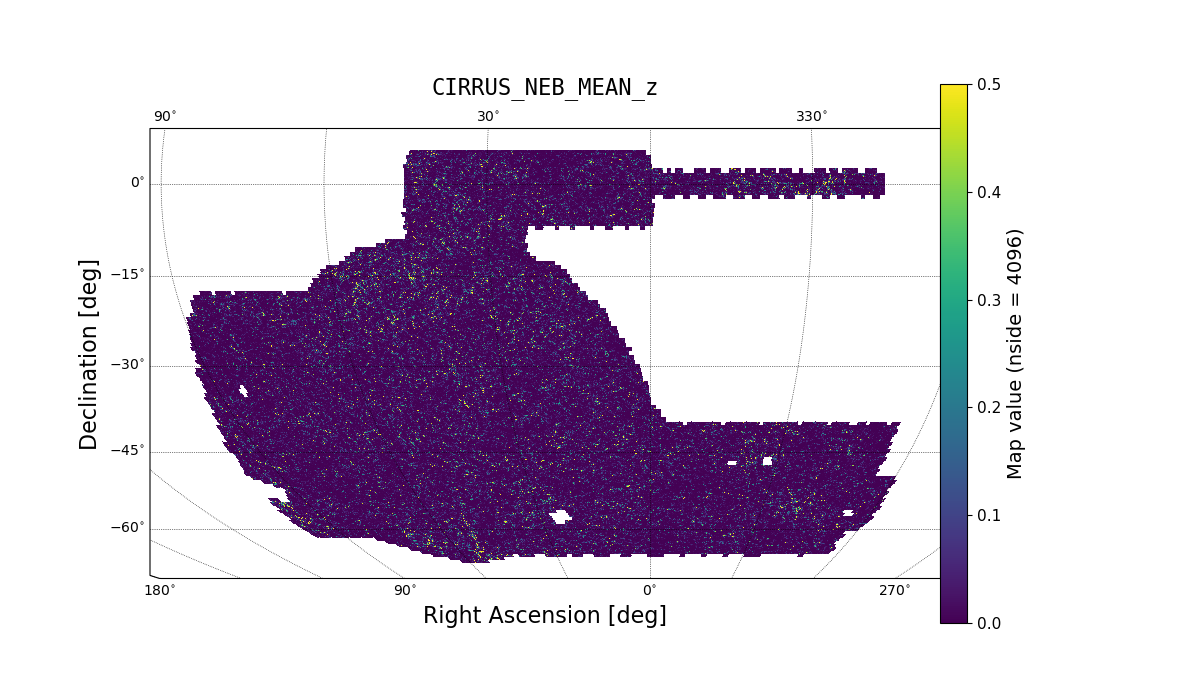}
    \vspace{-0.5cm}
    \caption{Sky projections of the \texttt{CIRRUS\_NEB\_MEAN\_g} (left) and \texttt{CIRRUS\_NEB\_MEAN\_z} (right) maps. It can be seen how the values greater than $0.5$ (yellow pixels) that we use to flag outliers are evenly scattered in \emph{z}-band, while the \emph{g}-band shows an additional concentration on the edges of the footprint that correlates with the most extreme values of \texttt{EBV\_CSFD23} (see upper plots of Figure \ref{fig:outlier_sky_maps}). }
    \label{fig:outlier_sky_maps_cirrus}
\end{figure*}

\begin{figure*}
    \centering 
    \includegraphics[width=0.49\linewidth]{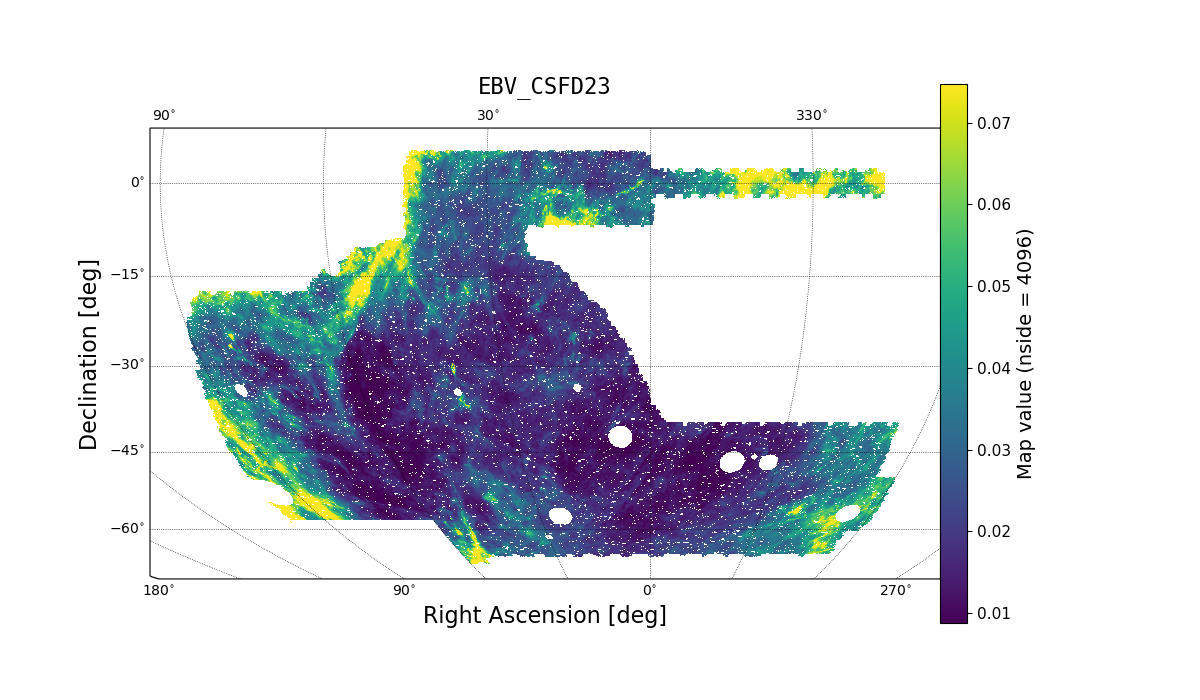}
    \hspace{-1.2cm}
    \includegraphics[width=0.49\linewidth]{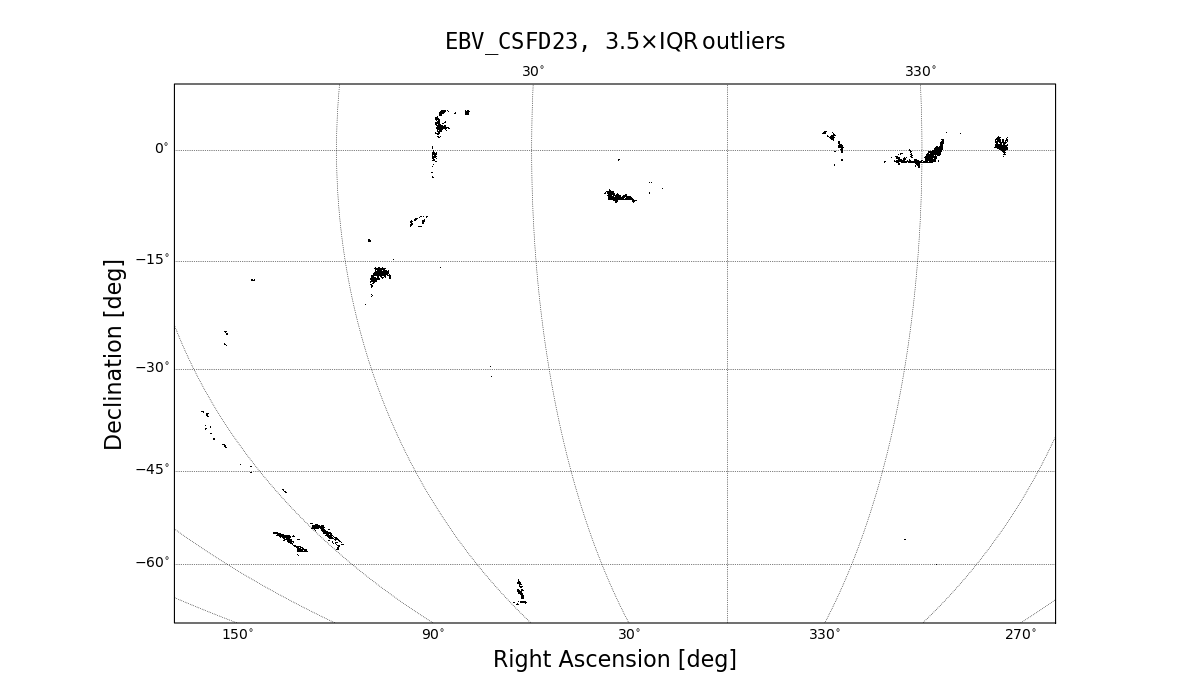}
    \includegraphics[width=0.49\linewidth]{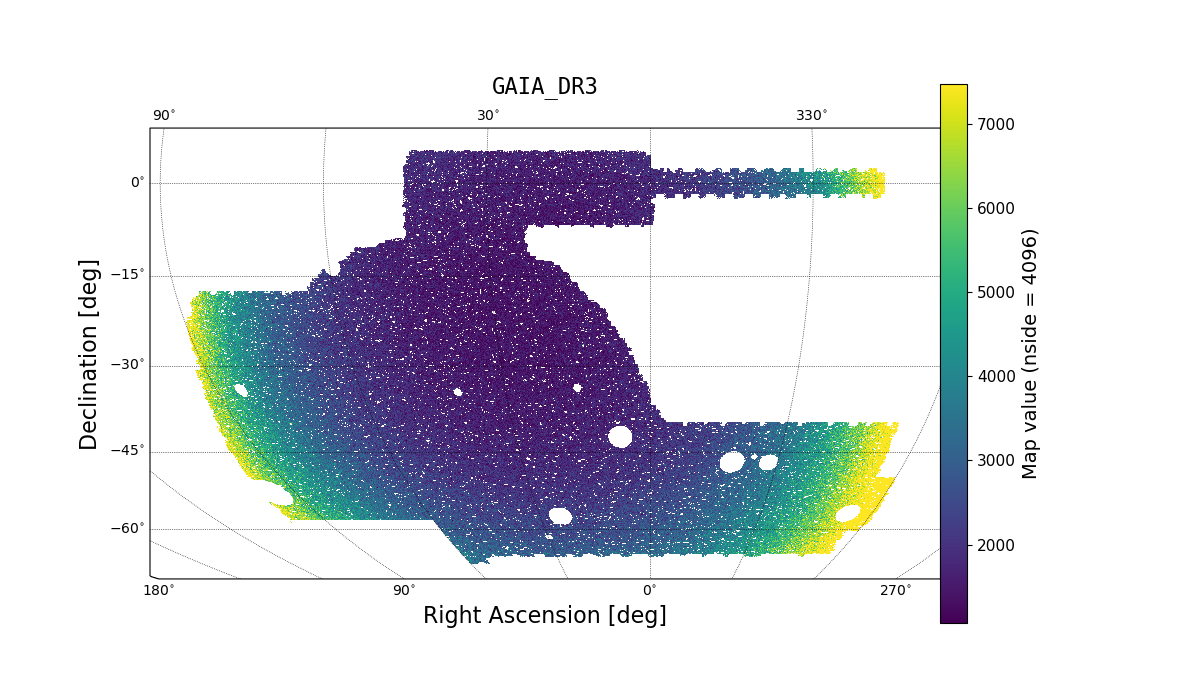}
    \hspace{-1.2cm}
    \includegraphics[width=0.49\linewidth]{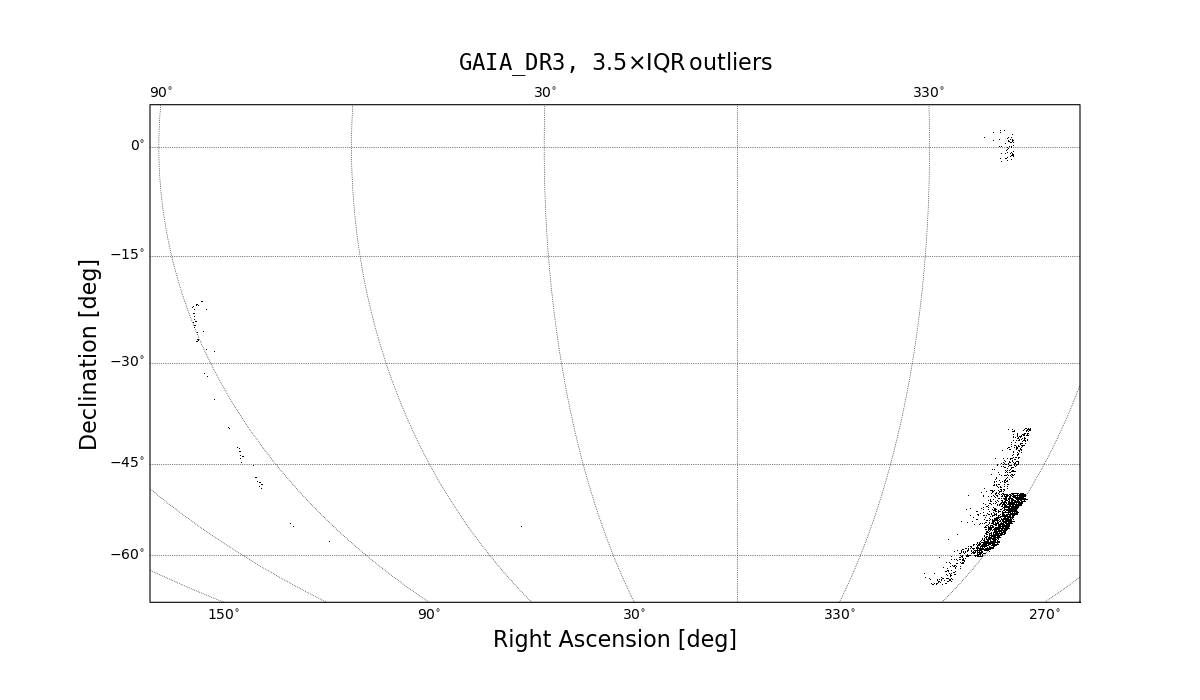} 
    \includegraphics[width=0.49\linewidth]{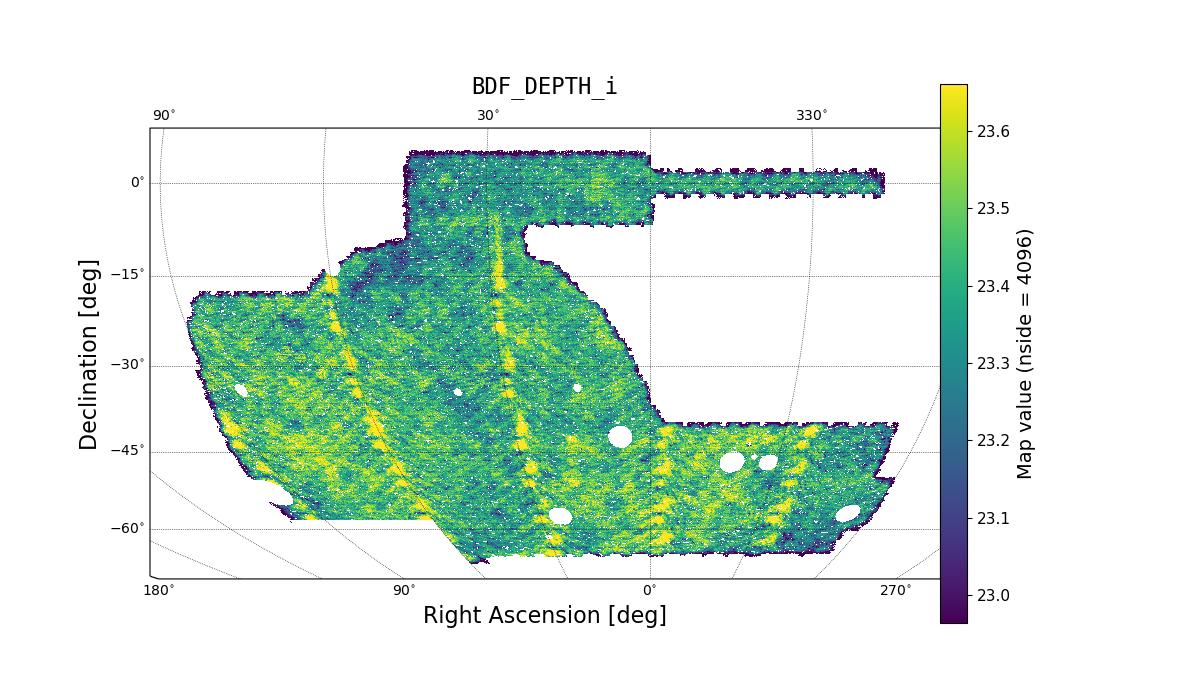}
    \hspace{-1.2cm}
    \includegraphics[width=0.49\linewidth]{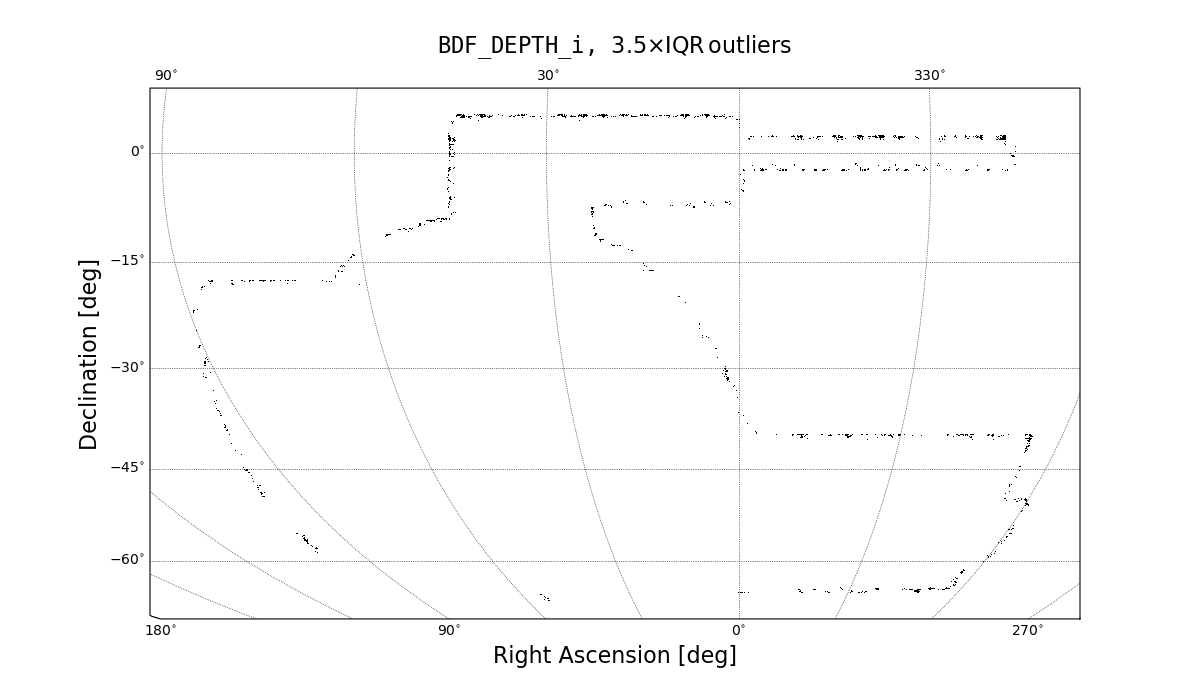}
    \includegraphics[width=0.49\linewidth]{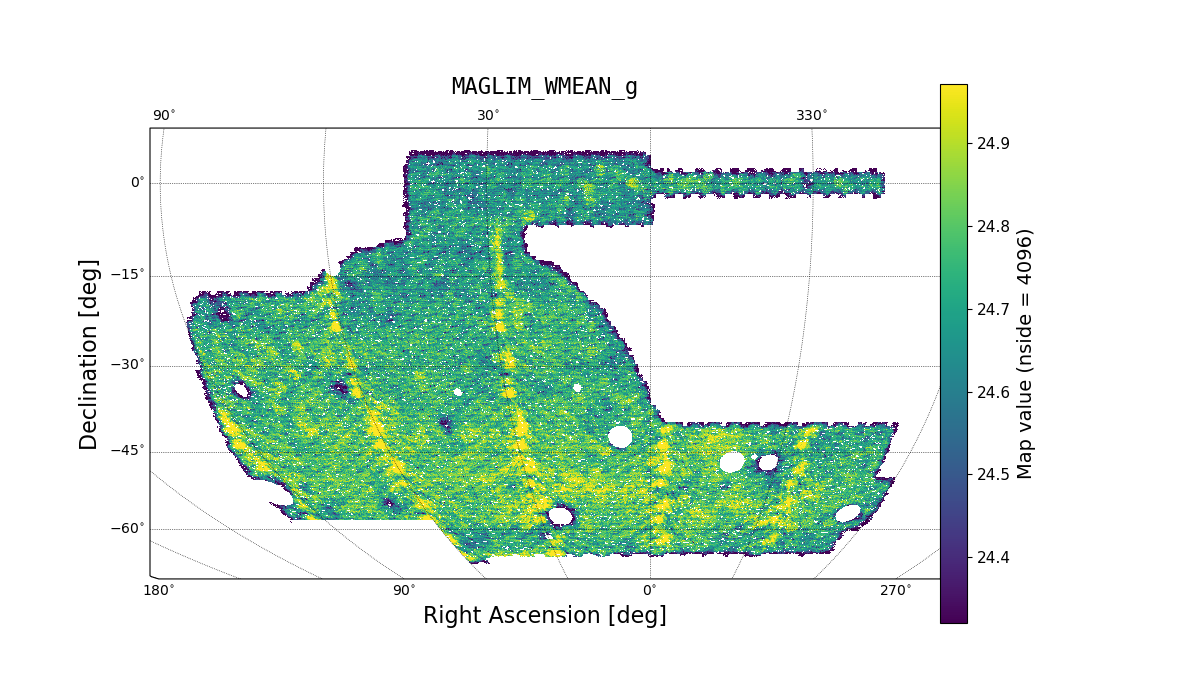}
    \hspace{-1.2cm}
    \includegraphics[width=0.49\linewidth]{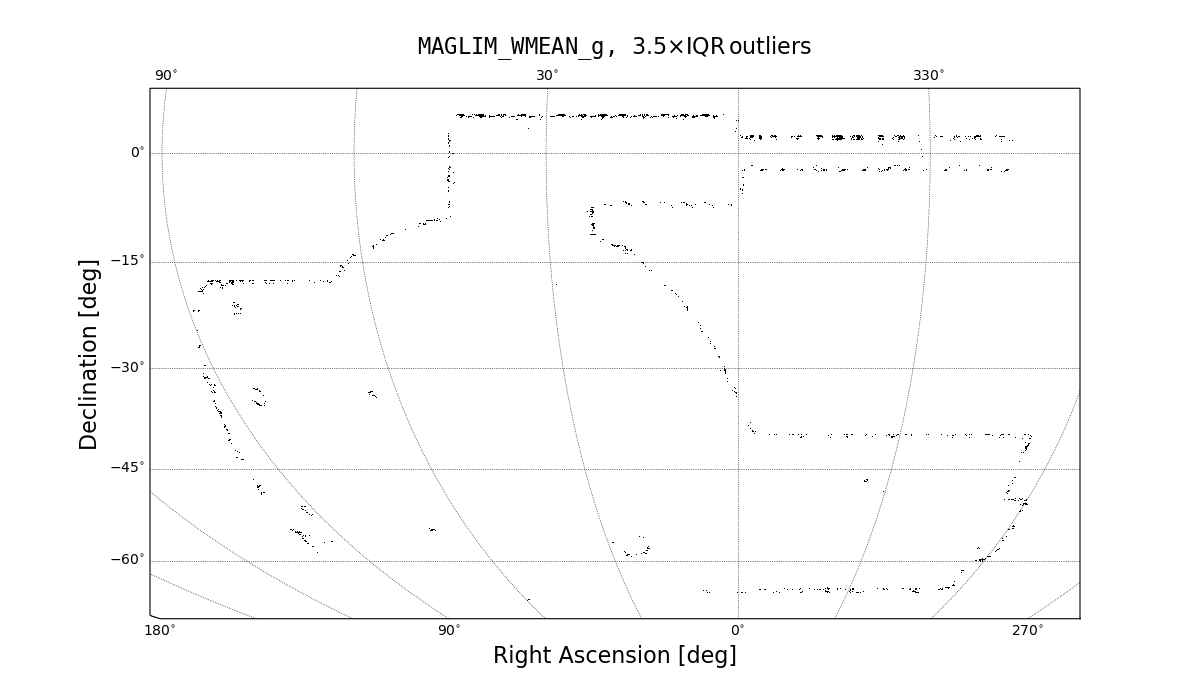} 
    \caption{Sky projections of the \texttt{EBV\_CSFD23}, \texttt{GAIA\_DR3}, \texttt{BDF\_DEPTH\_i} and \texttt{MAGLIM\_WMEAN\_g} maps (left) and positions on the footprint of their corresponding outlier values according to our $3.5\times \rm IQR$ criterion (right). In all cases, the outliers are located on the edges of the initial footprint. }
    \label{fig:outlier_sky_maps}
\end{figure*}

The \systmask mask is built by combining the \emph{SP outlier} and \emph{leverage} masks, with the former resulting from the combination of the \texttt{CIRRUS\_NEB\_MEAN}$>0.5$, the \texttt{CIRRUS\_SB\_MEAN}$>99.99\%$ and the SP map $3.5\times \rm IQR$ outlier cuts. As we show on the top left panel of Figure \ref{fig:syst_mask}, out of those, the main contributors to the \emph{SP outlier} mask are the cuts from \texttt{CIRRUS\_NEB\_MEAN} and  $3.5\times \rm IQR$, with \texttt{CIRRUS\_SB\_MEAN}$>99.99\%$ removing very little area with respect to the cuts from the other cirrus estimate maps. This is due to the high correlation between these two types of maps, as it can be seen in Figure \ref{fig:overlap_matrix} in terms of the overlap between the two masks derived from the cuts applied to them. Given its definition (see subsection \ref{sec:leverage}), we find similar conclusions for the additional area removed by the \emph{leverage} mask with respect to the $3.5\times \rm IQR$ outlier mask. \\
\\
Regarding the area removal from the individual \texttt{CIRRUS\_NEB\_MEAN} cuts in each photometric band, we identify two blocks: one formed by \emph{gri}-bands and the other by the \emph{z}-band. This is due to a higher spatial correlation between the first three bands compared with that with the redder one. We quantify this correlation in Figure \ref{fig:spearman_corr_cirrus}, where we show the Spearman correlation coefficient between the pixel values of the four \texttt{CIRRUS\_NEB\_MEAN} maps (we use this correlation coefficient to avoid relying on the assumption of linear correlation between maps). Looking at the sky projection of these maps in Figure \ref{fig:outlier_sky_maps_cirrus}, we find that the pixels with values $>0.5$ (yellow pixels), which are flagged as outliers according to our criterion for these maps, are evenly distributed on the footprint in the four bands, but for the first three they tend to concentrate on the edges, as we illustrate with the map in \emph{g}-band. Taken separately, the different cuts applied to \texttt{CIRRUS\_NEB\_MEAN} and \texttt{CIRRUS\_SB\_MEAN} remove $77.54 \rm \, deg^2$ from the initial area given by the \seed footprint. \\
\begin{figure*}
    \centering
    \includegraphics[width=\linewidth]{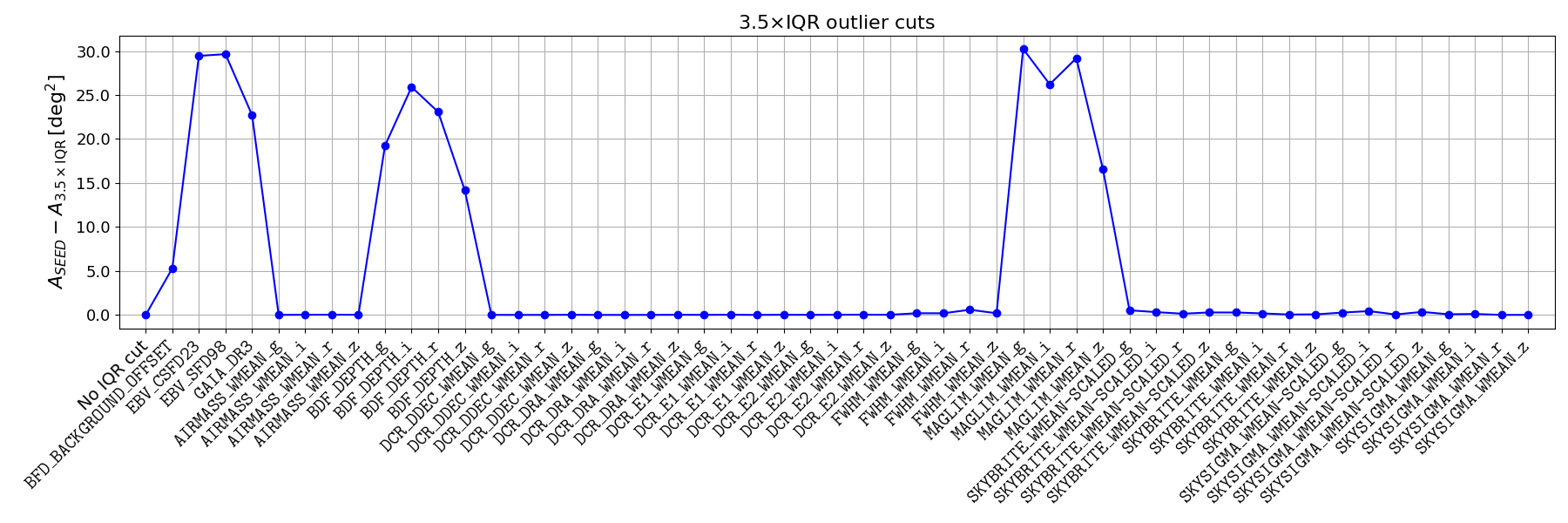}
    \vspace{-0.5cm}
    \caption{Area removed from the \seed footprint by each $3.5\times \rm IQR$ cut separately. The main contributors to area loss are the astrophysical foregrounds, the \texttt{BDF\_DEPTH} and the \texttt{MAGLIM\_WMEAN} maps. }
    \label{fig:iqr_individual}
\end{figure*}
\begin{figure}
    \centering
    \includegraphics[width=\linewidth]{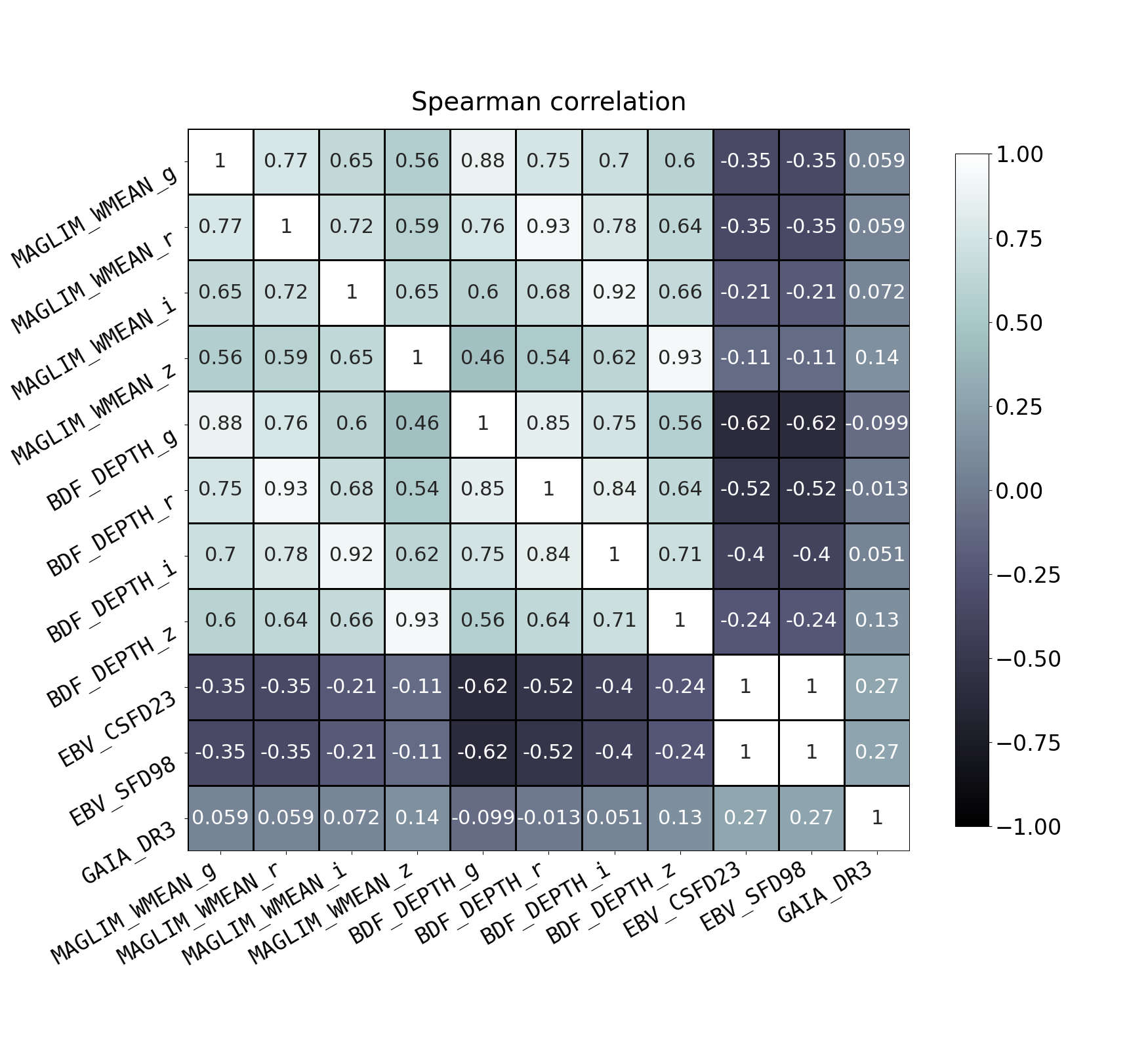}
    \vspace{-0.5cm}
    \caption{Spearman’s correlation coefficient matrix the astrophysical foregrounds, the \texttt{BDF\_DEPTH} and the \texttt{MAGLIM\_WMEAN} maps. The different survey depth maps show high, positive correlations, similar to the two Galactic dust extinction maps. }
    \label{fig:spearman_corr}
\end{figure}\\
Looking at the components of the $3.5\times \rm IQR$ outlier mask, we find in Figure \ref{fig:iqr_individual} that the main contributors are the astrophysical foreground maps, namely \texttt{EBV\_CSFD23}, \texttt{EBV\_SFD98} and \texttt{GAIA\_DR3}, and the maps for the survey's depth, \texttt{BDF\_DEPTH} and \texttt{MAGLIM\_WMEAN}, in \emph{griz}-bands. Taken separately from the other masks, the area removed from the \seed footprint by the $3.5\times \rm IQR$ outlier mask is $100.45 \rm \, deg^2$. This value is smaller than the sum of the individual $3.5\times \rm IQR$ cuts from each SP map. This is due to the different levels of spatial correlation between them. We quantify this by means of the Spearman correlation matrix shown in Figure \ref{fig:spearman_corr}, where it can be seen that there are high, positive correlations between the pairs of maps within the \texttt{BDF\_DEPTH} and the \texttt{MAGLIM\_WMEAN} families. For a given photometric band, the pairs \texttt{BDF\_DEPTH} - \texttt{MAGLIM\_WMEAN} show even higher correlations, given that both quantities are different estimates of the same property at the same band. In the case of \texttt{EBV\_CSFD23} and \texttt{EBV\_SFD98} maps, they are completely correlated. We also note the mild anti-correlations between both Galactic dust extinction maps and the depth maps, which is expected. In the different panels of Figure \ref{fig:outlier_sky_maps} we showcase the sky projection of some of these maps (left) together with the positions of the sky of the pixels flagged as outliers according to our $3.5\times \rm IQR$ criterion. In all cases, the outliers are located on the edges of the initial footprint, closer to the Galactic plane (i.e. where the presence of dust and stars is more important) and where reached depth is swallower. \\ 
\\
Accounting for all these contributions and the correlation / overlap between them, once the \emph{SP outlier} and the \emph{leverage} masks are combined the resulting \systmask mask removes $169.80 \rm \, deg^2$ from the \seed footprint, making it the main masking element of this analysis. 

\section{Bias on $w(\theta)$ due to LSS masking on the \texttt{CIRRUS} maps}\label{app:cirrus_mask}
\begin{figure*}
    \centering
    \includegraphics[width=\linewidth]{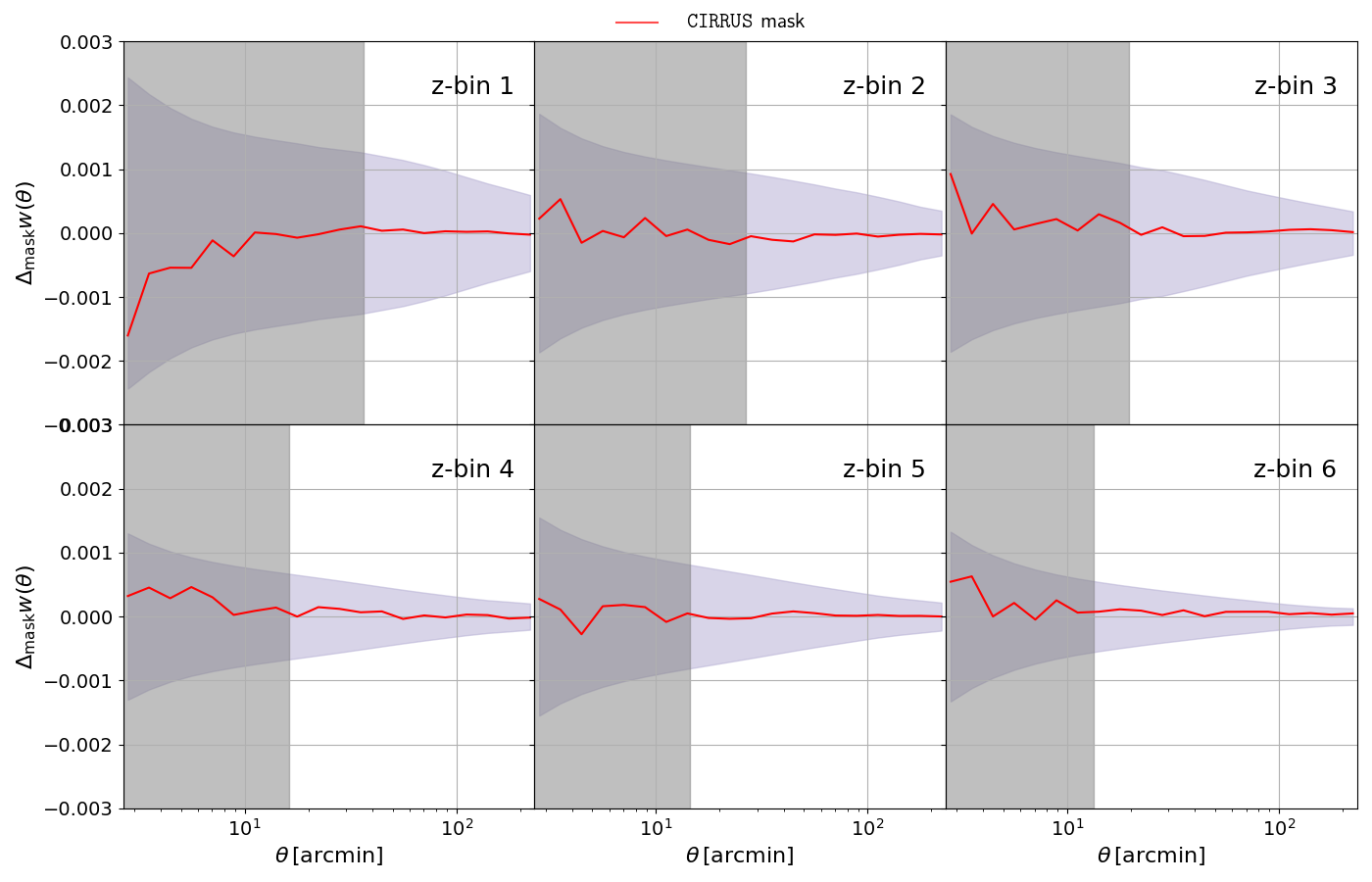}
    \caption{Difference between $w(\theta)$ measured on the \seed footprint and on the \seed footprint + \texttt{CIRRUS} mask. The blue shadowed region represents the errors of the diagonal from the galaxy clustering part of the DES Y6 covariance matrix. We find that $\Delta_{\rm mask} w(\theta) = w_{\rm SEED+CIRRUS}(\theta) - w_{\rm SEED}(\theta)$ is compatible with zero at all redshift bins within the statistical uncertainty. This indicates that the masking of extreme \texttt{CIRRUS} values that may trace real LSS does not significantly bias the measurement of $w(\theta)$. }
    \label{fig:wtheta_cirrus_mask}
\end{figure*}

As mentioned in Section \ref{sec:syst_mask}, the \texttt{CIRRUS} maps may contain actual LSS, so we exclude them from our list of template maps to obtain corrective weights from. On the other hand, we use them for masking by removing their most extreme values. However, if these maps do trace real LSS, masking based on their values could bias the measurement of $w(\theta)$ due to the removal of such LSS. To study this potential issue, we a) create a CIRRUS mask that contains the cuts only from these maps, b) apply it to the SEED footprint and c) evaluate the difference in $\Delta w(\theta) = w_{\rm SEED+CIRRUS}(\theta)-w_{\rm SEED}(\theta)$. These results are shown in Figure \ref{fig:wtheta_cirrus_mask}. As it can be seen, this $\Delta w(\theta)$ is compatible with zero for all redshift bins within the statistical errors and for the angular scales of interest. The blue shadowed region depicts the errors from the diagonal of the galaxy clustering part of the DES Y6 covariance matrix (\citet*{y6modelling}). Using the full covariance matrix and excluding the angles below the scale cuts (grey shadowed regions on the Figure), we obtain $\chi^2 / N_{\rm dof}$ values of 0.020, 0.049, 0.043, 0.115, 0.016 and 0.085 with $N_{dof} = 8, 10, 11, 12, 12, 13$ for each redshift bin, respectively. We conclude that the masking of extreme values from the \texttt{CIRRUS} maps does not bias the measurement of $w(\theta)$, even if they trace real LSS, at least to the level of masking considered in this work. 

\section{Contaminating SP maps according to \isd}\label{app:chi2_model_vals}
\begin{figure}
    \centering
    \includegraphics[width=\linewidth]{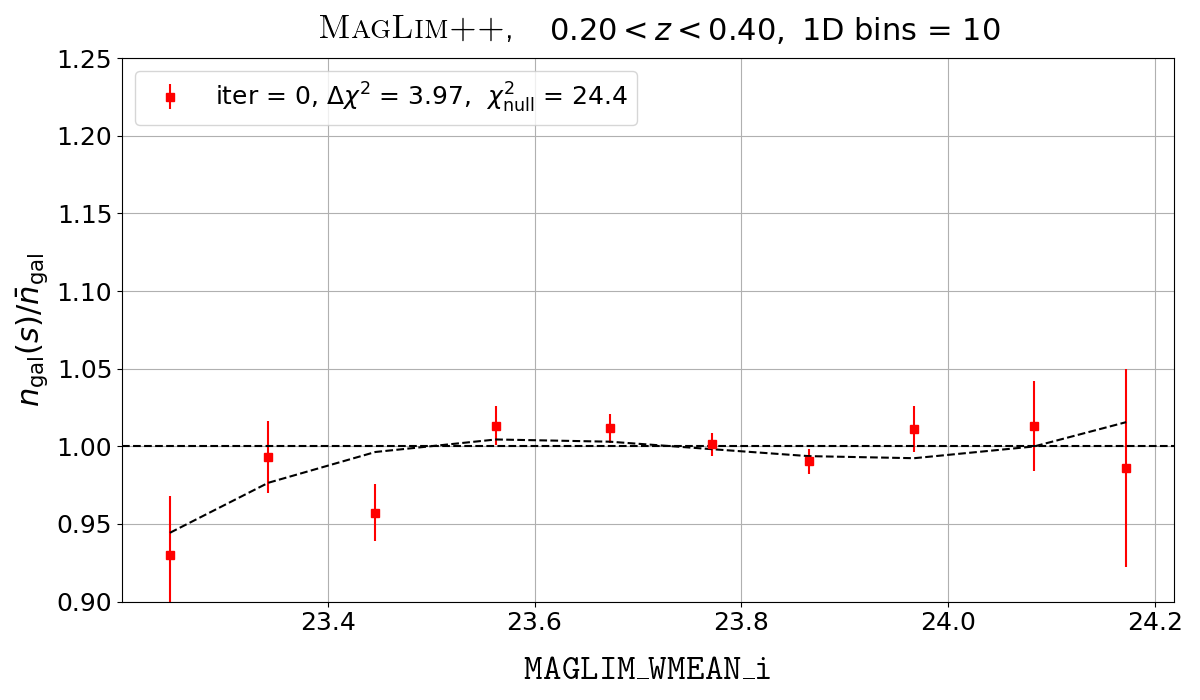}
    \includegraphics[width=\linewidth]{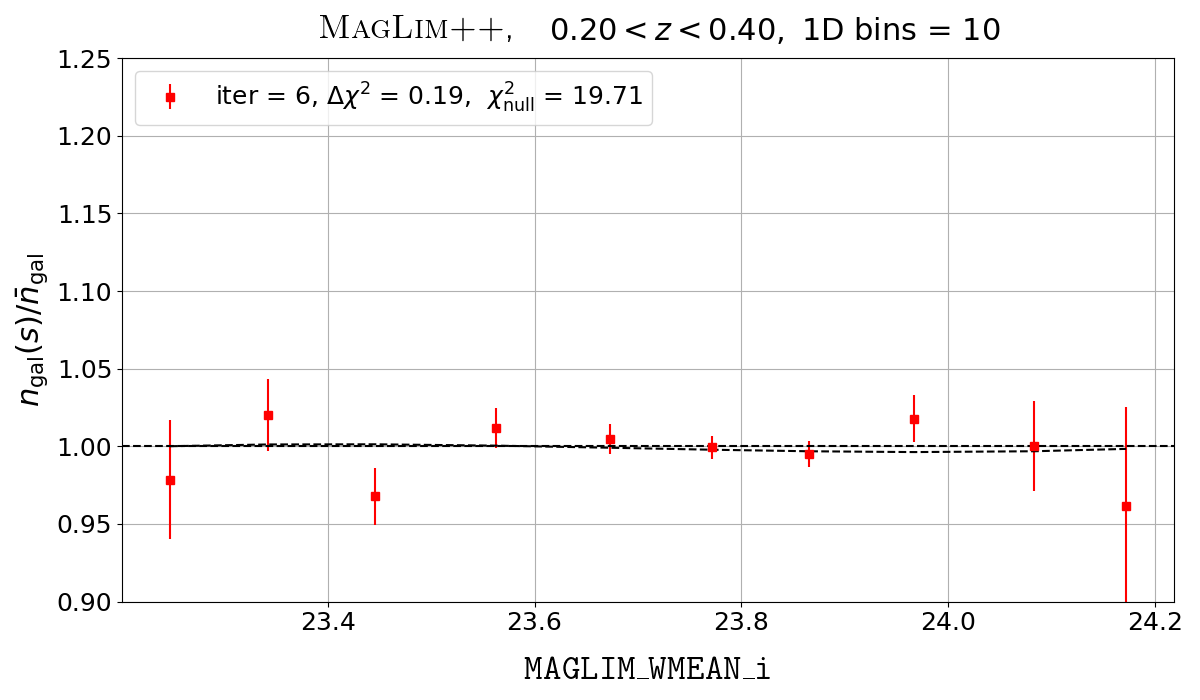}
    \caption{\emph{Top:} 1D relation of \texttt{MAGLIM\_WMEAN\_i} at iteration 0 of \isd, that is before applying any correction. The trend is properly fitted by a cubic function except for the third 1D bin. \emph{Bottom:} 1D relation of the same map after correcting for it. In our fiducial run of \isd this map is not flagged as contaminant for a threshold of 2, being necessary to go to $T_{1D} = 1$ to find it significantly contaminant. }
    \label{fig:1d_maglim_i}
\end{figure}

In Section \ref{sec:impact_1ds} we discuss how the usage of the \seed and \jointmask masks affects the fitting of the SP maps and the computation of the $\chi^2_{\rm model}$. For the first four cases of Figure \ref{fig:chi2_boxplot}, we do not consider the covariance between 1D bins, therefore these $\chi^2_{\rm model}$ values should be seen as a qualitative estimation of the true ones. On the other hand, for the last two cases we do take into account the 1D covariances as in a fiducial \isd run. There, we found that cubic fits seems to have more outliers than the linear case. In particular, there is one map, \texttt{MAGLIM\_WMEAN\_i}, which lies beyond the $\rm Q3 + 1.5IQR$ limit depicted by the whiskers when evaluated at the first redshift bin of \maglimpp. Looking at the 1D relation of this map at iteration 0, that is, when no corrective weights have been applied, we see in the upper panel of Figure \ref{fig:1d_maglim_i} that its third 1D bins triggers its higher $\chi^2_{\rm model}$ value. However, the significance of the impact of this map is very low, being 5\% at maximum, as it can be seen in the figure. In fact, this map is not flagged as significant by \isd for a threshold of $T_{1D} = 2$, as we show in Table \ref{tab:spmaps_weighted}. Nevertheless, it is necessary to weigh for \texttt{MAGLIM\_WMEAN\_i} if the threshold was 1 (this value was considered during the validation process of the corrective weights in \citet*{y6weights}, but was discarded as it leads to over-correction), so we show its 1D relation after applying corrective weights for it. It can be seen how the third 1D bin is mildly corrected, while the rest of them are compatible with one.

\begin{table*}[]
    \centering
    \begin{tabular}{c|c}
        Redshift bin & SP maps - Cubic fits, $T_{1D} = 2$ \\
        \hline \hline
        0.20 < z < 0.40 & \texttt{FWHM\_WMEAN\_z} \\ \hline
        0.40 < z < 0.55 & \texttt{SKYSIGMA\_WMEAN\_r}, \texttt{AIRMASS\_WMEAN\_r}, \texttt{GAIA\_DR3} \\ \hline
        0.55 < z < 0.70 & \texttt{GAIA\_DR3}, \texttt{FWHM\_WMEAN\_r}, \texttt{MAGLIM\_WMEAN\_r}, \texttt{SKYSIGMA\_WMEAN\_g}, \texttt{AIRMASS\_WMEAN\_i} \\ \hline
        0.70 < z < 0.85 & \texttt{FWHM\_WMEAN\_z}, \texttt{EBV\_CSFD}, \texttt{FWHM\_WMEAN\_r}, \texttt{MAGLIM\_WMEAN\_i} \\ \hline
        0.85 < z < 0.95 & \texttt{GAIA\_DR3}, \texttt{MAGLIM\_WMEAN\_r}, \texttt{FWHM\_WMEAN\_z} \\ \hline
        0.95 < z < 1.05 & \texttt{GAIA\_DR3}, \texttt{MAGLIM\_WMEAN\_i}, \texttt{AIRMASS\_WMEAN\_g}, \texttt{FWHM\_WMEAN\_i}, \texttt{FWHM\_WMEAN\_z} \\ \hline
        \hline
        \\
          & SP maps - Linear fits, $T_{1D} = 2$ \\
        \hline \hline
        0.20 < z < 0.40 & \texttt{MAGLIM\_WMEAN\_g}, \texttt{SKYSIGMA\_WMEAN\_i}, \texttt{FWHM\_WMEAN\_z}, \texttt{FWHM\_WMEAN\_g} \\ \hline
        0.40 < z < 0.55 & \texttt{FWHM\_WMEAN\_i}, \texttt{AIRMASS\_WMEAN\_r}, \texttt{EBV\_CSFD}, \texttt{MAGLIM\_WMEAN\_r} \\ \hline
        0.55 < z < 0.70 & \texttt{GAIA\_DR3}, \texttt{MAGLIM\_WMEAN\_r}, \texttt{FWHM\_WMEAN\_z} \\ \hline
        0.70 < z < 0.85 & \texttt{FWHM\_WMEAN\_z}, \texttt{EBV\_CSFD}, \texttt{FWHM\_WMEAN\_r}, \texttt{MAGLIM\_WMEAN\_z}, \texttt{GAIA\_DR3}, \texttt{AIRMASS\_WMEAN\_z}, \texttt{MAGLIM\_WMEAN\_g} \\ \hline
        0.85 < z < 0.95 & \texttt{GAIA\_DR3}, \texttt{EBV\_CSFD}, \texttt{FWHM\_WMEAN\_g}, \texttt{FWHM\_WMEAN\_z} \\ \hline
        0.95 < z < 1.05 & \texttt{GAIA\_DR3}, \texttt{MAGLIM\_WMEAN\_i}, \texttt{FWHM\_WMEAN\_z}, \texttt{AIRMASS\_WMEAN\_g}, \texttt{EBV\_CSFD}, \texttt{GAIA\_DR3} \\ \hline
    \end{tabular}
    \caption{List of SP maps that \isd finds it has to correct for with a significance threshold of $T_{1D} = 2$ and cubic (upper part) and linear (lower part) fits. As it can be seen, \texttt{MAGLIM\_WMEAN\_i} is not among the maps to obtain weights from at the first redshift bin for this threshold. We also note how in the linear case there are redshift bins where it is necessary to weigh for more maps than in the cubic case. This happens because the cubic fits are more efficient at capturing contamination, while the linear ones make the \isd method to iterate more to achieve convergence. For more information, see \citet*{y6weights}. }
    \label{tab:spmaps_weighted}
\end{table*}

\end{document}